\documentclass[preprint,12pt,authoryear]{elsarticle} 

\usepackage{color}
\usepackage{fullpage}
\usepackage{amsmath}
\usepackage{mathrsfs}
\usepackage{newfloat}
\usepackage{lineno}
\usepackage{subfig}
\usepackage{multirow}
\usepackage{caption}
\usepackage{arydshln}
\usepackage{tikz}
\usepackage{pstool}
\usepackage[hidelinks]{hyperref}
\hypersetup{colorlinks,bookmarksopen,linkcolor=blue,citecolor=blue,urlcolor=blue}

\newcommand{\bs}[1]{{\boldsymbol{#1}}}
\newcommand{\sign}{\mathrm{sign}}
\newcommand{\ignore}[1]{}
\DeclareFloatingEnvironment{algorithm}

\usepackage{amssymb}

\makeatletter
\def\ps@pprintTitle{
 \let\@oddhead\@empty
 \let\@evenhead\@empty
 \def\@oddfoot{\footnotesize\itshape Computer Methods in Applied Mechanics and Engineering \hfill\today}
 \let\@evenfoot\@oddfoot}
\makeatother

\begin{document}

\begin{frontmatter}



\title{eXtended Variational Quasicontinuum Methodology for Lattice Networks with Damage and Crack Propagation\tnoteref{t1}}


\author[affiliation1]{O. Roko\v{s}\corref{mycorrespondingauthor}}
\cortext[mycorrespondingauthor]{Corresponding author, presently at Department of Mechanical Engineering, Eindhoven University of Technology, P.O. Box~513, 5600~MB Eindhoven, The Netherlands.}
\ead{o.rokos@tue.nl}

\author[affiliation2]{R.H.J. Peerlings}

\author[affiliation1,affiliation3]{J. Zeman}

\address[affiliation1]{Department of Mechanics, Faculty of Civil Engineering, Czech Technical University in Prague, Th\'{a}kurova~7, 166~29 Prague~6, Czech Republic.}

\address[affiliation2]{Department of Mechanical Engineering, Eindhoven University of Technology, P.O. Box~513, 5600~MB Eindhoven, The Netherlands.}

\address[affiliation3]{Department of Decision-Making Theory, Institute of Information Theory and Automation, Czech Academy of Sciences, Pod Vod\'{a}renskou v\v{e}\v{z}\'{i}~4, 182~08 Prague~8, Czech Republic}

\tnotetext[t1]{This is the accepted version of the following article: O. Roko\v{s}, R.H.J. Peerlings, and J. Zeman, eXtended variational quasicontinuum methodology for lattice networks with damage and crack propagation, Comput. Methods in Appl. Mech. Engrg.~320 (2017) 769--792, DOI: \href{http://www.sciencedirect.com/science/article/pii/S0045782516315997}{10.1016/j.cma.2017.03.042}. This manuscript version is made available under the \href{https://creativecommons.org/licenses/by-nc-nd/4.0/}{CC-BY-NC-ND 4.0} license.}

\begin{abstract}
Lattice networks with dissipative interactions are often employed to analyze materials with discrete micro- or meso-structures, or for a description of heterogeneous materials which can be modelled discretely. They are, however, computationally prohibitive for engineering-scale applications. The (variational) QuasiContinuum~(QC) method is a concurrent multiscale approach that reduces their computational cost by fully resolving the (dissipative) lattice network in small regions of interest while coarsening elsewhere. When applied to damageable lattices, moving crack tips can be captured by adaptive mesh refinement schemes, whereas fully-resolved trails in crack wakes can be removed by mesh coarsening. In order to address crack propagation efficiently and accurately, we develop in this contribution the necessary generalizations of the variational QC methodology. First, a suitable definition of crack paths in discrete systems is introduced, which allows for their geometrical representation in terms of the signed distance function. Second, special function enrichments based on the partition of unity concept are adopted, in order to capture kinematics in the wakes of crack tips. Third, a summation rule that reflects the adopted enrichment functions with sufficient degree of accuracy is developed. Finally, as our standpoint is variational, we discuss implications of the mesh refinement and coarsening from an energy-consistency point of view. All theoretical considerations are demonstrated using two numerical examples for which the resulting reaction forces, energy evolutions, and crack paths are compared to those of the direct numerical simulations.
\end{abstract}


\end{frontmatter}



%
%
\section{Introduction}
\label{Sect:Introduction}
The mechanical response of materials with discrete micro- or meso-structures such as 3D-printed structures, woven textiles, paper, or foams can be modelled with dissipative lattice networks, cf. e.g.~\cite{RidruejoFiberGlass,LiuPaper,KulachenkoPaper,BeexTextile,Bosco:2015:Paper2,Bosco:2015:Paper}. The main advantage of these models consists in their conceptual simplicity, because lattice springs or beams can be identified as individual fibres or yarns of the underlying structure. Therefore, material parameters such as Young's or hardening moduli, and constitutive damage or plasticity laws can be determined in a relatively straightforward manner. Furthermore, lattice networks incorporate large deformations and yarn reorientations rather easily compared to phenomenological continuum models, see e.g.~\cite{Peng:2005}. For heterogeneous cohesive-frictional materials such as concrete, lattice networks are capable of capturing distributed microcracking, the heterogeneity at the microscale, and size effects; applications and further discussions can be found, e.g., in~\cite{SCHLANGENconcrete,Cusatis:2006,GrasslConcreteLattice,Elias:2015}.

The main drawback of lattice structures is their considerable computational cost---which may well be prohibitive for engineering applications, in which the lattice spacing is generally several orders of magnitude smaller than the problem size. Ergo, multiscale, or reduced-order modelling methods are required to tackle realistic applications.

A reduced-order modelling method with a concurrent multiscale character is the QuasiContinuum~(QC) method. It was originally designed for conservative atomistic systems at the nano-scale level, see~\cite{Tadmor:1996:QAD} for its initial formulation and~\cite{Curtin:2003:ACC,MillTad2002,MillTad2009,Iyer:QC:2011,Luskin:QC:2013} for various extensions. This approach was further generalized to tackle materials at the meso-scale level by introducing dissipation along with internal variables in a virtual-power-based format by~\cite{BeexDisLatt,BeexFiber} and in a variational format by~\cite{VarQCDiss,VarQCDamage}. In its essence, the QC methodology resolves the underlying lattice only in small regions of interest, whereas it coarsens it elsewhere. This results in a considerable reduction of the number of Degrees Of Freedom~(DOFs), internal variables, and effort to construct the governing equations.

\begin{figure}
	\centering
	\includegraphics[scale=1]{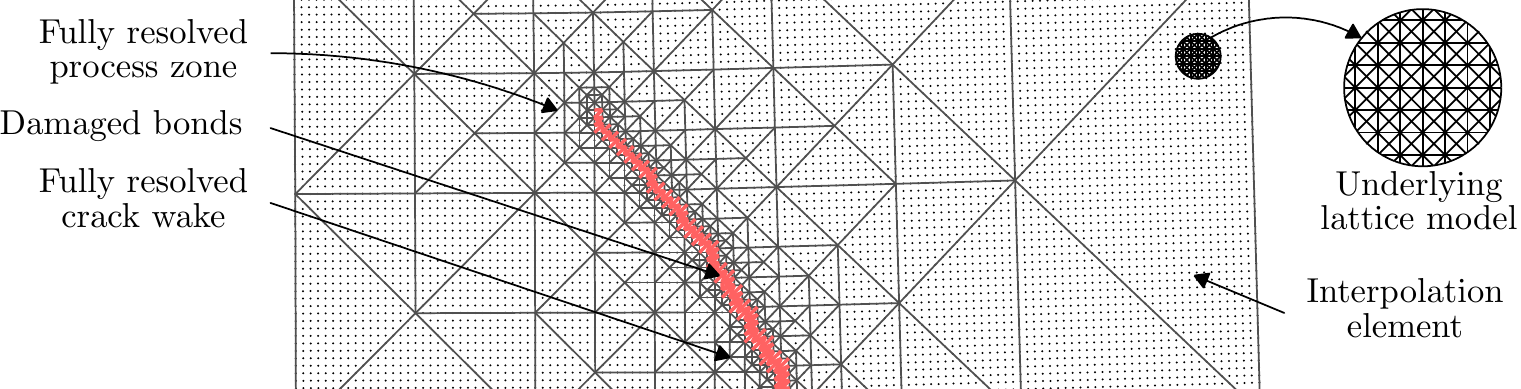}
	\caption{Sketch of a crack propagating in a damageable lattice using an adaptive QC method that only includes mesh refinement. The lattice at the crack tip is fully resolved because large fibre deformations and dissipation occur there. Elsewhere, the displacements are locally homogeneous and no damage occurs, allowing for efficient coarsening and large elements.}
	\label{Fig:motivation}
\end{figure}

The problem of interest in this contribution is the combination of the QC method and crack propagation in damageable lattice structures, as depicted in Fig.~\ref{Fig:motivation}. Because most of the dissipation and fibre reorientation occurs near the crack tip, this region must be fully resolved. To this end, a mesh indicator that follows the evolution of the crack tip needs to be provided. Such an approach typically leaves a trail of the fully resolved region in the wake of the crack, cf.~\cite{VarQCDamage}. It may be clear that, apart from allowing the crack to open up, these fully resolved regions do not contribute to the overall physics and accuracy. It is therefore desirable to coarsen them.  However, this coarsening should take into account the effect which the crack has on the local, coarse-scale kinematics of the problem---i.e. it should allow an arbitrary amount of crack opening without any mechanical resistance.

The aim of this contribution is to develop an extended version of the variational QC methodology, building thereby an efficient approach to model "strong discontinuities" in discrete systems with fully-nonlocal representations. This is possible because the damage in the crack wake is highly localized, and can be treated as a "discontinuity", although the system itself is fully discrete. One can introduce, therefore, techniques known from the Extended Finite Element Method~(XFEM), \citep{Belytschko:XFEM:1999,Moes:XFEM:1999}, or Generalized Finite Element Method~(GFEM), \citep{Strouboulis:GFEM:2000,Strouboulis:GFEM:2000a}, that rely on the Partition of Unity~(PU) concept~\citep{MELENK:1996,Babuska:1997}. In particular, the space of interpolation functions is enriched to include jumps in the kinematics as well as in the internal variables. Such an enrichment requires the development of several theoretical concepts and generalizations to our previous adaptive version of the variational QC methodology~\cite{VarQCDamage}, which was designed for materials with a higher degree of distributed cracking where effective coarsening is not possible. First, an appropriate geometric definition of crack path, identified through the state variable of a lattice system, is required. Second, cracks are geometrically described in terms of the level set function, which in turn provides the means to build special enrichments capturing the kinematics in the wakes of cracks. In order to sample the incremental energy of the system accurately, a suitable summation rule needs to be provided. Finally, to obtain energy-consistent solutions, certain energy quantities must be treated in a special way due to the coarsening. As will be explained below, and demonstrated in the examples section, the developed methodology offers an accurate yet efficient framework to model cracks in damageable lattices. In what follows, this approach will be referred to as the \emph{eXtended QC} (in analogy to continuum systems), or \emph{X-QC} for short.

Several previous works have focused on generalizations of the QC methodology incorporating XFEM-type of enrichments as well, cf.~\cite{QC:XFEM:Belytschko,Aubertin:QCXFEM:2009,Aubertin:QCXFEM:2010,QC:XFEM:Bordas}. Nevertheless, these works mainly dealt with conservative atomistic lattice models within the scope of molecular dynamics and the bridging domain method (concurrent coupling between the atomistic and continuum regions). In contrast to these previous works, this contribution focuses on dissipative QC methodologies, specifically designed for dissipative structural lattice networks at the meso-scale level. Moreover, our approach relies on the fully non-local formulation, meaning that no continuum is used and that the developed framework is fully discrete.

The paper is organized as follows. In Section~\ref{Sect:VarForm}, we briefly recall the theory of rate-independent systems on which the variational QC is built. For simplicity, only time-discrete versions of all equations are provided. Having specified the theoretical background, we can introduce the model at hand, its geometry, state variables, and energies driving the evolution. For simplicity, our considerations are confined to 2D, although entire framework extends to 3D as well. The main developments of this contribution, i.e. the interpolation and summation QC steps revisited from the point of view of lattice structures with propagating cracks, are discussed in Section~\ref{Sect:QC}. In particular, the enrichment functions together with the required changes in the summation rule are introduced in detail. The numerical solution of the resulting governing equations is not discussed as it can be found elsewhere, cf.~\cite{VarQCDamage}, Section~4. Instead, we directly proceed to examples in Section~\ref{Sect:Examples}, where the additional efficiency of the proposed methodology is demonstrated. The results show that the X-QC approach allows one to reduce the number of DOFs to approximately 25~-- 75\,\% of that of an adaptive QC, with no significant increase in error. The number of DOFs reduces to 1~-- $15 \, \%$ of that of the Direct Numerical Simulations (DNS). The corresponding computing times are reduced by a factor of 4~-- 20, depending on the particular example. The paper closes with a summary and conclusions in Section~\ref{Sect:Conclusion}.
%
%
\section{Variational Formulation of Lattice Networks with Localized Damage}
\label{Sect:VarForm}
In this section, the basic principles of the variational formulation of rate-independent systems in the context of lattice networks are recalled. For the sake of brevity, we limit ourselves to the time-discrete setting; the general theory is discussed in~\cite{MieRou:2015}, whereas applications to continuous systems can be found, e.g., in~\cite{MuhlhausAifantis,HanReddy,BourdinVar,Levitas,MultPlastMie,Bourdin:2007,Bourdin:VarBook:2008,burke_adaptive_2010,StabPhaMarMaur,Hofacker2012,JiZe:Damage,Mesgarnejad:2015}. Applications to lattice networks and QC are presented in~\cite{VarQCDiss,VarQCDamage} for isotropic hardening plasticity and localized damage, respectively.
%
%
\subsection{General Considerations}
\label{SubSect:GenCon}
An admissible configuration of a rate-independent system of interest is specified by a state variable~$\widehat{\bs{q}} = (\widehat{\bs{r}}, \widehat{\bs{z}})$, where~$\widehat{\bs{r}}$ stores the kinematic variables and~$\widehat{\bs{z}}$ the internal variables. All admissible configurations are specified by the state space~$\mathscr{Q} = \mathscr{R} \times \mathscr{Z}$, $\widehat{\bs{r}} \in \mathscr{R}$, $\widehat{\bs{z}} \in \mathscr{Z}$, which incorporates, e.g., prescribed displacements or constraints on damage variables.

The energetic solution~$\bs{q}$ is determined using the following incremental minimization problem at time~$t_k$
\begin{equation}
\bs{q}(t_k) \in \underset{\widehat{\bs{q}}\in\mathscr{Q}}{\mbox{arg min }}
\Pi^k(\widehat{\bs{q}};\bs{q}(t_{k-1})), \quad k = 1, \ldots, n_T,
\tag{IP}
\label{IP}
\end{equation}
with an initial condition~$\bs{q}(0)=\bs{q}_0$, where~$0=t_0<t_1<\dots<t_{n_T} = T$ describes a discrete time horizon~$[0, T]$, $\Pi^k$ denotes an incremental energy of the form
\begin{equation}
\Pi^k(\widehat{\bs{q}};\bs{q}(t_{k-1}))=\mathcal{E}(t_k,\widehat{\bs{q}})+\mathcal{D}(\widehat{\bs{z}},\bs{z}(t_{k-1})),
\tag{IE}
\label{IE}
\end{equation}
and the inclusion sign~$\in$ indicates that the potential~$\Pi^k$ is in general nonsmooth or may have multiple minima. The incremental energy consists of the potential (Gibbs type) energy~$\mathcal{E} : [0, T] \times \mathscr{Q} \rightarrow \mathbb{R}$, and the dissipation distance~$\mathcal{D} : \mathscr{Z} \times \mathscr{Z} \rightarrow \mathbb{R}^+ \cup \{+\infty\}$.

The dissipation distance~$\mathcal{D}(\bs{z}_2, \bs{z}_1)$ measures the minimum dissipation by a continuous transition between two consecutive states~$\bs{z}_1$ and~$\bs{z}_2$; for further details see~\cite{MieRou:2015}, Section~3.2. The potential energy reads
\begin{equation}
\mathcal{E}(t_k,\widehat{\bs{q}}) = \mathcal{V}(\widehat{\bs{q}}) - \bs{f}_\mathrm{ext}^\mathsf{T}(t_k) \widehat{\bs{r}},
\label{Sect:VarForm:Eq:3}
\end{equation}
where~$\mathcal{V} : \mathscr{Q} \rightarrow \mathbb{R}$ is the internal free energy, $\bs{f}_\mathrm{ext} : [0,T] \rightarrow \mathscr{R}^*$ represents the column matrix with external loads, and~$\mathscr{R}^*$ is the space dual to~$\mathscr{R}$.

Each time the minimization problem~\eqref{IP} is solved in this contribution, a local minimum that satisfies the energy balance
\begin{equation}
\mathcal{V}(\bs{q}(t_k)) + \mathrm{Var}_{\mathcal{D}}(\bs{q}; 0, t_k) = \mathcal{V}(\bs{q}(0)) + \mathcal{W}_\mathrm{ext}(\bs{q};0, t_k), \quad k = 1, \ldots, n_T,
\tag{E}\label{E}
\end{equation}
is searched. The energy balance~\eqref{E} equates the internally stored energy plus the \emph{dissipated energy}
\begin{equation}
\mathrm{Var}_{\mathcal{D}}(\bs{q};0,t_k) =  \sum_{l=1}^{k}\mathcal{D}(\bs{z}(t_{l}),\bs{z}(t_{l-1}))
\label{Sect:VarForm:Eq:1}
\end{equation}
with the work performed by the external forces
\begin{equation}
\mathcal{W}_\mathrm{ext}(\bs{q};0,t_k) = \sum_{l=1}^{k}\frac{1}{2}[\bs{f}(t_l)+\bs{f}(t_{l-1})]^\mathsf{T}[\bs{r}(t_l)-\bs{r}(t_{l-1})].
\label{Sect:VarForm:Eq:2}
\end{equation}
In Eqs.~\eqref{E}, \eqref{Sect:VarForm:Eq:1}, and~\eqref{Sect:VarForm:Eq:2}, the symbol~$(\bs{q};0,t_k)$ indicates the dependence on~$\bs{q}(t_l)$ for~$l = 0, \dots, k$. 
%
%
\subsection{Geometry and State Variables}
\label{SubSect:Geometry}
In this section, we briefly specify the geometry, underlying lattice, and the kinematic as well as the internal variables. In what follows, lattice nodes or particles will be referred to as "atoms", in order to be consistent with the original QC terminology developed for atomistic systems.

\begin{figure}
	\centering
	\subfloat[configurations and kinematics]{\includegraphics[scale=1]{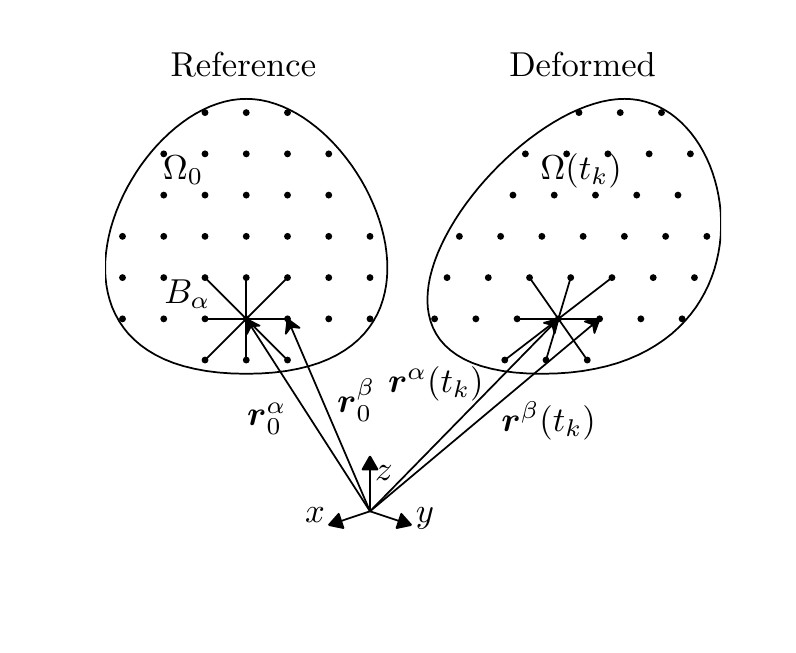}\label{SubSect:Geometry:Fig:1a}}\hspace{2em}
	\subfloat[damageable interaction]{\includegraphics[scale=1]{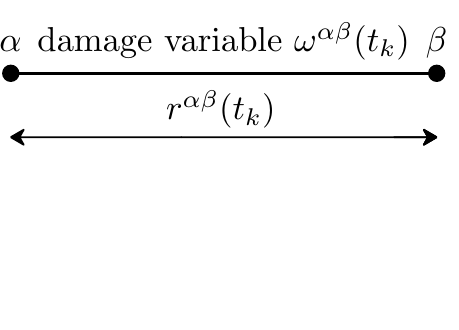}\label{SubSect:Geometry:Fig:1b}}
	\caption{Sketch of system configurations and kinematic as well as internal variables.}
	\label{SubSect:Geometry:Fig:1}
\end{figure}

First, the reader is referred to Fig.~\ref{SubSect:Geometry:Fig:1}, where a sketch of the kinematic as well as internal variables is presented. As indicated, we assume an X-braced lattice with the nearest-neighbour interactions. The original location vector of atom~$\alpha$ is denoted as~$\bs{r}^\alpha_0 \in \mathbb{R}^2$. For all~$n_\mathrm{ato}$ atoms (collected in an index set~$N_\mathrm{ato}$), they are stored in the column~$\bs{r}_0 = [\bs{r}^1_0, \dots, \bs{r}^{n_\mathrm{ato}}_0]^\mathsf{T}$, $\bs{r}_0 \in \mathbb{R}^{2 \, n_\mathrm{ato}}$. By analogy, the current location vectors~$\bs{r}^\alpha(t_k) \in \mathbb{R}^2$ are stored in~$\bs{r}(t_k) = [\bs{r}^1(t_k), \dots, \bs{r}^{n_\mathrm{ato}}(t_k)]^\mathsf{T}$. 

The entire system contains~$n_\mathrm{int}$ interactions stored in an index set~$N_\mathrm{int}$. Each interaction~$\alpha\beta \in N_\mathrm{int}$ that connects atoms~$\alpha$ and~$\beta$ is of the length
\begin{equation}
r^{\alpha\beta}(t_k) = || \bs{r}^\beta(t_k) - \bs{r}^\alpha(t_k) ||_2, \quad \alpha\beta = 1, \dots, n_\mathrm{int},
\label{SubSect:Geometry:Eq:1}
\end{equation}
where~$|| \bullet ||_2$ denotes the Euclidean norm.

For \emph{damaging} interactions, cf. Fig.~\ref{SubSect:Geometry:Fig:1b}, the only internal variable is the damage variable~$\omega(t_k) \in [0,1]$. All of these damage variables are stored in the column~$\bs{z}(t_k)$, which is hence of length~$n_\mathrm{int}$.
%
%
\subsection{Definition of Energies}
\label{SubSect:LNDamage}
The internal energy associated with damaging lattices reads
\begin{equation}
\mathcal{V}( \widehat{\bs{r}}, \widehat{\bs{z}} ) = \frac{1}{2}\sum_{\alpha,\beta \in B_\alpha} \left[( 1 - \widehat{\omega}^{\alpha\beta} ) \phi^{\alpha\beta}( \widehat{r}^{\alpha\beta}_+ ) + \phi^{\alpha\beta}( \widehat{r}^{\alpha\beta}_- )\right],
\label{SubSect:LNDamage:Eq:1}
\end{equation}
where~$B_\alpha \subseteq N_\mathrm{ato}$ denotes the set of the nearest neighbours associated with atom~$\alpha$. The elastic energy of a bond~$\phi^{\alpha\beta}$ is reduced by the damage variable~$\omega$ only if the bond is loaded in tension to better reflect physical behaviour of damageable materials. In Eq.~\eqref{SubSect:LNDamage:Eq:1}, we have introduced~$\widehat{r}^{\alpha\beta}_+ = \max(\widehat{r}^{\alpha\beta},r_0^{\alpha\beta})$, $\widehat{r}^{\alpha\beta}_- = \min(\widehat{r}^{\alpha\beta},r_0^{\alpha\beta})$, and assumed~$\phi^{\alpha\beta}(r_0^{\alpha\beta}) = 0$. 

The dissipation distance for bond~$\alpha\beta$ for two consecutive states~$\widehat{\bs{z}}_1$ and~$\widehat{\bs{z}}_2$ is defined as
\begin{equation}
\mathcal{D}^{\alpha\beta}(\widehat{\bs{z}}_2,\widehat{\bs{z}}_1)=
\left\{
\begin{array}{ll}
{\displaystyle D^{\alpha\beta}(\widehat{\omega}_2^{\alpha\beta})-D^{\alpha\beta}(\widehat{\omega}_1^{\alpha\beta})} &\mbox{ if }\widehat{\omega}_2^{\alpha\beta}\geq\widehat{\omega}_1^{\alpha\beta}\\
+\infty & \mbox{ otherwise,}
\end{array}
\right.\quad\alpha\beta \in N_\mathrm{int},
\label{SubSect:LNDamage:Eq:4}
\end{equation}
and the total dissipation distance is specified as
\begin{equation}
\mathcal{D}(\widehat{\bs{z}}_2,\widehat{\bs{z}}_1) = \frac{1}{2}\sum_{\alpha,\beta \in B_\alpha}\mathcal{D}^{\alpha\beta}(\widehat{\bs{z}}_2,\widehat{\bs{z}}_1).
\label{SubSect:LNDamage:Eq:5}
\end{equation}
In definition~\eqref{SubSect:LNDamage:Eq:4}, $D^{\alpha\beta}(\omega^{\alpha\beta})$ is the energy dissipation of a single bond during a unidirectional damage process up to a damage level~$\omega^{\alpha\beta}$. This function therefore increases from~$D^{\alpha\beta}(0) = 0$ to~$D^{\alpha\beta}(1) = g_{f,\infty}$, where~$g_{f,\infty}$ is the energy dissipated at complete failure.

The incremental interaction energy, $\widetilde{\pi}_{\alpha\beta}^k$, and incremental site energy, $\pi_{\alpha}^k$, then read:
\begin{subequations}
	\label{SubSect:LNDamage:Eq:6}
	\begin{align}
	\widetilde{\pi}_{\alpha\beta}^k(\widehat{\bs{q}};\bs{q}(t_{k-1})) & = (1-\widehat{\omega}^{\alpha\beta})\phi^{\alpha\beta}(\widehat{r}^{\alpha\beta}_+) + \phi^{\alpha\beta}(\widehat{r}^{\alpha\beta}_-) + \mathcal{D}^{\alpha\beta}(\widehat{\omega}^{\alpha\beta},\omega^{\alpha\beta}(t_{k-1})), \quad {\alpha\beta}=1,\dots,n_\mathrm{int},\label{SubSect:DissLatt:Eq:7a}\\
	\pi_\alpha^k(\widehat{\bs{q}};\bs{q}(t_{k-1})) & =\frac{1}{2}\sum_{\beta \in B_\alpha}\widetilde{\pi}_{\alpha\beta}^k(\widehat{\bs{q}},\bs{q}(t_{k-1})), \quad \alpha=1,\dots,n_\mathrm{ato}.\label{SubSect:DissLatt:Eq:7b}
	\end{align}
\end{subequations}
%
%
\section{Variational Quasicontinuum Methodology with Refinement and Coarsening}
\label{Sect:QC}
In this section we specify the two steps of the standard QC methodology, interpolation and summation, as well as additional procedures that must be adopted to extend it to an efficient description for localized damage. We start with a geometric crack description for lattice systems in Section~\ref{SubSect:Crack}, proceeding to the interpolation step in Section~\ref{SubSect:Interpolation}, where the standard framework as well as the mesh refinement and the mesh coarsening are treated. The first part closes with the description of the enrichment functions. The second part is devoted to summation, starting with the standard framework in Section~\ref{SubSect:Summation}, which is later extended to incorporate also the enrichment functions. Finally, the complete X-QC procedure is summarized in Section~\ref{SubSect:XQC} in terms of a general algorithm, and energy implications are discussed in Section~\ref{SubSect:EnImplications} in order to allow for proper verification of the energy equality~\eqref{E}.
%
%
\subsection{Description of a Crack}
\label{SubSect:Crack}
In what follows, a crack is defined as an ordered set of points collected in a set~$C$, cf. also~\cite{Fries:Cracks:2012}:
\begin{equation}
\mbox{If }\omega^{\alpha\beta} \geq \eta \quad \mbox{then} \quad \frac{1}{2}( \bs{r}^\alpha_0 + \bs{r}^\beta_0 ) \in C, \quad \mbox{and} \quad \alpha, \beta \in N_\mathrm{cw}.
\label{SubSect:Crack:Eq:1}
\end{equation}
The set~$N_\mathrm{cw}$ collects all crack wake atoms, whereas~$C$ stores the midpoints of all fully damaged bonds. Upon ordering, the set~$C$ defines an oriented polygon~$\bs{\Gamma}_C(s)$ (a curve parametrized by a curvilinear coordinate~$s$) geometrically representing the crack. The threshold~$\eta \in (0,1)$ should be close enough to~$1$, e.g.~$\eta = 0.95$. Naturally, any duplicates in~\eqref{SubSect:Crack:Eq:1} are eliminated, and the ordering may be implemented in various ways, e.g. by sorting the points in~$C$ with respect to their $x$-coordinates.\footnote{Note that in the case of extensive distributed cracking, definition~\eqref{SubSect:Crack:Eq:1} may not be sufficient and the X-QC presented here cannot be employed. In those cases, it is reasonable to fully resolve such regions according to the strategy presented in~\cite{VarQCDamage}.}

\begin{figure}
	\centering
	\includegraphics[scale=1]{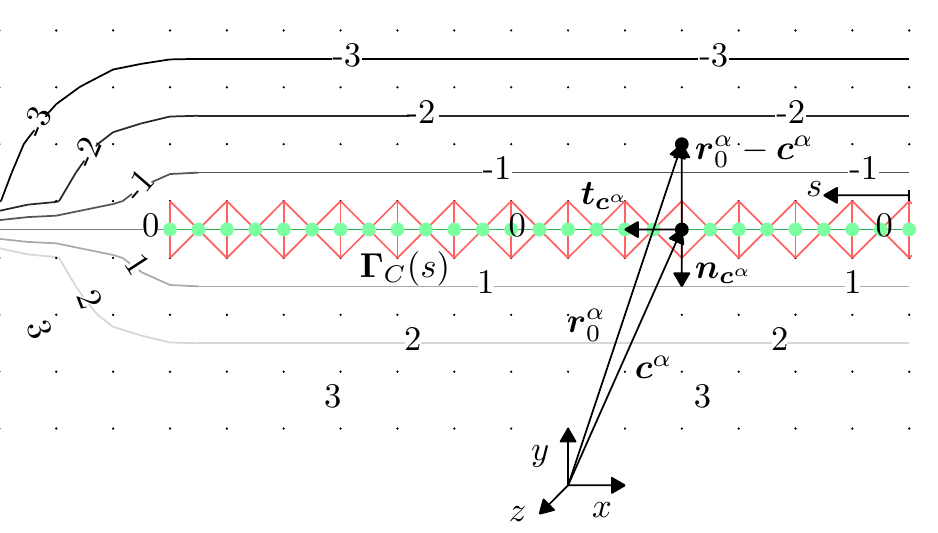}
	\caption{Representation of an existing crack in a discrete lattice network. Damaged bonds (for which~$\omega^{\alpha\beta} \geq \eta$, cf. definition~\eqref{SubSect:Crack:Eq:1}) are depicted by red lines (other interactions are omitted for clarity) and set~$C$ is depicted using green dots. The signed distance function~$\psi$ is represented by contour lines, and the crack polygon~$\bs{\Gamma}_C$ is presented as a green line. Atom sites are shown as black dots.}
	\label{SubSect:Crack:Fig:1}
\end{figure}

For future use, it is convenient to introduce the signed distance function~$\psi$, cf. e.g.~\cite{Fries:XFEM}, Section~3. Its definition reads
\begin{equation}
\begin{aligned}
\bs{c}^\alpha &= \underset{\widehat{\bs{c}} \in \bs{\Gamma}_C}{\mbox{arg min }}|| \bs{r}_0^\alpha - \widehat{\bs{c}} ||,\\
\psi(\bs{r}_0^\alpha) &= || \bs{r}_0^\alpha - \bs{c}^\alpha || \cdot \sign(\bs{n}_{\bs{c}^\alpha}^\mathsf{T} (\bs{r}_0^\alpha - \bs{c}^\alpha)),
\end{aligned}
\quad \alpha \in N_\mathrm{ato}, 
\label{SubSect:Crack:Eq:2}
\end{equation}
where~$\bs{n}_{\bs{c}^\alpha}$ denotes the column storing the unit normal to the crack~$\bs{\Gamma}_C$ at the closest point~$\bs{c}^\alpha$. For instance, $\bs{n}_{\bs{c}^\alpha}$ can be defined such that~$\bs{t}_{\bs{c}^\alpha} \times \bs{n}_{\bs{c}^\alpha} = \bs{e}_z$, where~$\bs{t}_{\bs{c}^\alpha}$ is the unit tangent vector to~$\bs{\Gamma}_C$ at~$\bs{c}^\alpha$, and~$\bs{e}_z$ is the unit basis vector pointing in the $z$-direction of the adopted coordinate system. If~$\bs{c}^\alpha$ is situated at a kink of~$\bs{\Gamma}_C$, entire cone of normals must be considered; then, the sign of~$\psi(\bs{r}_0^\alpha)$ is positive if the vector~$\bs{r}_0^\alpha - \bs{c}^\alpha$ belongs to the cone of normals and negative otherwise. The function~$\psi$ is therefore positive on one side of the crack and negative on the other. Because the crack tip is always located inside the fully refined region, no second level set function is required to describe the crack tip, cf.~\cite{Fries:XFEM}, Section~3.2. For a pictorial representation of the introduced quantities see Fig.~\ref{SubSect:Crack:Fig:1}.

Within the X-QC framework, crack path bifurcations are included automatically as a result of the full refinement in the process zone. Subsequent coarsening in the wakes of cracks through special function enrichments would require further development. Because crack branching designed for continuous systems is rather technical, see e.g.~\cite{crackBranching:NME}, one can expect a similar degree of technical complexity also for discrete systems. These kinds of generalizations lie outside the scope of the current manuscript and are left as possible future challenges. Only isolated cracks are treated in the remainder of this manuscript. 
%
%
\subsection{Interpolation}
\label{SubSect:Interpolation}
In the four following subsections, the interpolation step of the QC method and its effect on the incremental energy of the full system ($\Pi^k$ in~\eqref{IE}) are detailed.
%
%
\subsubsection{General Framework}
\label{SubSect:GenInt}
Interpolation introduces to the minimization problem~\eqref{IP} the following equality constraints:
\begin{equation}
\widehat{\bs{r}} = \bs{\Phi}\widehat{\bs{g}}, \quad \mbox{for} \quad \widehat{\bs{g}} \in \mathscr{G}(t_k),
\label{SubSect:GenInt:Eq:1}
\end{equation}
where~$\widehat{\bs{g}}$ denotes a column of generalized DOFs located in a kinematically constrained subspace~$\mathscr{G}(t_k)$ of the fully dimensional space~$\mathscr{R}(t_k)$.

After substitution of Eq.~\eqref{SubSect:GenInt:Eq:1}, the incremental energy~\eqref{IE} becomes a function of~$\widehat{\bs{g}}$, i.e.
\begin{equation}
\Pi^k(\widehat{\bs{r}},\widehat{\bs{z}};\bs{r}(t_{k-1}),\bs{z}(t_{k-1})) = \Pi^k(\bs{\Phi}\widehat{\bs{g}},\widehat{\bs{z}};\bs{\Phi}\bs{g}(t_{k-1}),\bs{z}(t_{k-1})),
\label{SubSect:GenInt:Eq:2}
\end{equation}
which can be minimized over the linear subspace~$\mathscr{G}(t_k)$. This reduces the computational effort if~$\dim(\mathscr{G}(t_k)) \ll \dim(\mathscr{R}(t_k))$.

In QC approaches, the generalised DOFs~$\widehat{\bs{g}}$ are chosen as the DOFs of a small set of atoms, the so-called \emph{repatoms}. Usually, repatoms are chosen as nodes of a mesh~$\mathcal{T}$ that triangulates the domain~$\Omega_0$, and the interpolation is implemented via piecewise affine ($\mathrm{P}_1$) FE shape functions constructed between the repatoms. Consequently, $\bs{\Phi}$ contains standard FEM shape function evaluations at all atom positions, cf. e.g.~\cite{TadmorModel}. Note that higher-order polynomial approximations inside elements can be adopted as well, see e.g.~\cite{BeexBeams}, \cite{YangMMM}, or~\cite{BeexHO}. 

In order to distinguish classical QC approach from its extended version presented below, we use the following notation for the classical QC: all~$n_\mathrm{rep}$ repatoms are collected in an index set~$N_\mathrm{rep} \subseteq N_\mathrm{ato}$, whereas their admissible positions~$\widehat{\bs{r}}_\mathrm{rep}^\alpha \in \mathbb{R}^2$, $\alpha \in N_\mathrm{rep}$, are stored in the column~$\widehat{\bs{r}}_\mathrm{rep} \in \mathscr{R}_\mathrm{rep}(t_k)$ by analogy to~$\widehat{\bs{r}}$. Associated interpolation matrix is denoted~$\bs{\Phi}_\mathrm{FE}$. 
%
%
\subsubsection{Mesh Refinement}
\label{SubSect:Refinement}
In this contribution, the mesh refinement indicator of~\cite{VarQCDamage}, is adopted with minor changes. It proceeds as follows: a triangle~$K$ of the current triangulation~$\mathcal{T}_k$ is endowed with a set of sampling interactions~$S_\mathrm{int}^K$ specified as
\begin{equation}
\text{$S_\mathrm{int}^K$ stores those interactions~$\alpha\beta \in S_\mathrm{int}$ for which~$\frac{1}{2}( \bs{r}^\alpha + \bs{r}^\beta ) \in K$.}\footnote{Note that since all elements~$K \in \mathcal{T}_k$ are considered as closed sets, bonds lying on edges may belong to two triangles.}
\end{equation}
Then, the following energy condition is introduced
\begin{equation}
(1-\omega^{\alpha\beta})\phi^{\alpha\beta}(r^{\alpha\beta}_+) \geq \theta_\mathrm{r}\,\phi^{\alpha\beta}_\mathrm{th}, \quad \alpha\beta \in S_\mathrm{int}^K, \quad \theta_\mathrm{r} \in (0,1),
\label{SubSect:Refinement:Eq:1}
\end{equation}
which specifies interactions that are likely to be damaged. In Eq.~\eqref{SubSect:Refinement:Eq:1}, $\phi^{\alpha\beta}$ denotes the pair potential, $\theta_\mathrm{r}$ a refinement safety parameter, and~$\phi^{\alpha\beta}_\mathrm{th}$ the stored threshold energy at which the internal variable starts to evolve. The threshold energy is therefore obtained as~$\phi_\mathrm{th}^{\alpha\beta} = \phi^{\alpha\beta}(r_0(1+\varepsilon_0))$, where~$\varepsilon_0$ is the limit elastic strain, one of the input parameters of the employed constitutive model, see Tab.~\ref{Sect:Examples:Tab:1}. The mesh refinement indicator then reads
\begin{equation}
\begin{aligned}
&&\mbox{If condition~\eqref{SubSect:Refinement:Eq:1}}\mbox{ holds at least for one interaction }\alpha\beta \in S_\mathrm{int}^K\\
&&\Longrightarrow\mbox{ mark~$K$ for refinement, i.e. add $K$ to~$\mathcal{I_\mathrm{r}}$.}
\end{aligned}	
\label{SubSect:Refinement:Eq:2}
\end{equation}

Having specified a set of triangles marked for refinement, $\mathcal{I}_\mathrm{r} \subseteq \mathcal{T}_k$, the current triangulation~$\mathcal{T}_k$ must be refined. To this end, the backward-longest-edge-bisection algorithm by~\cite{Rivara:LEPP} is used.
%
%
\subsubsection{Mesh Coarsening}
\label{SubSect:Coarsening}
Before coarsening the current triangulation~$\mathcal{T}_k$, first an indicator is presented that marks which triangles need to be coarsened. In analogy to Eq.~\eqref{SubSect:Refinement:Eq:1} we introduce the following energy condition
\begin{equation}
(1-\omega^{\alpha\beta})\phi^{\alpha\beta}(r_+^{\alpha\beta}) + \phi^{\alpha\beta}(r_-^{\alpha\beta}) \leq \theta_\mathrm{c}\,\phi^{\alpha\beta}_\mathrm{th}, \quad \alpha\beta \in S_\mathrm{int}^K, \quad \theta_\mathrm{c} \in (0,1), \quad \theta_\mathrm{c} < \theta_\mathrm{r},
\label{SubSect:Coarsening:Eq:1}
\end{equation}
where the compressive part is now included as well. The rationale behind condition~\eqref{SubSect:Coarsening:Eq:1} is that in the crack wake the stress is low, and hence also the energy. The mesh coarsening indicator then reads
\begin{equation}
\begin{aligned}
&&\mbox{If condition~\eqref{SubSect:Coarsening:Eq:1}}\mbox{ holds for all interactions }\alpha\beta \in S_\mathrm{int}^K\\
&&\Longrightarrow\mbox{ mark~$K$ for coarsening, i.e. add $K$ to~$\mathcal{I_\mathrm{c}}$.}
\end{aligned}	
\label{SubSect:Coarsening:Eq:2}
\end{equation}
Verifying condition~\eqref{SubSect:Coarsening:Eq:2} for all elements yields a set of triangles marked for coarsening, $\mathcal{I}_\mathrm{c} \subseteq \mathcal{T}_k$. 

\begin{figure}
	\centering
	\includegraphics[scale=1]{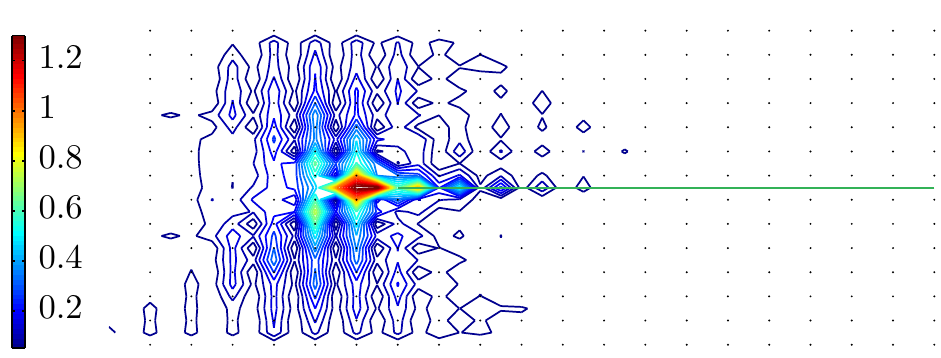}
	\caption{An example of contours of normalized energy~$\phi^{\alpha\beta}/\phi^{\alpha\beta}_\mathrm{th}$ around a crack tip (based on linearly interpolated data) of a QC system. Regular atom sites are shown as black dots and the existing crack as a green line.}
	\label{SubSect:NodeProtection:Fig:1}
\end{figure}

Note that because the damage remains localized, it yields steep spatial (interaction) energy variations in the vicinity of the crack tip. For instance, in Fig.~\ref{SubSect:NodeProtection:Fig:1}, the normalized interaction energy can be seen to drop from~$1.2$ to~$0.2$ over just two to three lattice spacings. From this observation it may be clear that if the values of~$\theta_\mathrm{c}$ and~$\theta_\mathrm{r}$ are too close, or if~$\theta_\mathrm{c}$ is too high, repetitive mesh refinement and coarsening of the same regions in subsequent time steps may occur. In order to achieve good performance, $\theta_\mathrm{c}$ should be significantly smaller than~$\theta_\mathrm{r}$. In particular, we choose~$\theta_\mathrm{r} = 0.5$ or~$0.25$\footnote{Note that in the case of quadratic potentials~$\phi^{\alpha\beta}$, values~$\theta_\mathrm{r} = 0.5$ and~$0.25$ correspond to stress levels~$70.7\,\%$ and~$50\,\%$ of the tensile strength~$E\varepsilon_0$, cf. Eqs.~\eqref{Examples:Eq:1} and~\eqref{Examples:Eq:3}.} and~$\theta_\mathrm{c} = 0.05$ in both examples of Section~\ref{Sect:Examples}. 

\begin{figure}
	\centering
	\includegraphics[scale=1]{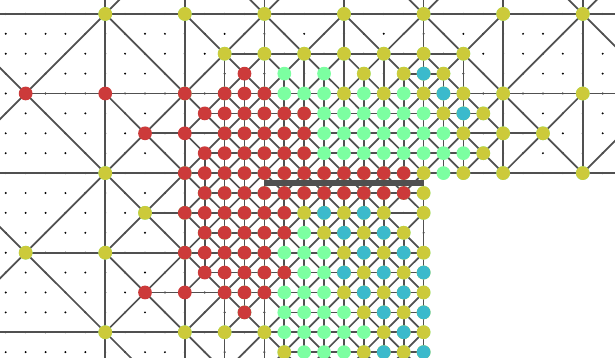}
	\caption{A typical situation before coarsening. The red points define the set~$N_\mathrm{r}$. They are found at, and ahead of, the crack tip, and are protected from coarsening. The yellow along with the blue points define the set~$N_\mathrm{c}$, and can be coarsened. The green points remain unchanged and correspond to the set~$N_\mathrm{u}$. Only the blue points satisfy the valence condition and can be removed. Regular atoms are shown as black dots and the current triangulation~$\mathcal{T}_k$ is depicted in black thin lines. The existing crack is indicated as a thick black line. All interactions are omitted for clarity.}
	\label{SubSect:Coarsening:Fig:1}
\end{figure}

In order to coarsen the current triangulation~$\mathcal{T}_k$, a similar algorithm to those proposed by~\cite{Chen:Coarsening} and~\cite{Funken:Coarsening} is adopted. Because we use the two-triangle refinement scheme, the entire binary-tree structure of the triangulation history must be stored. Coarsening of a patch of triangles is decided based on their nodes, cf. Fig.~\ref{SubSect:Coarsening:Fig:1}. All nodes (repatoms) of the current triangulation are stored in a set~$N_\mathrm{rep}$, which is divided into three disjoint sets: $N_\mathrm{r}$, $N_\mathrm{c}$, and~$N_\mathrm{u}$. The set~$N_\mathrm{r}$ stores nodes associated with triangles in~$\mathcal{I}_\mathrm{r}$ (and also other protected nodes, see Alg.~\ref{SubSect:XQC:Alg:1}). The set~$N_\mathrm{c}$ collects nodes associated with triangles in~$\mathcal{I}_\mathrm{c}$, and~$N_\mathrm{u}$ complements~$N_\mathrm{rep}$. As a consequence of the mesh hierarchy, only a subset of~$N_\mathrm{c}$ can be removed. These are identified through their valences~$v_\alpha$ (the number of triangles connected to a node): 
\begin{equation}
v_\alpha = \# \{K \in \mathcal{T}_k \, | \, \alpha \in K \}, \quad \alpha \in N_\mathrm{c}.
\label{SubSect:Coarsening:Eq:4}
\end{equation}
Only nodes with valence~$v_\alpha = 2$ (boundary nodes) or~$4$ (internal nodes) can be removed (the blue points in Fig.~\ref{SubSect:Coarsening:Fig:1}). The corresponding two or four child elements are subsequently replaced with one or two parent elements, shifting locally the binary-tree hierarchy of the mesh one level up. As valences change during the coarsening process, this procedure is repeated until convergence.

In what follows, we use a right-angled initial triangulation~$\mathcal{T}_0$, as already indicated in Fig.~\ref{SubSect:Coarsening:Fig:1}. Because the above described refinement and coarsening algorithms yield self-similar triangulations, all meshes~$\mathcal{T}_k$ are right-angled as well. This is desirable because the summation error is minimized, cf.~\cite{VarQCDiss}, and poorly shaped triangles are avoided.

An alternative approach to the mesh refinement and coarsening indicators, proposed in Eqs.~\eqref{SubSect:Refinement:Eq:2} and~\eqref{SubSect:Coarsening:Eq:2}, would be to use \emph{error estimators} such as the \emph{goal-oriented} error estimator reported e.g. in~\cite{Becker:Rannacher:2001,OdenGoal,Prudhomme_2006Goal,MemarnahavandiGoal}. Generalizations of this type will not be developed in this contribution and are left as a possible future challenge.
%
%
\subsubsection{Enrichment Functions}
\label{SubSect:Enrichment}
In Sections~\ref{SubSect:GenInt}~-- \ref{SubSect:Coarsening} we discussed the QC interpolation and mesh coarsening independently of an existing crack---the crack served merely to identify the fully resolved region at its tip and a coarse region elsewhere. In this subsection, we describe how to extend the QC interpolation when the current triangulation~$\mathcal{T}_k$ is such that a crack cuts through a coarse triangle~$K$, and how it influences~$\bs{\Phi}$, $\widehat{\bs{r}}_\mathrm{rep}$, and~$\widehat{\bs{g}}$. Similarly to the previous sections, the discussion considers the system at time step~$t_k$. 

\begin{figure}
	\centering
	\subfloat[reproducing and blending elements]{\includegraphics[scale=1]{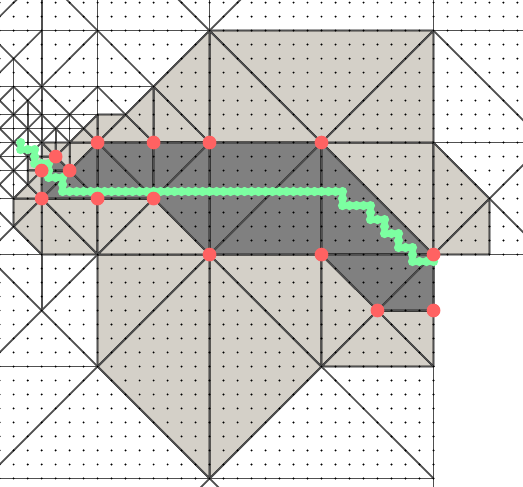}\label{SubSect:Enrichment:Fig:1a}}\hspace{2em}
	\subfloat[enrichment function, $\frac{1}{2}\sign (\psi(\bs{r}_0))$]{\includegraphics[scale=1]{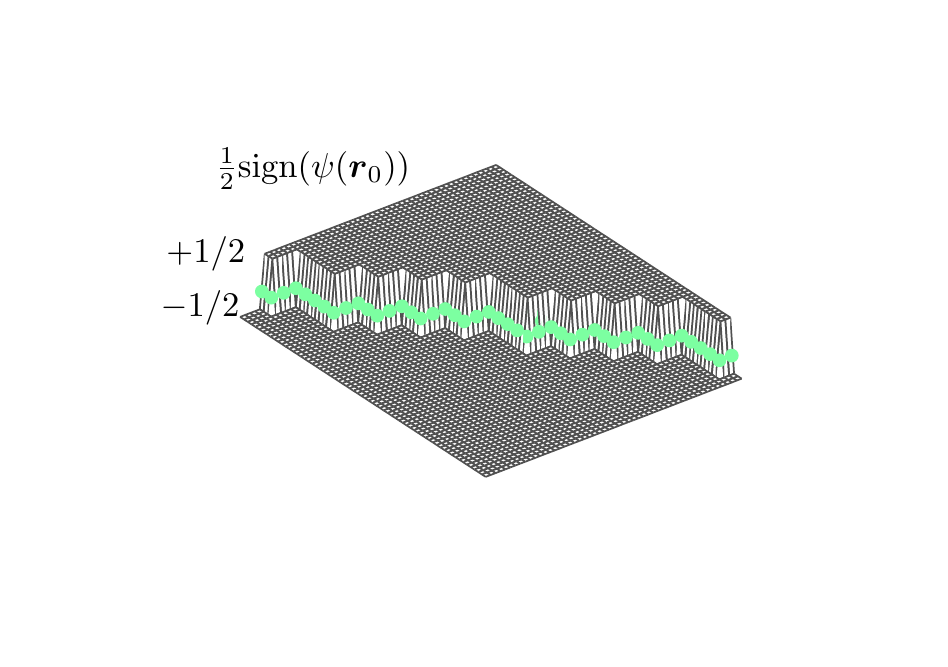}\label{SubSect:Enrichment:Fig:1b}}
	\caption{(a)~An example of cut or reproducing (dark grey) and blending (light grey) elements. The set of enriched nodes~$N^\star$ is shown in red dots, and regular atom sites as small black dots. (b)~An example of the enrichment function, $\frac{1}{2}\sign (\psi(\bs{r}_0^\alpha))$; crack points~$C$ are shown as green dots.}
	\label{SubSect:Enrichment:Fig:1}
\end{figure}

We first specify the cut elements (also called reproducing, \citealt{Fries:XFEM}), cf. Fig.~\ref{SubSect:Enrichment:Fig:1a}. A coarse element~$K\in\mathcal{T}_k$ such that
\begin{equation}
\# (C \cap K) > 0 \quad \mathrm{and} \quad
\min_{\bs{r}_0^\alpha \in K}\psi(\bs{r}_0^\alpha) \cdot \max_{\bs{r}_0^\alpha \in K}\psi(\bs{r}_0^\alpha)<0
\label{SubSect:Enrichment:Eq:1}
\end{equation}
is considered to be cut by a crack. Conditions~\eqref{SubSect:Enrichment:Eq:1} require that at least one point in~$C$ belongs to~$K$ and that the signed distance function~$\psi$ changes its sign inside~$K$. Let us assume in what follows that a crack cannot cut directly through any of the nodes, atoms, or along domain boundaries.

Assume then that our system is approximated with a standard QC framework in which the positions of the repatoms are collected in a column~$\widehat{\bs{r}}_\mathrm{rep}$. Then, the reconstruction in Eq.~\eqref{SubSect:GenInt:Eq:1} can be rewritten as
\begin{equation}
\widehat{\bs{r}}^\alpha_\mathrm{qc} =
\sum_{\beta \in N_\mathrm{rep}} \varphi_\beta(\bs{r}_0^\alpha) \, \widehat{\bs{r}}_\mathrm{rep}^\beta
, \quad \alpha \in N_\mathrm{ato},
\label{SubSect:Enrichment:Eq:2}
\end{equation}
where~$\varphi_\beta(\bs{r}_0^\alpha)$ represents the $\mathrm{P}_1$ FE shape function associated with the repatom~$\beta$ (this function is evaluated at the undeformed configuration of atom~$\alpha$), and~$\widehat{\bs{r}}_\mathrm{rep}^\beta \in \mathbb{R}^2$ denotes an admissible position of a repatom~$\beta$. This definition does not include an existing crack yet.

In order to incorporate a crack, an enrichment using the local PU concept (standard local XFEM approximation) is introduced. Enrichment terms are included as follows
\begin{equation}
\widehat{\bs{r}}^\alpha =
\widehat{\bs{r}}^\alpha_\mathrm{qc}
+ \sum_{j = 1}^{n^\star} \underbrace{\varphi_{\beta_j}(\bs{r}_0^\alpha) \, \frac{1}{2}[\sign(\psi(\bs{r}_0^\alpha)) - \sign(\psi(\bs{r}_0^{\beta_j}))]}_{\varphi_j^\star(\bs{r}_0^\alpha)} \, \widehat{\bs{g}}^\star_j
, \quad \alpha \in N_\mathrm{ato}, \quad \beta_j \in N^\star,
\label{SubSect:Enrichment:Eq:3}
\end{equation}
where the same $\mathrm{P}_1$ FE shape functions for the extrinsic enrichment terms are used as well.

In Eq.~\eqref{SubSect:Enrichment:Eq:3}, $N^\star \subseteq N_\mathrm{rep}$ is a set of~$n^\star$ enriched repatoms associated with the cut elements, $\varphi_j^\star$ denotes the enrichment function of the~$\beta_j$-th repatom, and~$\widehat{\bs{g}}^\star_j  \in \mathbb{R}^2$ denotes an admissible vector of generalized DOFs associated with $j$-th enriched repatom. Let us emphasize that the only enrichment function is~$\frac{1}{2}\sign(\psi(\bs{r}_0))$, depicted in Fig.~\ref{SubSect:Enrichment:Fig:1b}, and that in accordance with Eq.~\eqref{SubSect:Enrichment:Eq:3} the magnitude of the jump across the crack is approximated in a continuous, piecewise linear manner along the crack path~$\bs{\Gamma}_C$. Note also that no crack tip enrichments are required, since the crack tip is always located inside the fully resolved region, and that the shifting adopted in Eq.~\eqref{SubSect:Enrichment:Eq:3} guarantees the Kronecker-$\delta$ property of the resulting approximation, cf.~\cite{Belytschko:XFEM:2001} or \cite{Fries:XFEM}. Furthermore, the shifting implies that there are actually no blending elements, as shifted enrichment functions vanish outside the cut elements. Even when shifting is not adopted, no difficulties with blending elements occur, because the interpolation is based on $\mathrm{P}_1$ FE shape functions and the enrichment function is piecewise constant. For further details see~\cite{Fries:XFEM}, Sections~4.3 and~5.1.2.

\begin{figure}
	\centering
	\includegraphics[scale=1]{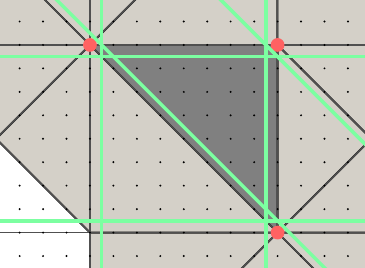}
	\caption{An example of several crack paths (shown as green lines) associated with a cut element (shown in dark grey) that render singular stiffness matrices. Enriched nodes are presented as red dots.}
	\label{SubSect:Enrichment:Fig:2}
\end{figure}

In continuum XFEM, poorly conditioned stiffness matrices occur if the crack path is close to triangle edges and nodes, cf. e.g.~\cite{WellsXFEM:2001,FlatTop:Laborde}, or~\cite{Fries:XFEM}. For lattice networks, the measure for closeness is the lattice spacing~$a$. Hence, if a crack path is closer than~$a$ to an element edge or node, the resulting system matrix is rendered singular, cf. Fig.~\ref{SubSect:Enrichment:Fig:2}.

Two approaches are used for continua to remedy this situation. The first approach is the use of clustering (or node gathering) implemented through flat-top functions, cf. e.g.~\cite{Fries:2008:CorrXFEM,FlatTop:Belytschko,Babuska:SGFEM,Kostas:1,Kostas:2}. These techniques usually increase the polynomial order of the resulting approximations if the Kronecker-$\delta$ property is required. Hence, this is not desirable within the QC framework, as more extensive changes in the summation rule would be needed.

The second approach is to neglect those~$\varphi_j^\star$ that would render the stiffness matrix singular altogether~\cite[Section~11 and references therein]{Fries:XFEM}, which is adopted in this contribution. The reason is that such shifted enrichments are identical to zero vector for lattice networks and hence, they provide no additional information to the system.

Finally, introducing interpolation matrices~$\bs{\Phi}_\mathrm{FE}$ and~$\bs{\Phi}_\star$, with individual components expressed as
\begin{align}
(\Phi_\mathrm{FE})_{(2\alpha-1)(2j-1)} &= (\Phi_\mathrm{FE})_{(2\alpha)(2j)} = \left\{
\begin{aligned}
&\varphi_{\beta_j}(\bs{r}_0^\alpha) && \mbox{for}\ \alpha\in N_\mathrm{ato},\ \beta_j\in N_\mathrm{rep},\ j=1,\dots,n_\mathrm{rep},\\
&0 && \mbox{otherwise,}
\end{aligned}\right.\label{SubSect:Enrichment:Eq:6}\\
(\Phi_\star)_{(2\alpha-1)(2j-1)} &= (\Phi_\star)_{(2\alpha)(2j)} = \left\{
\begin{aligned}
&\varphi_j^\star(\bs{r}_0^\alpha) &&\mbox{for}\ \alpha\in N_\mathrm{ato},\ j=1,\dots,n^\star,\\
&0 && \mbox{otherwise,}
\end{aligned}\right.\label{SubSect:Enrichment:Eq:7}
\end{align}
one can define a global interpolation matrix~$\bs{\Phi}$ along with associated kinematic variable~$\widehat{\bs{g}}$ as
\begin{equation}
\bs{\Phi} = [\bs{\Phi}_\mathrm{FE},\bs{\Phi}_\star]
\quad \mbox{and} \quad
\widehat{\bs{g}} = [\widehat{\bs{r}}_\mathrm{rep},\widehat{\bs{g}}_1^\star,\dots,\widehat{\bs{g}}_{n^\star}^\star]^\mathsf{T},
\label{SubSect:Enrichment:Eq:8}
\end{equation}
and cast the approximation of Eq.~\eqref{SubSect:Enrichment:Eq:3} in the form of Eq.~\eqref{SubSect:GenInt:Eq:1}.
%
%
\subsection{Summation}
\label{SubSect:Summation}
The second step of QC frameworks, summation, is the QC counterpart to numerical integration in the finite element method. It serves two purposes. The first one is to estimate the exact system's incremental energy~$\Pi^k$ in an efficient, yet accurate way. This reduces the computational effort related to determining the energies, gradients, and Hessians. The second purpose is to reduce the dimensionality of the internal variable~$\bs{z}$, which in combination with the interpolation step realises the reduction of the entire state variable~$\bs{q}(t_k)$.
%
%
\subsubsection{General Framework}
\label{SubSect:GenSum}
The exact incremental energy~$\Pi^k$ is approximated using a limited number of site- or interaction-energies and their weight factors. These are typically chosen such that the error of the induced approximation is limited, but a low number of sampling atoms/interactions is required. The interested reader is referred for further details to~\cite{BeexQC,BeexCSumRule} for atom-based summation rules, to~\cite{Beex:textileReliability} for an interaction-based summation rule, or to~\cite{Amelang:2015} for an overview of summation rules in atomistic systems.

Here, we employ the central site-energy-based summation rule~\citep{BeexCSumRule,BeexDisLatt}, which considers only the atoms at the element corners plus one near the center. The central one is taken to be representative also of atoms which interact across the element boundaries. If no internal atoms exist, all boundary atoms are considered. The central rule involves a certain degree of approximation, but is significantly more efficient than the exact summation rule not discussed here; for details see~\citep{BeexQC}. The summation rule is implemented by introducing a set of~$n_\mathrm{sam}^\mathrm{ato}$ sampling atoms collected in an index set~$S_\mathrm{ato} \subseteq N_\mathrm{ato}$. The~$n_\mathrm{sam}^\mathrm{int}$ interactions connected to these sampling atoms are stored in an index set~$S_\mathrm{int}$. Then, the summation step yields the following approximation,
\begin{equation}
\begin{aligned}
\Pi^k(\widehat{\bs{q}};\bs{q}(t_{k-1})) \approx \widehat{\Pi}^k(\widehat{\bs{q}};\bs{q}(t_{k-1})) 
=& \sum_{\alpha\in S_\mathrm{ato}}w_\alpha \pi_\alpha^k(\widehat{\bs{q}};\bs{q}(t_{k-1})) - \bs{f}_\mathrm{ext}^\mathsf{T}\widehat{\bs{r}} \\
=& \sum_{\alpha\beta \in S_\mathrm{int}} \widetilde{w}_{\alpha\beta}\widetilde{\pi}^k_{\alpha\beta}(\widehat{\bs{q}};\bs{q}(t_{k-1})) - \bs{f}_\mathrm{ext}^\mathsf{T}\widehat{\bs{r}},
\end{aligned}
\label{SubSect:Summation:Eq:3}
\end{equation}
where~$w_\alpha$ are the weight factors associated with atom sites, and~$\widetilde{w}_{\alpha\beta}$ are corresponding weight factors associated with interactions.

Because the approximate incremental energy in Eq.~\eqref{SubSect:Summation:Eq:3} only depends on the internal variables associated with~$n_\mathrm{sam}^\mathrm{int}$ interactions, one can introduce a reduced state variable of the QC system in the form~$\bs{q}_\mathrm{red}(t_k) = (\bs{g}(t_k),\bs{z}_\mathrm{sam}(t_k)) \in \mathscr{Q}_\mathrm{red}(t_k)$, where~$\bs{z}_\mathrm{sam} \in \mathbb{R}^{n_\mathrm{sam}^\mathrm{int}} = \mathscr{Z}_\mathrm{sam}$ is a column that stores the internal variables of all sampling interactions. The reduced state space then reads~$\mathscr{Q}_\mathrm{red}(t_k) = \mathscr{G}(t_k) \times \mathscr{Z}_\mathrm{sam}$.
%
%
\subsubsection{Summation Rule for Enrichment Functions}
\label{SubSect:XSumRule}
If the enrichment of Section~\ref{SubSect:Enrichment} is adopted and a coarse triangle is cut by a crack, Eq.~\eqref{SubSect:Summation:Eq:3} still holds, but the selection of the individual sampling atoms~$S_\mathrm{ato}$ and their weights~$w_\alpha$ changes. 

\begin{figure}
	\centering
	\includegraphics[scale=1]{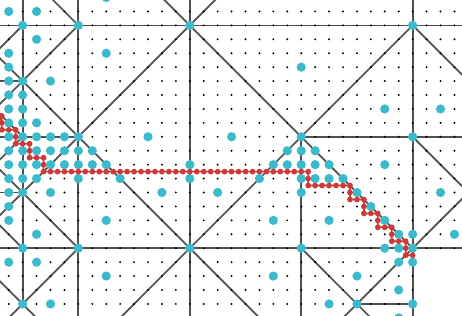}
	\caption{Schematic representation of the set of sampling atoms contained in~$S_\mathrm{ato}$ for the central summation rule. Large blue dots represent the sampling atoms, and the crack points~$C$ from~\eqref{SubSect:Crack:Eq:1} are shown as red dots.}
	\label{SubSect:XSumRule:Fig:1}
\end{figure}

In the central summation rule, all crack wake atoms are treated as additional triangle edges. Therefore, a cut triangle is split into two regions, bounded by triangle edges and one crack face. For each of the two subregions, the standard central summation rule is used. Hence, if no internal atoms exist, all edge atoms are added as discrete sampling atoms. Otherwise, a central sampling atom is selected with the weight corresponding to the sum of the number of internal atoms plus one half of the number of edge atoms. Atoms at edge intersections are added as discrete sampling atoms. For the detailed algorithmic description and pictorial representation see Alg.~\ref{SubSect:XSumRule:Alg:2} and Fig.~\ref{SubSect:XSumRule:Fig:1}.

Note that the above-described summation rule, which in its current form does not take into account crack closure (recall Eq.~\eqref{SubSect:LNDamage:Eq:1} and the discussion therein), can be easily generalized to deal also with non-monotonous loading and crack closure merely by adding all crack wake atoms~$N_\mathrm{cw}$ as discrete sampling atoms. Special summation rule that samples crack faces can also be devised, but this is omitted from our further considerations for the sake of simplicity. Because only monotonous loading and crack opening are observed in examples Section~\ref{Sect:Examples}, the summation rule presented in Alg.~\ref{SubSect:XSumRule:Alg:2} is adequate yet efficient.
\begin{algorithm}
  \begin{minipage}{\linewidth}
	\caption{Central summation rule for cut elements.}
	\label{SubSect:XSumRule:Alg:2}
	\centering
	\vspace{-\topsep}
	\begin{enumerate}[1:]
		\item Identify crack wake atoms~$N_\mathrm{cw} = \{\alpha, \beta \,|\, \omega^{\alpha\beta} \geq \eta\}$.
		\item \textbf{for each triangle~$K \in \mathcal{T}_k$}	
		\begin{enumerate}[(I):]
			\item \textbf{If} triangle is not cut by a crack, use standard central summation rule~\citep{BeexCSumRule}.
			\item \textbf{Else}
			\begin{enumerate}[(i):]
					\item Identify triangle's internal~$N_\mathrm{i}^K$, edge~$N_\mathrm{e}^K$, vertex~$N_\mathrm{v}^K$, and all~$N^K$, atoms.
					\item Select vertex atoms~$N^K_\mathrm{v}$ as discrete sampling atoms ($w_\alpha = 1$).
					\item \textbf{If} no internal atoms exist, select all atoms~$N^K$ as discrete sampling atoms~($w_\alpha = 1$).
					\item \textbf{Else}
					\begin{enumerate}[(a):]
						\item Split~$N^K$ in sets~$N^K_+$ and~$N^K_-$ according to the sign of~$\psi$.
						\item Define~$\widetilde{N}^K_\mathrm{i} = N^K_\mathrm{i} \backslash N^K_\mathrm{cw}$. Split~$\widetilde{N}^K_\mathrm{i}$ into two disjoint sets~$\widetilde{N}_{\mathrm{i},+}^K$ and~$\widetilde{N}_{\mathrm{i},-}^K$ according to the sign of~$\psi$ and sort both with respect to the number of neighbours inside triangle (in decreasing order).
						\item \textbf{If} $\widetilde{N}_{\mathrm{i},\pm}^K$ is empty, all atoms in~$N^K_\pm\backslash N^K_\mathrm{v}$ are discrete sampling atoms.
						\item \textbf{Else}
						\begin{itemize}	
							\item Choose~$\beta = \widetilde{N}_{\mathrm{i},\pm}^K(1)$ as central sampling atom.
							\item For each~$\alpha \in \widetilde{N}_{\mathrm{i},\pm}^K$, $w_\beta = w_\beta + 1$.\footnote{Note that this formula, and the one below ($w_\beta = w_\beta +0.5$), indicates that the weight factor of the central sampling atom~$\beta$ is incremented by~$1$ or by~$0.5$ for each~$\alpha$ from the corresponding set.}
							\item Add all~$\alpha \in (N_\mathrm{cw} \cap N^K_\mathrm{e})$ as discrete sampling atoms ($w_\alpha = 1$).
							\item Identify discrete sampling atoms on edges resulting from neighbouring triangles~$\widetilde{N}_\mathrm{e}^K$.
							\item For each~$\alpha \in (N_\pm^K \backslash (N_\mathrm{v}^K \cup \widetilde{N}_{\mathrm{i},\pm}^K \cup \widetilde{N}_\mathrm{e}^K \cup (N_\mathrm{cw} \cap N^K_\mathrm{e})))$, $w_\beta = w_\beta + 0.5$.
						\end{itemize}
					\end{enumerate}
			\end{enumerate}
		\end{enumerate}
		\item \textbf{end}
	\end{enumerate}
	\vspace{-\topsep}
  \end{minipage}
\end{algorithm}
%
%
\subsection{X-QC}
\label{SubSect:XQC}
In all previous Sections~\ref{SubSect:Crack}~-- \ref{SubSect:Summation}, we have assumed an existing crack and triangulation~$\mathcal{T}_k$. We are, however, interested in evolving crack tips and fully resolved regions. Therefore, we summarize the mesh refinement, coarsening, and additions of enrichments of the total X-QC framework in Alg.~\ref{SubSect:XQC:Alg:1}.
\begin{algorithm}
  \begin{minipage}{\linewidth}
	\caption{X-QC algorithm.}
	\label{SubSect:XQC:Alg:1}
	\centering
	\vspace{-\topsep}
	\begin{enumerate}[1:]
		\item Initialize the system: apply initial condition~$\bs{q}_0$ and construct initial (coarse) mesh~$\mathcal{T}_0$ with required information, e.g.~$N_\mathrm{rep}(t_0)$, $\bs{\Phi}(t_0)$, $S_\mathrm{ato}(t_0)$, $S_\mathrm{int}(t_0)$.
		\item \textbf{for~$k=1,\dots,n_T$}
		\begin{enumerate}[(i):]
			\item Apply the boundary conditions at time~$t_k$, $\mathcal{T}_k=\mathcal{T}_{k-1}$, $N_\mathrm{rep}(t_k) = N_\mathrm{rep}(t_{k-1})$, $\bs{\Phi}(t_k)=\bs{\Phi}(t_{k-1})$, $S_\mathrm{ato}(t_k) =S_\mathrm{ato}(t_{k-1})$, $S_\mathrm{int}(t_k) = S_\mathrm{int}(t_{k-1})$, etc.
			\item Equilibrate the unbalanced system, i.e. solve for~$\bs{q}_\mathrm{red}(t_k)\in\mathscr{Q}_\mathrm{red}(t_k)$ using~$\widehat{\Pi}^k(\widehat{\bs{q}}_\mathrm{red};\bs{q}_\mathrm{red}(t_{k-1}))$ of Eq.~\eqref{SubSect:Summation:Eq:3}. Update crack description~$\psi$, $C$, etc.
			\item For each coarse element~$K\in\mathcal{T}_k$ evaluate its refinement indicator in~\eqref{SubSect:Refinement:Eq:2} and construct~$\mathcal{I}_\mathrm{r}\subseteq\mathcal{T}_k$. If~$\mathcal{I}_\mathrm{r}\neq\emptyset$ refine and update~$\mathcal{T}_k$.
			\item Protect newly added repatoms from all previous refinements during the current time step~$t_k$; if necessary, protect also other repatoms, e.g. those near the crack tip. Use~\eqref{SubSect:Coarsening:Eq:2} to identify~$\mathcal{I}_\mathrm{c}\subseteq\mathcal{T}_k$. If possible, coarsen and update~$\mathcal{T}_k$.
			\item \textbf{If}~$\mathcal{T}_k$ changed in (iii)~-- (iv), update the system information~$N_\mathrm{rep}(t_k)$, $\bs{\Phi}(t_k)$, $S_\mathrm{ato}(t_k)$, $S_\mathrm{int}(t_k)$, etc., and return to~(ii) since the system is unbalanced.\\
			\textbf{Else if} the mesh has converged; proceed to~(vi).
			\item Store relevant outputs: $\bs{q}_\mathrm{red}(t_k)$, $\mathcal{T}_k$, $N_\mathrm{rep}(t_k)$, $\bs{\Phi}(t_k)$, $S_\mathrm{ato}(t_k)$, $S_\mathrm{int}(t_k)$, $C$, etc.
		\end{enumerate}
		\item \textbf{end}
	\end{enumerate}
	\vspace{-\topsep}
  \end{minipage}
\end{algorithm}
%
%
\subsection{Energy Implications}
\label{SubSect:EnImplications}
This section discusses the implications of the adaptive scheme summarized in Alg.~\ref{SubSect:XQC:Alg:1} from an energetic point of view, and extends the discussion presented in~\cite{VarQCDamage}, Section~3.5. Recall that the motivations for energy considerations are threefold. First, the entire theory presented so far is based on energy minimization and the adaptive procedure should be consistent with these principles. Second, validity of the solution~$\bs{q}(t_k)$, as well as of~$\bs{q}_\mathrm{red}(t_k)$, is decided based on the energy balance~\eqref{E}, which must hold during the entire evolution of the system. Finally, upon taking into account energy implications, the accuracy of the X-QC framework can be assessed from the energetic point of view.
\begin{figure}
	\centering
	\includegraphics[scale=1]{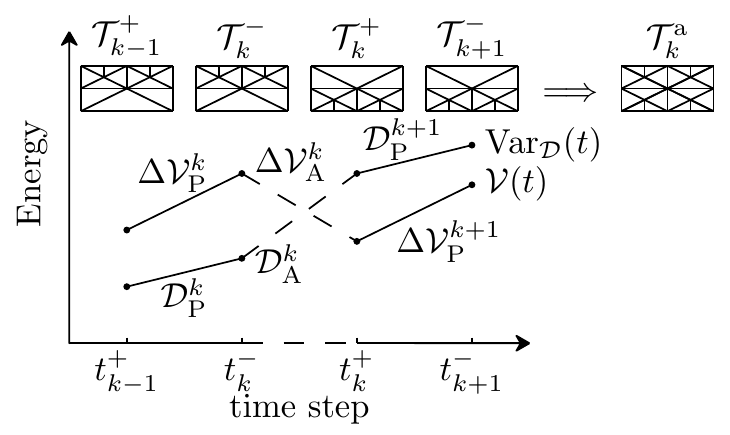}
	\caption{Sketch of possible energy evolutions during time step~$t_k$ when refinement and coarsening occur. Only the~$\mathcal{V}(t)$ and~$\mathrm{Var}_\mathcal{D}(t)$ components are shown for clarity.}
	\label{SubSect:EnImplications:Fig:1}
\end{figure}

The issue focused on here is illustrated in Fig.~\ref{SubSect:EnImplications:Fig:1}. At the end of the previous time step~$t_{k-1}^+$, the system is relaxed for some mesh~$\mathcal{T}_{k-1}^+$. After the next increment is applied for mesh~$\mathcal{T}_k^- = \mathcal{T}_{k-1}^+$ and the system is equilibrated (stage~(ii) of Alg.~\ref{SubSect:XQC:Alg:1}), refinement and coarsening may be required ((ii)~-- (iv)). Upon convergence, this yields a triangulation~$\mathcal{T}_k^+$. In order to accurately construct the energy increments, one should distinguish physical increments, computed on mesh~$\mathcal{T}^+_{k-1}$ and denoted with subscripts~"P", and changes in energy which are due to the mesh adaptation from~$\mathcal{T}^-_k$ to~$\mathcal{T}^+_k$, denoted with subscripts~"A". The artificial energy increments due to mesh adaptation are computed by projecting converged solutions onto an artificial mesh~$\mathcal{T}_k^\mathrm{a}$, which establishes communication between the two systems (note that these two systems have different numbers of DOFs and internal variables). The artificial mesh therefore contains the union of repatoms and sampling interactions present in both systems and hence, it is locally the finest triangulation of the two, cf. Fig.~\ref{SubSect:EnImplications:Fig:1}. Using the above-described procedure and evaluating energies at time instants~$t_k^+$ provides energy evolution paths that we will call \emph{reconstructed}. For further details see~\cite{VarQCDamage}, Section~3.5.

Let us note that we omit dissipative coarsening (i.e. the interpolation of the internal variables when the mesh is coarsened), and assume that the coarsening occurs only in elastic regions. This assumption is reasonable because damage is localized. Consequently, all damaged bonds are captured accurately by the special function enrichments and associated summation rule.
%
%
\section{Numerical Examples}
\label{Sect:Examples}
In this section the framework is applied to two examples discussed already in~\cite{VarQCDamage}, where coarsening was not considered. Consequently, we can make a direct comparison in terms of accuracy and efficiency between the adaptive QC frameworks with and without coarsening.

\begin{figure}
	\centering
	\includegraphics[scale=1]{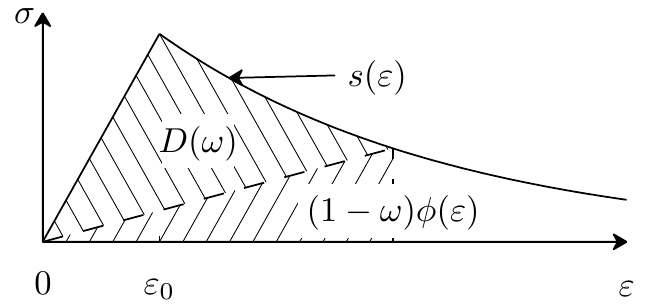}
	\caption{Exponential softening under tension: a sketch of the stress--strain diagram with corresponding quantities~$\varepsilon_0$ and~$s(\varepsilon)$, cf. Eq.~\eqref{Examples:Eq:3}. The elastically stored energy, $(1-\omega)\phi(\varepsilon)$, corresponds to the area of the dashed triangle, and the dissipated energy, $D(\omega)$, to the area of the upper triangle with the curved edge.}
	\label{Sect:Examples:Fig:1}
\end{figure}

The same material model is used for each lattice spring (the superscripts~$\alpha\beta$ are dropped for brevity):
\begin{equation}
\phi(\widehat{r}) = \frac{1}{2}E A r_0(\widehat{\varepsilon}(\widehat{r}))^2,
\label{Examples:Eq:1}
\end{equation}
where~$E$ is the Young's modulus, $A$ the cross-sectional area, and~$\widehat{\varepsilon} = (\widehat{r} - r_0)/r_0$ the bond strain. The dissipation function~$D$ in Eq.~\eqref{SubSect:LNDamage:Eq:4} is adopted from~\cite{VarQCDamage}, Appendix~A, providing an exponential softening rule
\begin{equation}
\widehat{\sigma} = s(\widehat{\varepsilon}) = E\varepsilon_0\exp{\left(-\frac{\widehat{\varepsilon}-\varepsilon_0}{\varepsilon_f}\right)}, \quad \varepsilon_0 \leq \widehat{\varepsilon}.
\label{Examples:Eq:3}
\end{equation}
In Eq.~\eqref{Examples:Eq:3}, $s(\widehat{\varepsilon})$ describes the softening branch of the associated stress-strain diagram, $\widehat{\sigma}$ is the bond stress, $\varepsilon_f$ measures inverse of the initial slope of the softening branch, and~$\varepsilon_0$ is the limit elastic strain, cf. Fig.~\ref{Sect:Examples:Fig:1}. The employed parameters are specified in Tab.~\ref{Sect:Examples:Tab:1}.
\begin{table}
	\caption{Dimensionless material and geometric parameters for both examples.}
	\centering
	\renewcommand{\arraystretch}{1.5}
	\begin{tabular}{l|r@{}lr@{}l}
		Physical parameters                                    & \multicolumn{2}{c}{Example~1} &                \multicolumn{2}{c}{Example~2}                \\ \hline
		Young's modulus, \hfill $E$                            & 1 &                           & 1 &  \\
		Cross-sectional area, \hfill $A$                       & 1 &                           & 1 &  \\
		Lattice spacing, \hfill $a_x$, $a_y$              & 1 &                           & 1 &  \\
		Limit elastic strain, \hfill $\varepsilon_0$           & 0 & .1                        & 0 & .01                                                     \\
		Inverse of initial slope, \hfill $\varepsilon_f$                  & 0 & .25                       & 0 & .025                                                    \\
		Indirect displacement increment, \hfill $\overline{\Delta\ell}$ & 0 & .025                      & \multicolumn{2}{c}{see Fig.~\ref{SubSect:ComplexEx:Fig:2a}}
	\end{tabular}
	\label{Sect:Examples:Tab:1}
\end{table}
%
%
\subsection{L-Shaped Plate Example}
\label{SubSect:Ex1}
\begin{figure}
	\centering
	\includegraphics[scale=1]{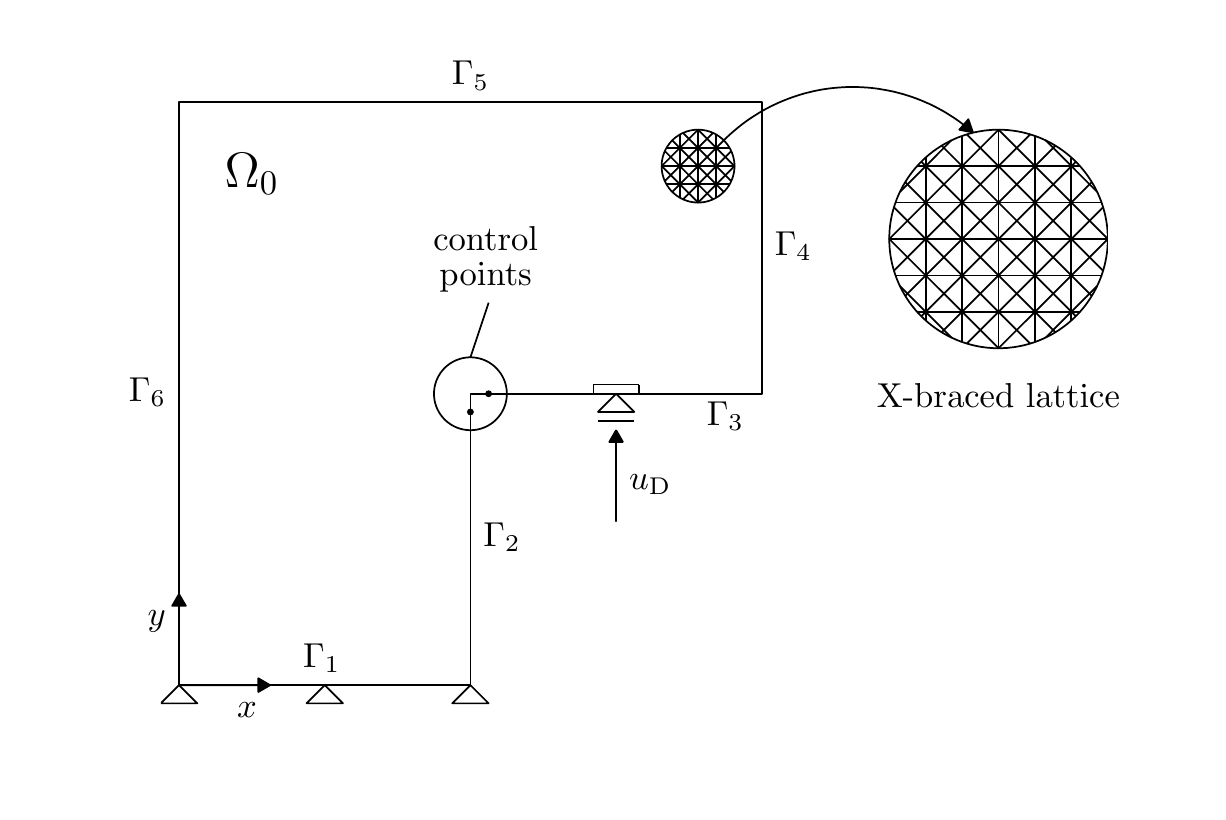}
	\caption{Sketch of the L-shaped plate example: geometry and boundary conditions. Variable~$u_\mathrm{D}$ denotes the applied vertical displacement.}
	\label{SubSect:SimpleEx:Fig:0}
\end{figure}
The first example focuses on an L-shaped plate test, see Fig.~\ref{SubSect:SimpleEx:Fig:0}. The lattice properties are homogeneous except for the vicinity of~$\Gamma_3/2$, where a vertical displacement is applied. Here, in order to prevent any damage, the Young's modulus is~$1,000$ times larger than elsewhere and the limit elastic strain~$\varepsilon_0$ is infinite. The reference domain comprises~$3,201$ atoms, $12,416$ interactions, and is fixed at the bottom part of the boundary, $\Gamma_1$. The evolution of the system is controlled by the vertical difference between the pair of atoms highlighted in Fig.~\ref{SubSect:SimpleEx:Fig:0}, which after crack initiation directly corresponds to the Crack Mouth Opening Displacement (CMOD) control.

The numerical simulation is performed for two X-QC systems, each with a different safety margin~$\theta_\mathrm{r}$, cf. Eqn.~\eqref{SubSect:Refinement:Eq:1}. The first system is referred to as the \emph{moderate X-QC} ($\theta_\mathrm{r} = 0.5$) and the second one to as the \emph{progressive X-QC} ($\theta_\mathrm{r} = 0.25$); in both cases, $\theta_\mathrm{c} = 0.05$, recall Eq.~\eqref{SubSect:Coarsening:Eq:1}. Relatively low safety margins are necessary due to the highly fluctuating interaction energies near the crack tip (recall Fig.~\ref{SubSect:NodeProtection:Fig:1} and the discussion therein). The results are compared to those of the adaptive version of the variational QC, reported in~\cite{VarQCDamage}, for the same choices of the refinement safety margins, i.e.~$\theta_\mathrm{r} = 0.5$ and~$\theta_\mathrm{r} = 0.25$, and for the central summation rule. These systems are referred to as the \emph{moderate adaptive QC} and the \emph{progressive adaptive QC}. For completeness, results of the DNS are also shown.

The deformed configuration computed by the DNS at~$u_\mathrm{D} = 14$ is depicted in Fig.~\ref{SubSect:SimpleEx:Fig:1a}. The crack path initiates at the inner corner ($\Gamma_2 \cap \Gamma_3$) and propagates to the left. This is also predicted by the QC frameworks. In Fig.~\ref{SubSect:SimpleEx:Fig:1b}, the reaction force~$F$ is presented as a function of~$u_\mathrm{D}$. The responses predicted by the X-QC are practically identical to those of the adaptive QC without coarsening.
\begin{figure}
	\centering
	\begin{tikzpicture}
	\linespread{1}
	
		\node[inner sep=0pt] (Ldeformed) at (0,0) {
			\subfloat[$\bs{r}(t_k)$ for~$u_\mathrm{D} = 14$]{\includegraphics[scale=1]{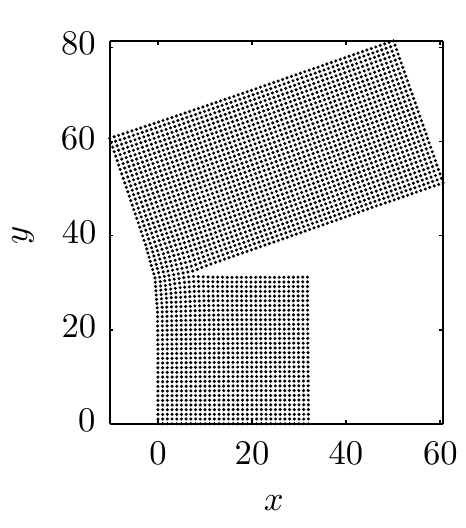}\label{SubSect:SimpleEx:Fig:1a}}
		};
		\node[inner sep=0pt] (Lreact) at (6.5,0) {
			\subfloat[force-displacement diagram]{\includegraphics[scale=1]{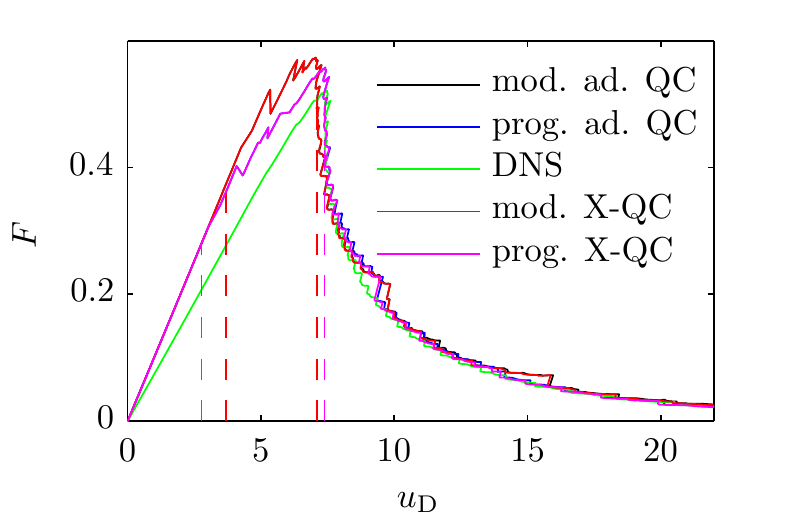}\label{SubSect:SimpleEx:Fig:1b}}
		};
		\node[inner sep=0pt] (Lzoom) at (11.5,-0.72) {
			\includegraphics[scale=1]{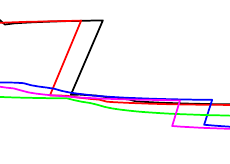}
		};
	
		\draw[black, thick, dashed, rounded corners] (8,-1.1) rectangle (8.7,-0.6);

		\draw[black] (10.3,-1.3) rectangle (12.7,0);

		\draw[black, thick, dashed] (8.35,-0.6) -- (10.3,0);
		\draw[black, thick, dashed] (8.35,-1.1) -- (10.3,-1.3);

		\node (text) at (11.5,0.2) {\footnotesize Zoom in};
	\end{tikzpicture}
	\caption{L-shaped plate test: (a)~the deformed (unscaled) configuration at~$u_\mathrm{D} = 14$ predicted by the DNS, (b)~the force-displacement diagram for the reaction force~$F$ as a function of~$u_\mathrm{D}$. The instants at which mesh refinement and coarsening occur for the first time are indicated by dashed vertical lines of the respective colours.}
	\label{SubSect:SimpleEx:Fig:1}
\end{figure}
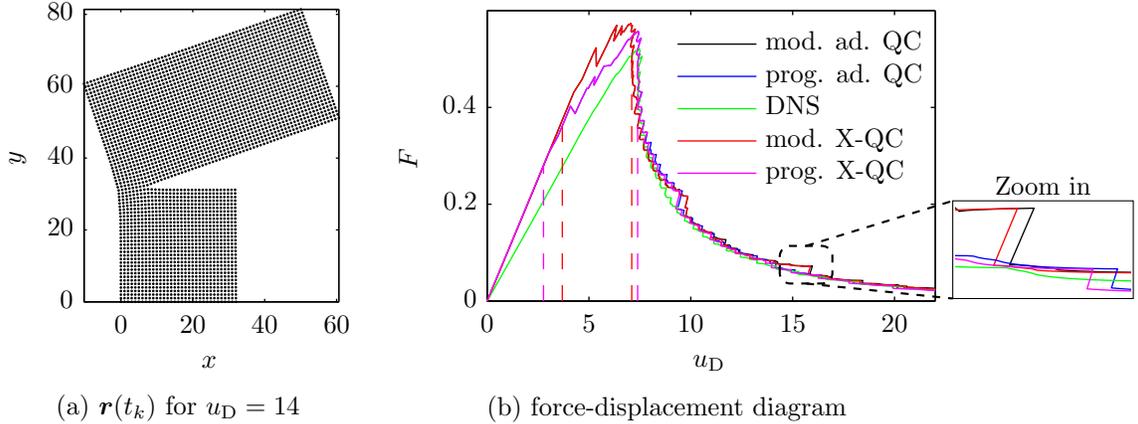
\begin{figure}
	\centering
	\subfloat[energy evolutions]{\includegraphics[scale=1]{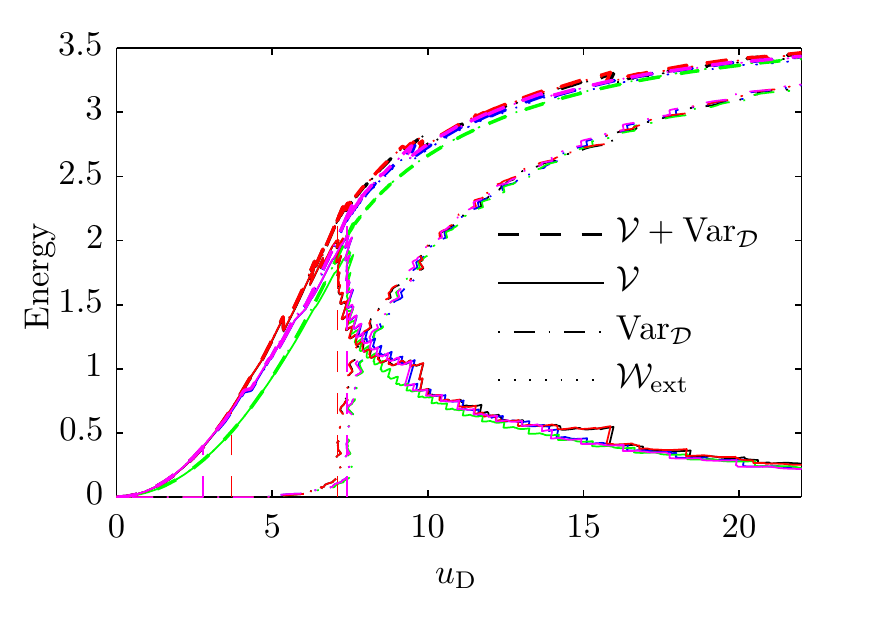}\label{SubSect:SimpleEx:Fig:2a}}\hfill
	\subfloat[energy components for moderate X-QC]{\includegraphics[scale=1]{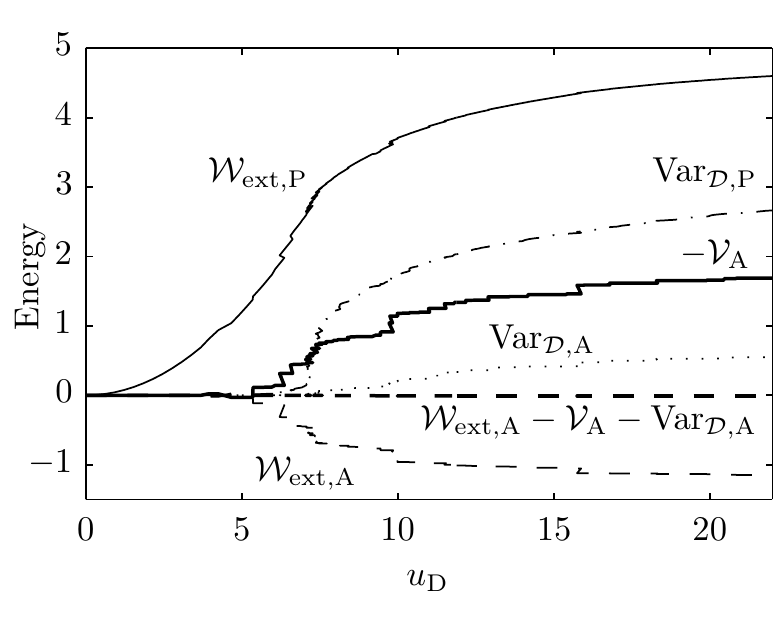}\label{SubSect:SimpleEx:Fig:2b}}
	\caption{Energy evolutions for the L-shaped plate test. (a)~The energy evolutions (black~-- the moderate adaptive QC, blue~-- the progressive adaptive QC, green~-- DNS, red~-- moderate X-QC, magenta~-- progressive X-QC). (b)~Energy exchanges due to mesh refinement and coarsening for the moderate X-QC approach, cf. Section~\ref{SubSect:EnImplications}.}
	\label{SubSect:SimpleEx:Fig:2}
\end{figure}

The energy evolutions appear in Fig.~\ref{SubSect:SimpleEx:Fig:2a} as functions of~$u_\mathrm{D}$. The energy balance~\eqref{E} is clearly satisfied along the entire loading path, i.e. the thin dotted lines presenting the work performed by external forces~$\mathcal{W}_\mathrm{ext}$ lie on top of the thick dashed lines representing~$\mathcal{V} + \mathrm{Var}_\mathcal{D}$. Second, similarly to the reaction forces, we notice that the energy paths are almost indistinguishable from those of adaptive QC, demonstrating good accuracy of the X-QC approach. In Fig.~\ref{SubSect:SimpleEx:Fig:2b}, we present the energy components that are exchanged (the artificial energies) in the moderate X-QC simulation. Their high magnitudes compared to the curves of Fig.~\ref{SubSect:SimpleEx:Fig:2a} indicate the importance of the energy reconstruction procedure, discussed in Section~\ref{SubSect:EnImplications}.

\begin{figure}
	\centering
	\includegraphics[scale=1]{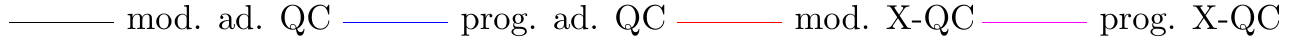}\\
	\subfloat[relative number of repatoms]{\includegraphics[scale=1]{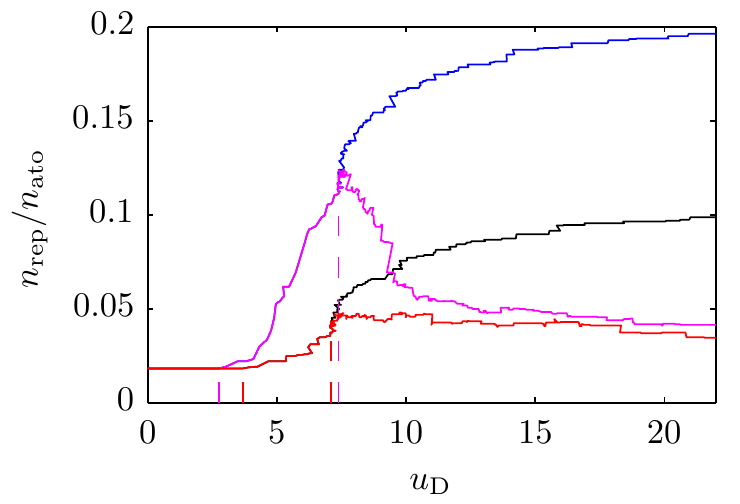}\label{SubSect:SimpleEx:Fig:3a}}\hspace{0.5em}
	\subfloat[relative number of sampling atoms]{\includegraphics[scale=1]{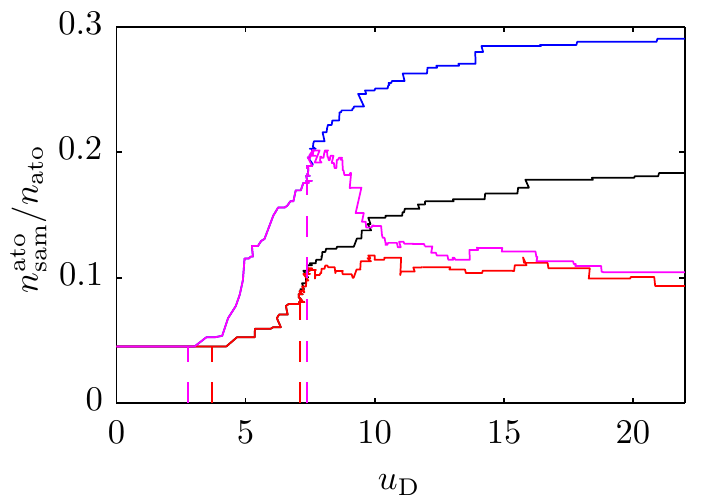}\label{SubSect:SimpleEx:Fig:3b}}
	\caption{(a)~Relative number of repatoms~$n_\mathrm{rep}/n_\mathrm{ato}$ as a function of~$u_\mathrm{D}$, and (b)~relative number of sampling atoms.}
	\label{SubSect:SimpleEx:Fig:3}
\end{figure}

Fig.~\ref{SubSect:SimpleEx:Fig:3a} presents the relative number of repatoms as a function of~$u_\mathrm{D}$. Note that the number of enriched repatoms~$n^\star$, defined in Eq.~\eqref{SubSect:Enrichment:Eq:3}, is negligible (in particular, $n^\star/n_\mathrm{rep} < 0.06$). Both X-QC systems initially follow the trends of the corresponding adaptive QC systems (the refinement stage and crack localization). After the crack localizes and starts to propagate however, the crack wake unloads and the mesh coarsens, resulting in a substantial drop in the number of repatoms of the progressive X-QC. At the end of the simulation, the specimen is almost stress-free, allowing for a very coarse representation. Note that the number of DOFs associated with the X-QC approach is less than one half compared to the adaptive QC in the case of the moderate approach, and approximately a quarter in the case of the progressive approach. In both cases, the final relative number of DOFs as well as repatoms remain below~$5\,\%$, whereas they never exceed~$15\,\%$ during the simulation. A similar behaviour can be observed also for the relative number of sampling atoms, presented in Fig.~\ref{SubSect:SimpleEx:Fig:3b}.

\begin{figure}
	\centering
	\subfloat[$u_\mathrm{D} = 0$]{\includegraphics[scale=0.5]{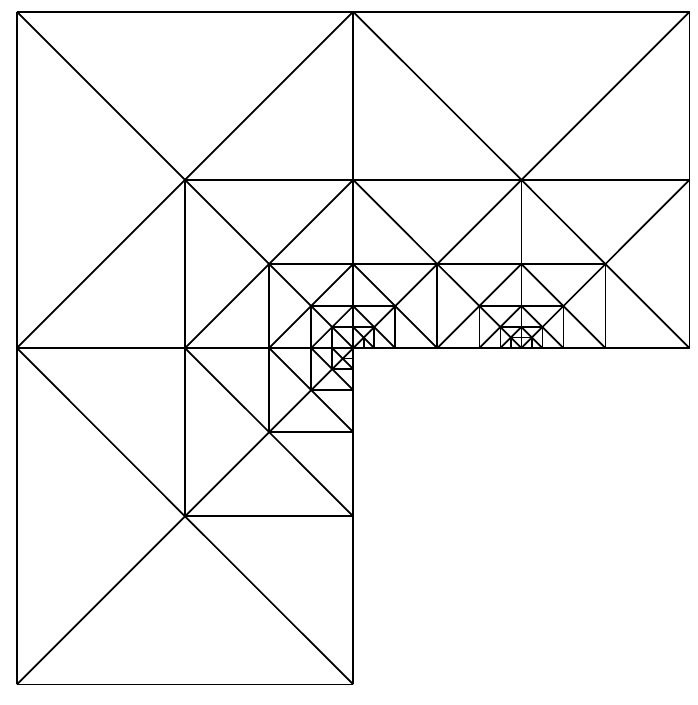}\label{SubSect:SimpleEx:Fig:4a}}\hspace{0.2em}
	\subfloat[$u_\mathrm{D} = 7$]{\includegraphics[scale=0.5]{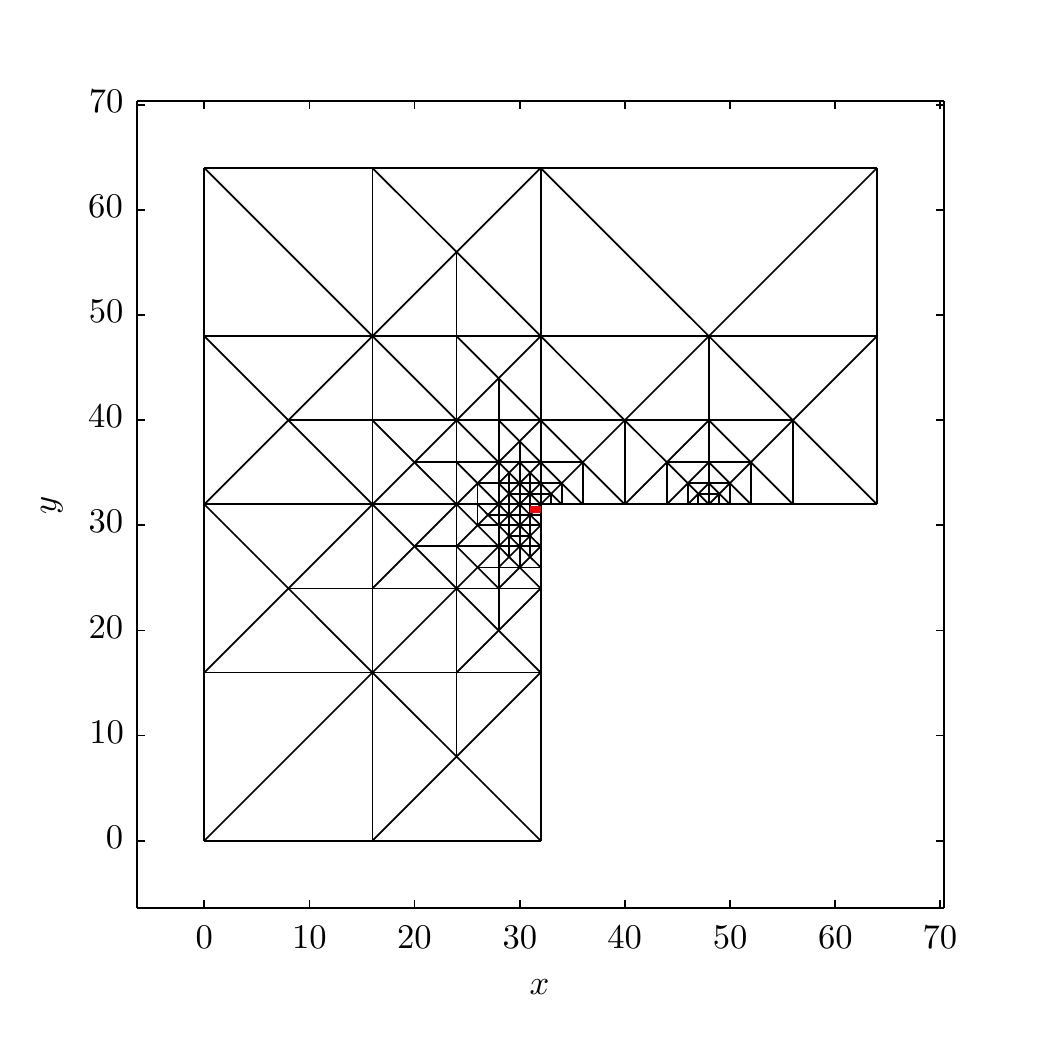}\label{SubSect:SimpleEx:Fig:4b}}\hspace{0.2em}
	\subfloat[$u_\mathrm{D} = 14$]{\includegraphics[scale=0.5]{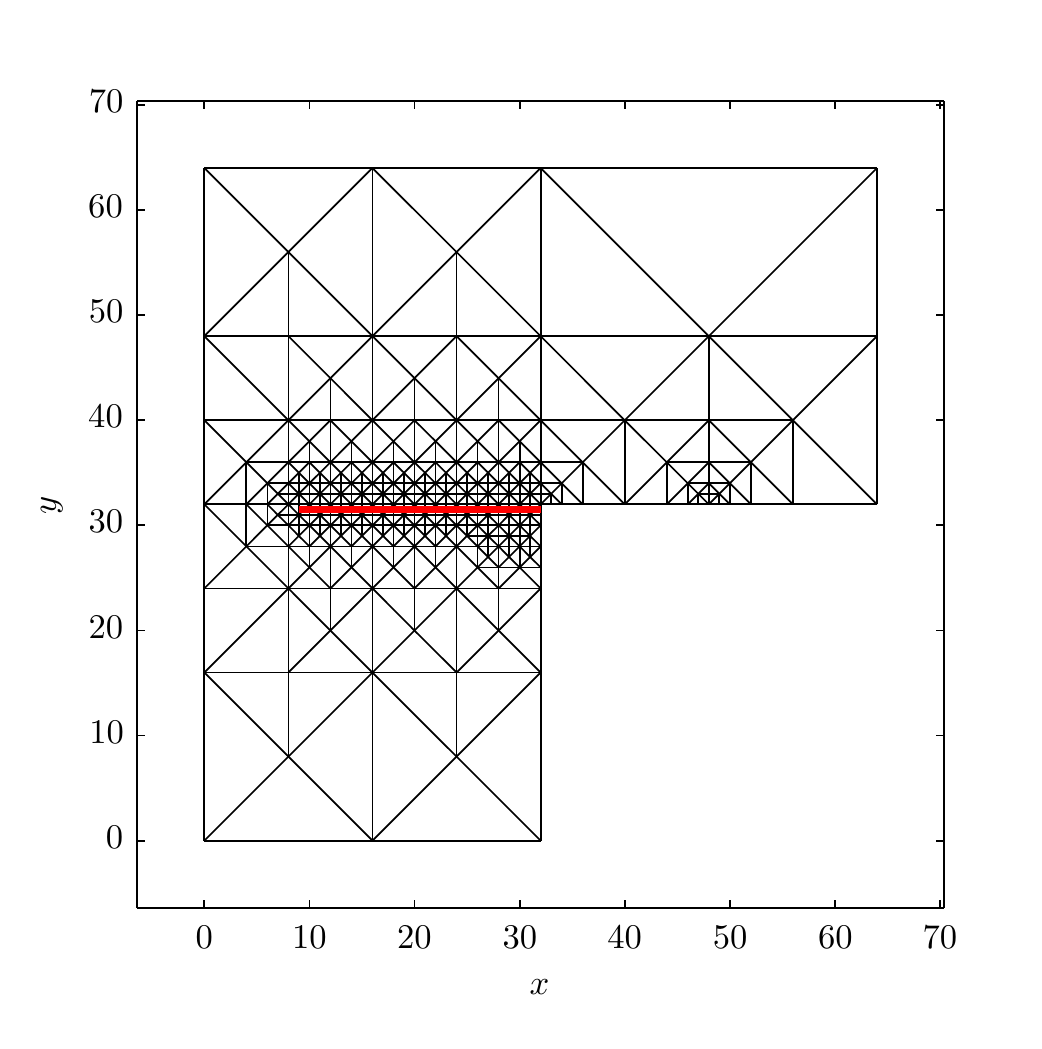}\label{SubSect:SimpleEx:Fig:4c}}\hspace{0.2em}
	\subfloat[$u_\mathrm{D} = 21$]{\includegraphics[scale=0.5]{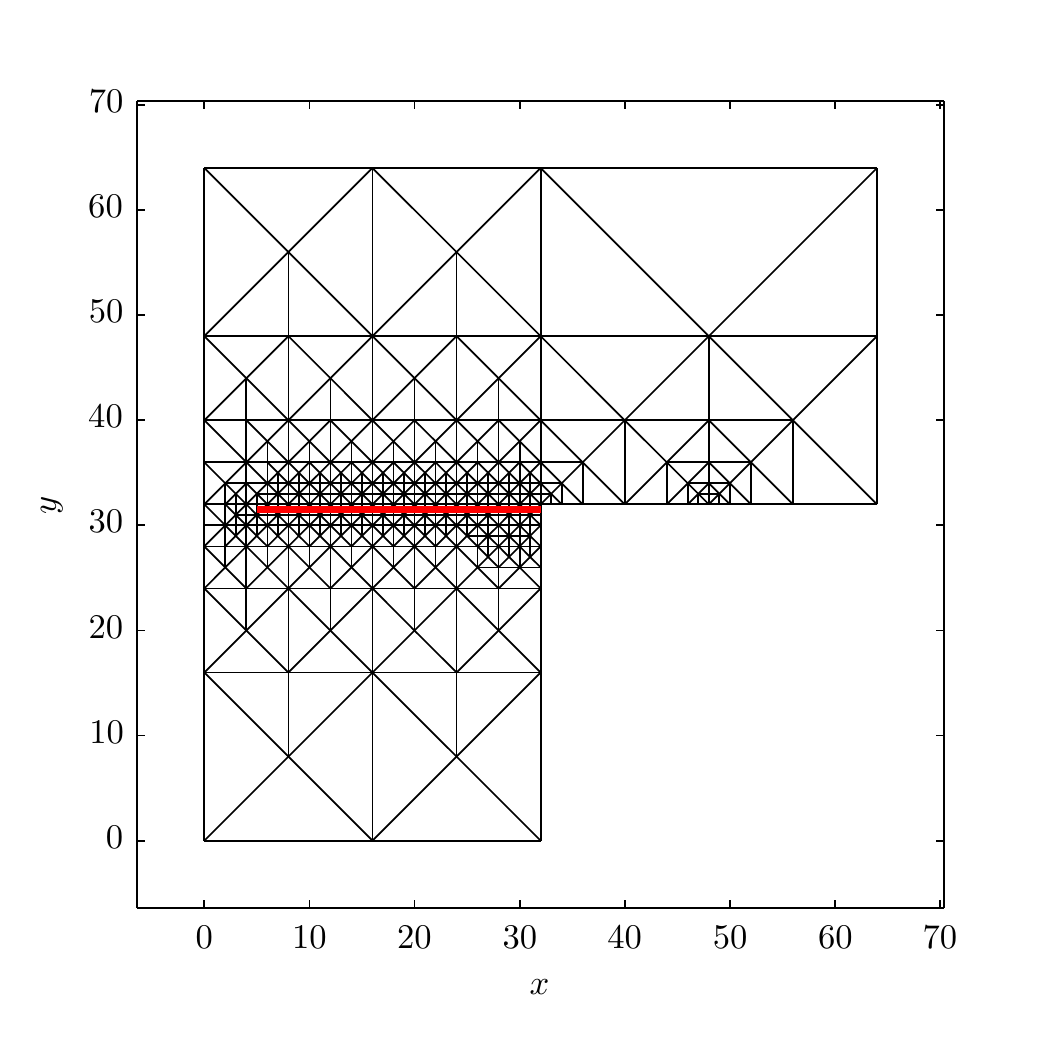}\label{SubSect:SimpleEx:Fig:4d}}\\	
	\subfloat[$u_\mathrm{D} = 0$]{\includegraphics[scale=0.5]{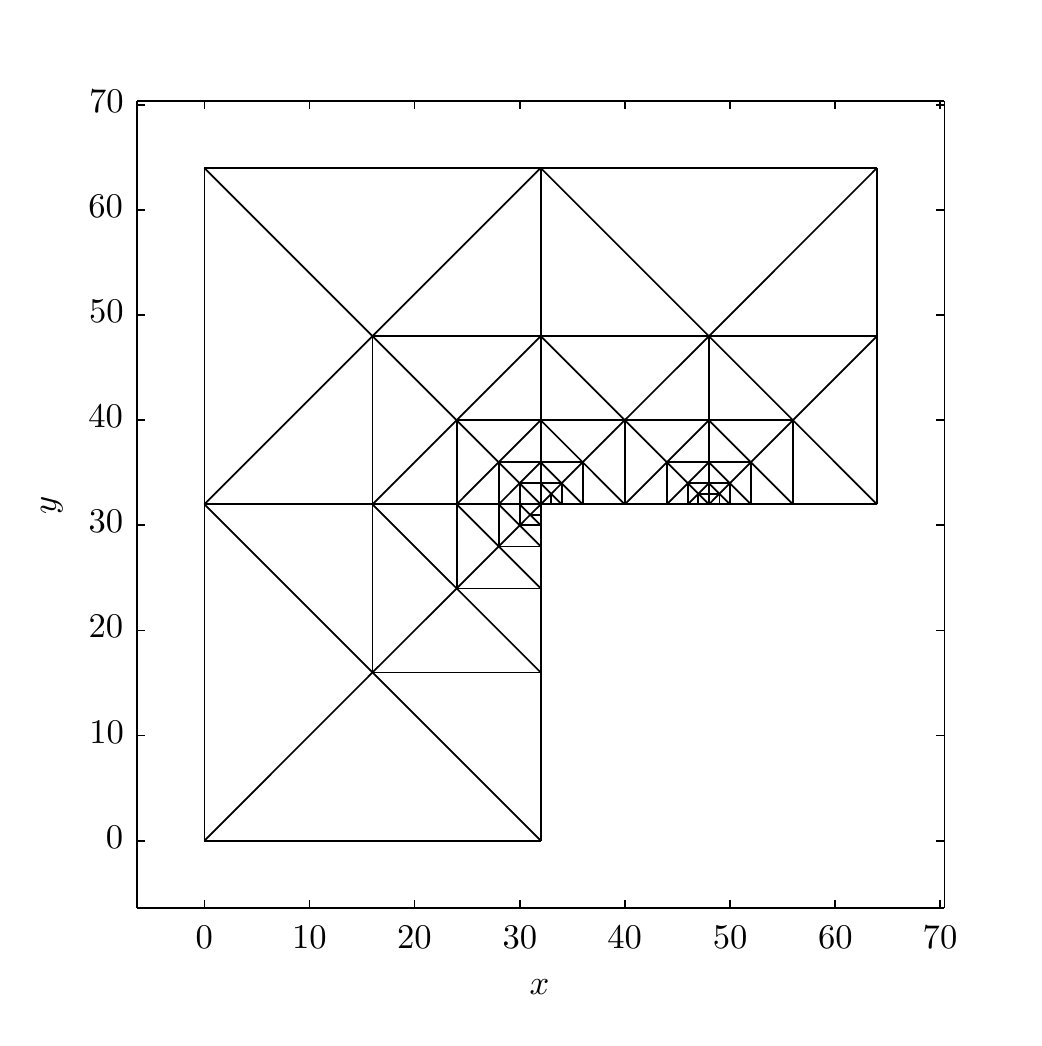}\label{SubSect:SimpleEx:Fig:4e}}\hspace{0.2em}
	\subfloat[$u_\mathrm{D} = 7$]{\includegraphics[scale=0.5]{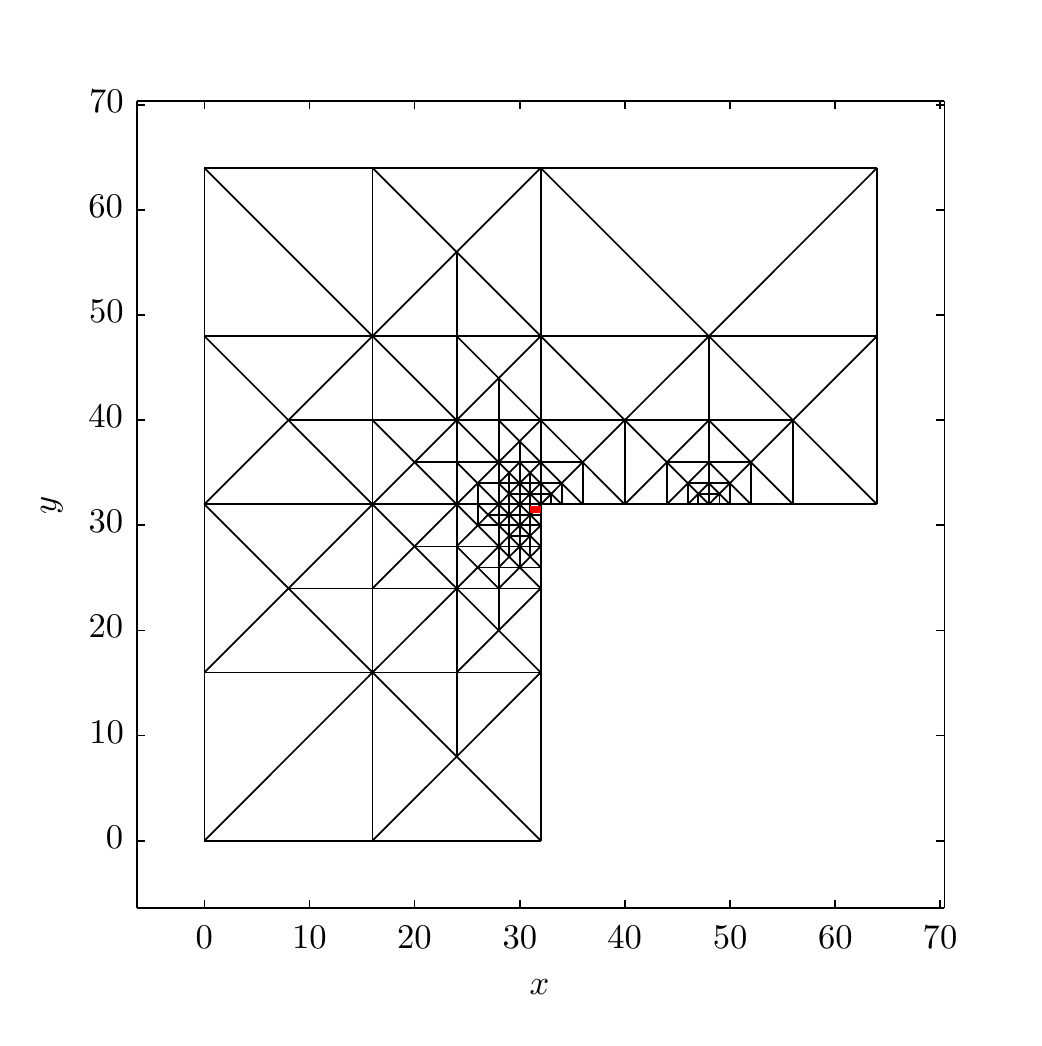}\label{SubSect:SimpleEx:Fig:4f}}\hspace{0.2em}
	\subfloat[$u_\mathrm{D} = 14$]{\includegraphics[scale=0.5]{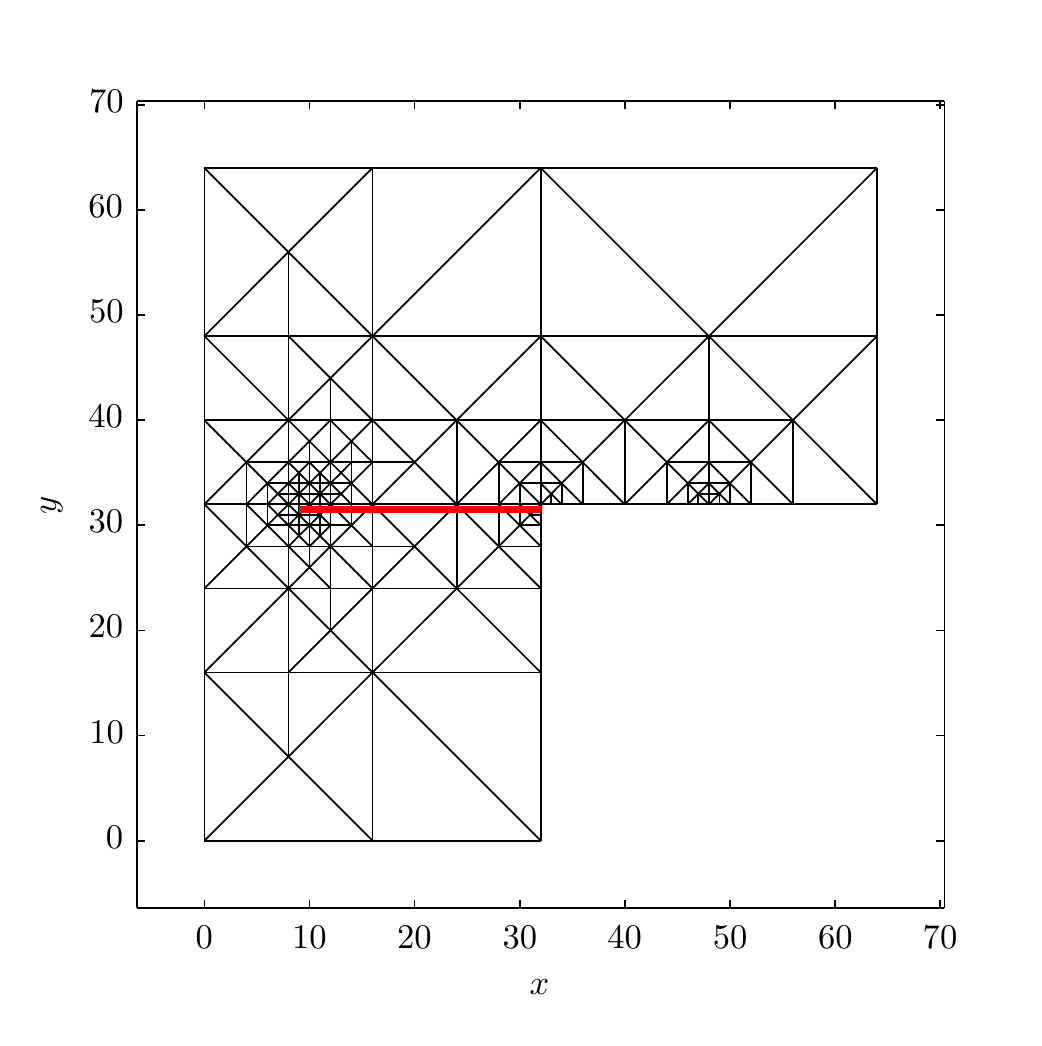}\label{SubSect:SimpleEx:Fig:4g}}\hspace{0.2em}
	\subfloat[$u_\mathrm{D} = 21$]{\includegraphics[scale=0.5]{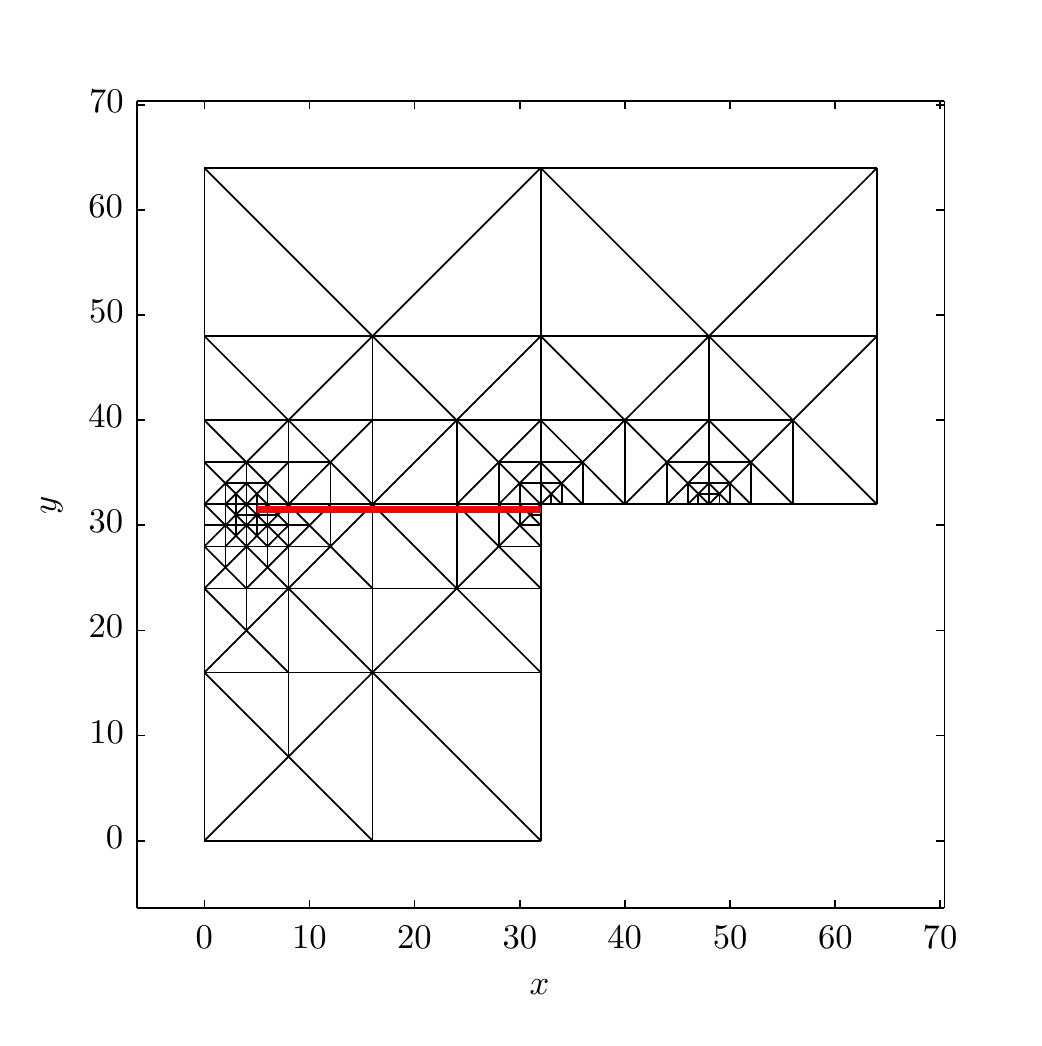}\label{SubSect:SimpleEx:Fig:4h}}
	\caption{Eight triangulations for the L-shaped plate test: (a)~-- (d) the moderate adaptive QC approach, (e)~-- (h) the moderate X-QC approach. The red line indicates the crack. The relative numbers of repatoms and sampling atoms corresponding to these meshes are shown in Fig.~\ref{SubSect:SimpleEx:Fig:3}.}
	\label{SubSect:SimpleEx:Fig:4}
\end{figure}

\begin{figure}
	\centering
	\subfloat[$u_\mathrm{D} = 0$]{\includegraphics[scale=0.5]{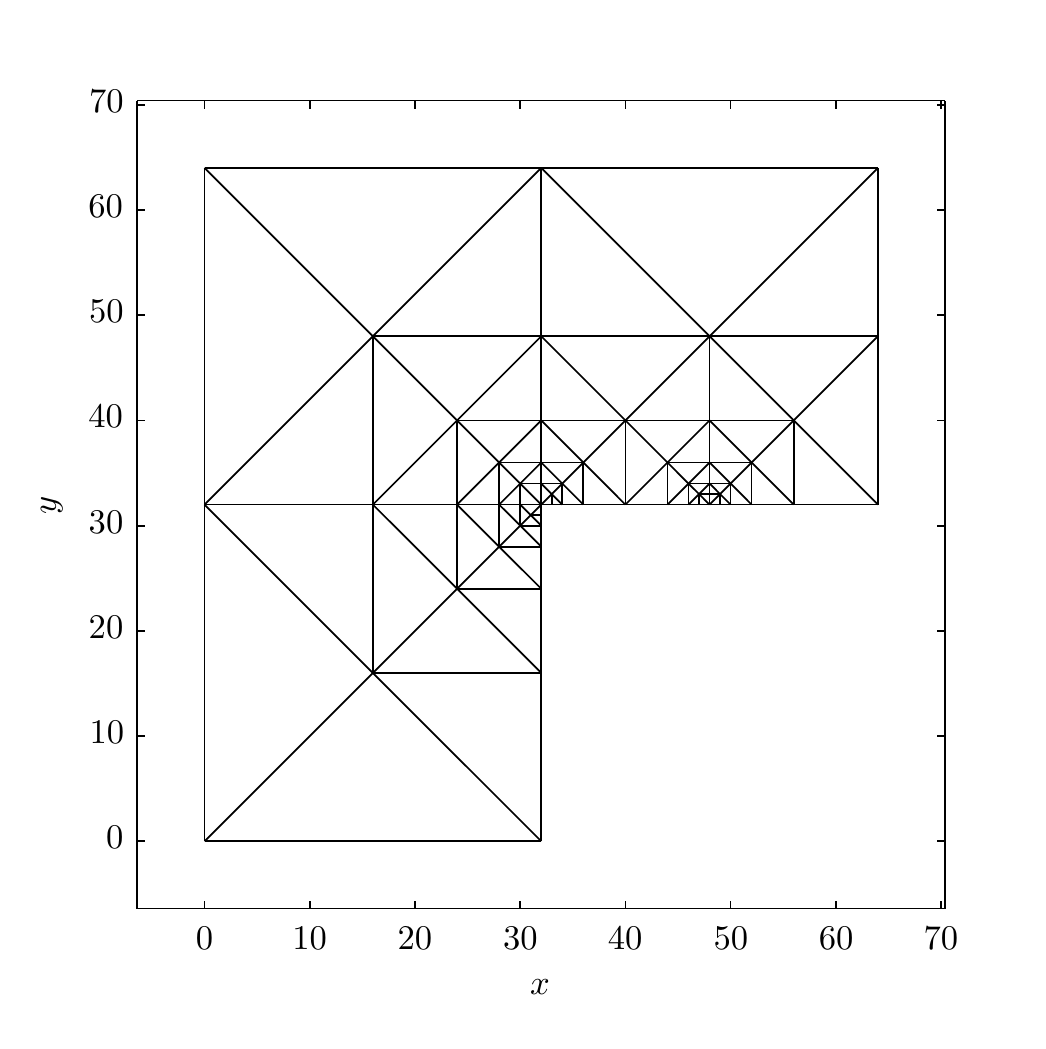}\label{SubSect:SimpleEx:Fig:5a}}\hspace{0.2em}
	\subfloat[$u_\mathrm{D} = 7$]{\includegraphics[scale=0.5]{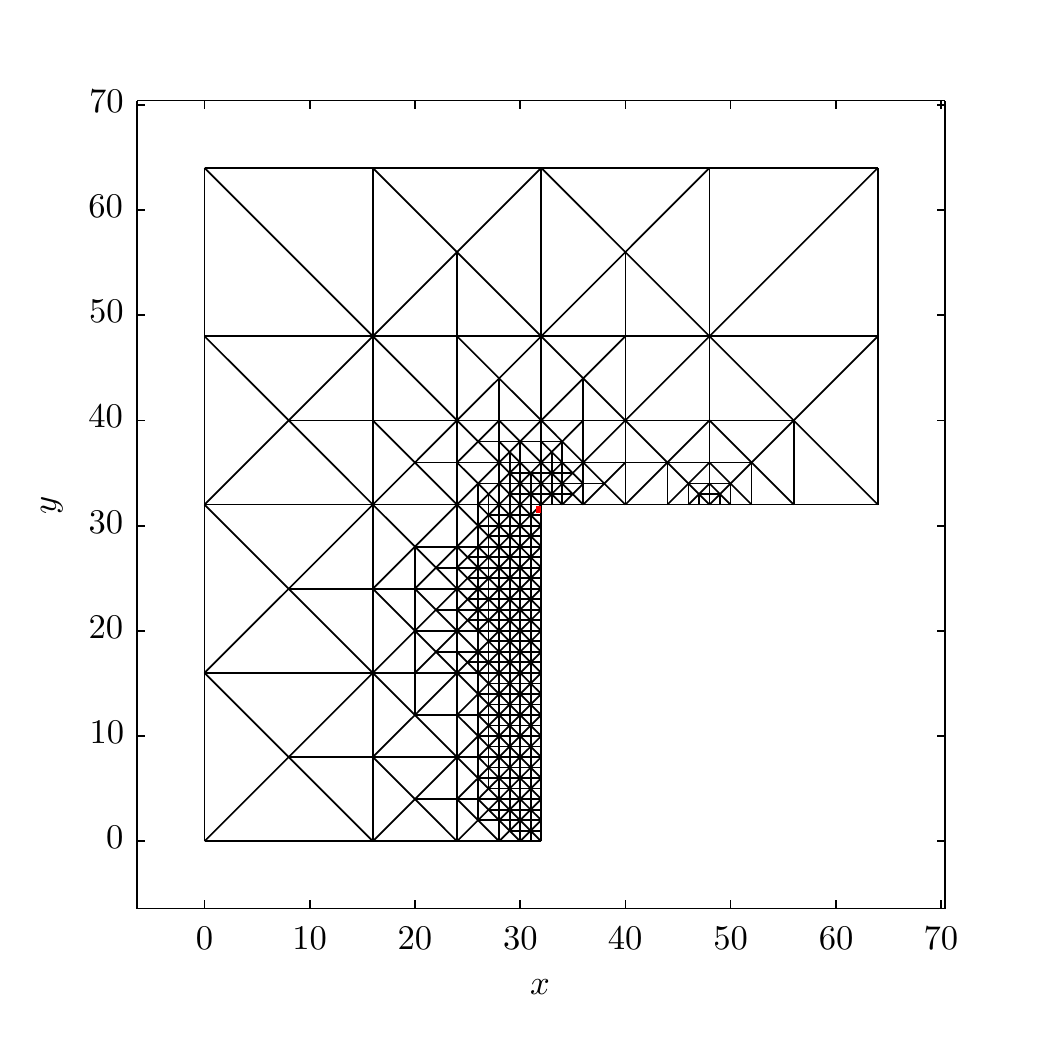}\label{SubSect:SimpleEx:Fig:5b}}\hspace{0.2em}
	\subfloat[$u_\mathrm{D} = 14$]{\includegraphics[scale=0.5]{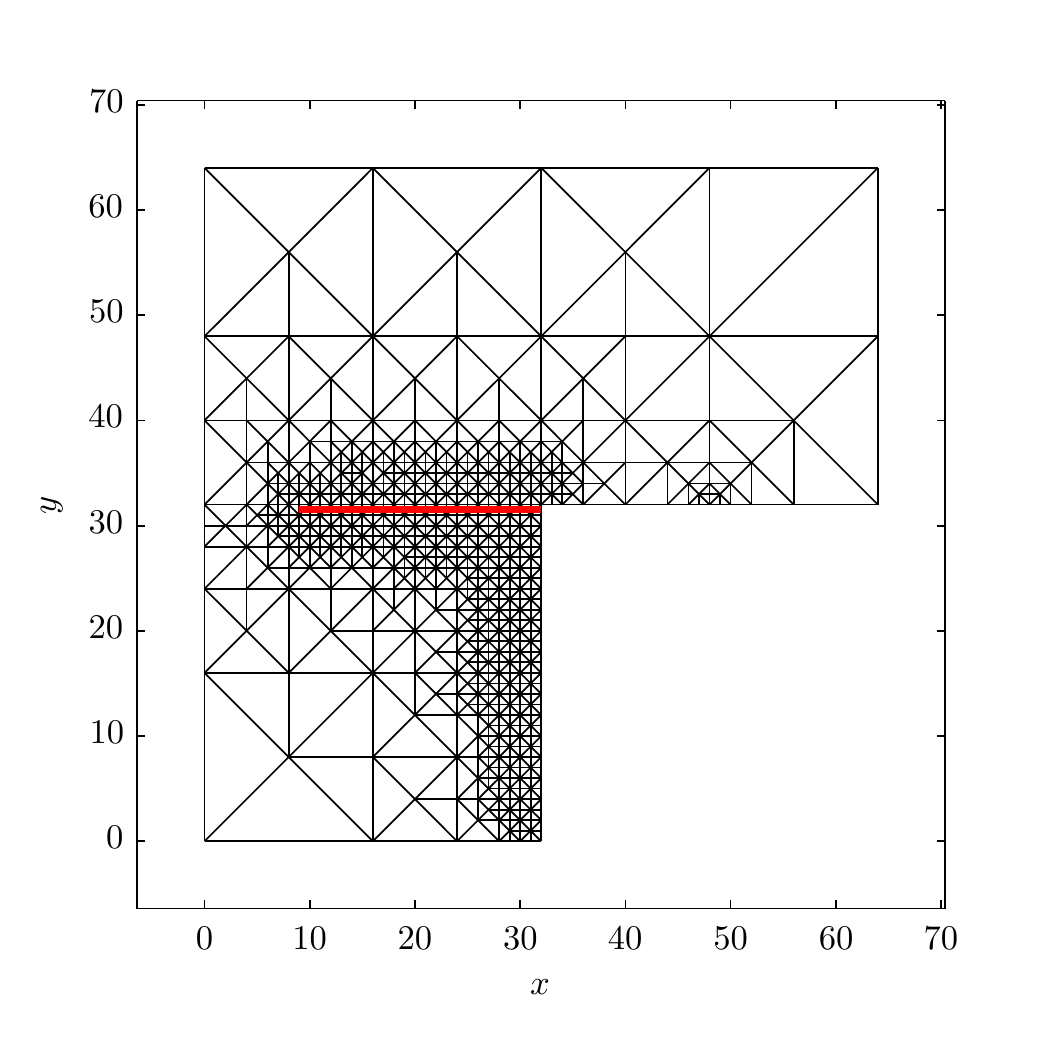}\label{SubSect:SimpleEx:Fig:5c}}\hspace{0.2em}
	\subfloat[$u_\mathrm{D} = 21$]{\includegraphics[scale=0.5]{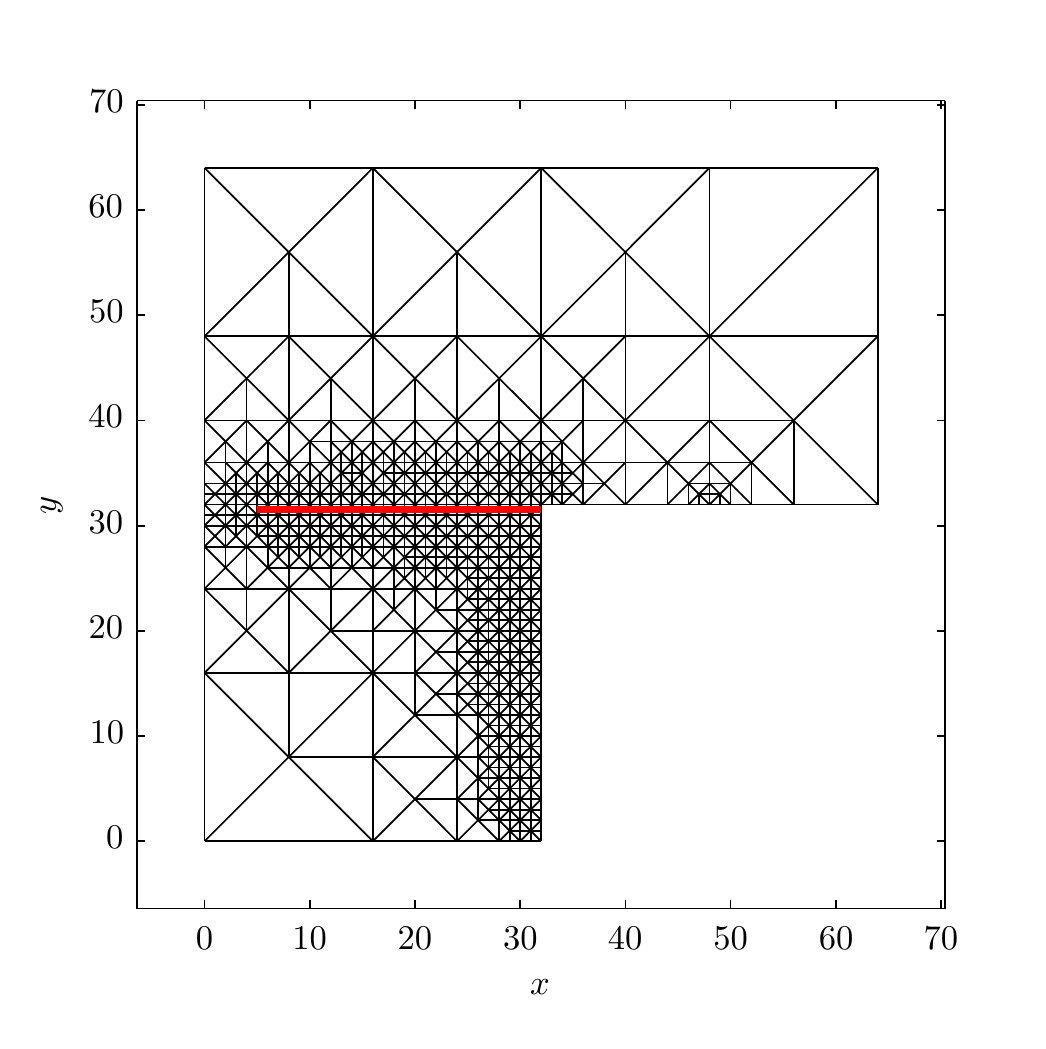}\label{SubSect:SimpleEx:Fig:5d}}\\	
	\subfloat[$u_\mathrm{D} = 0$]{\includegraphics[scale=0.5]{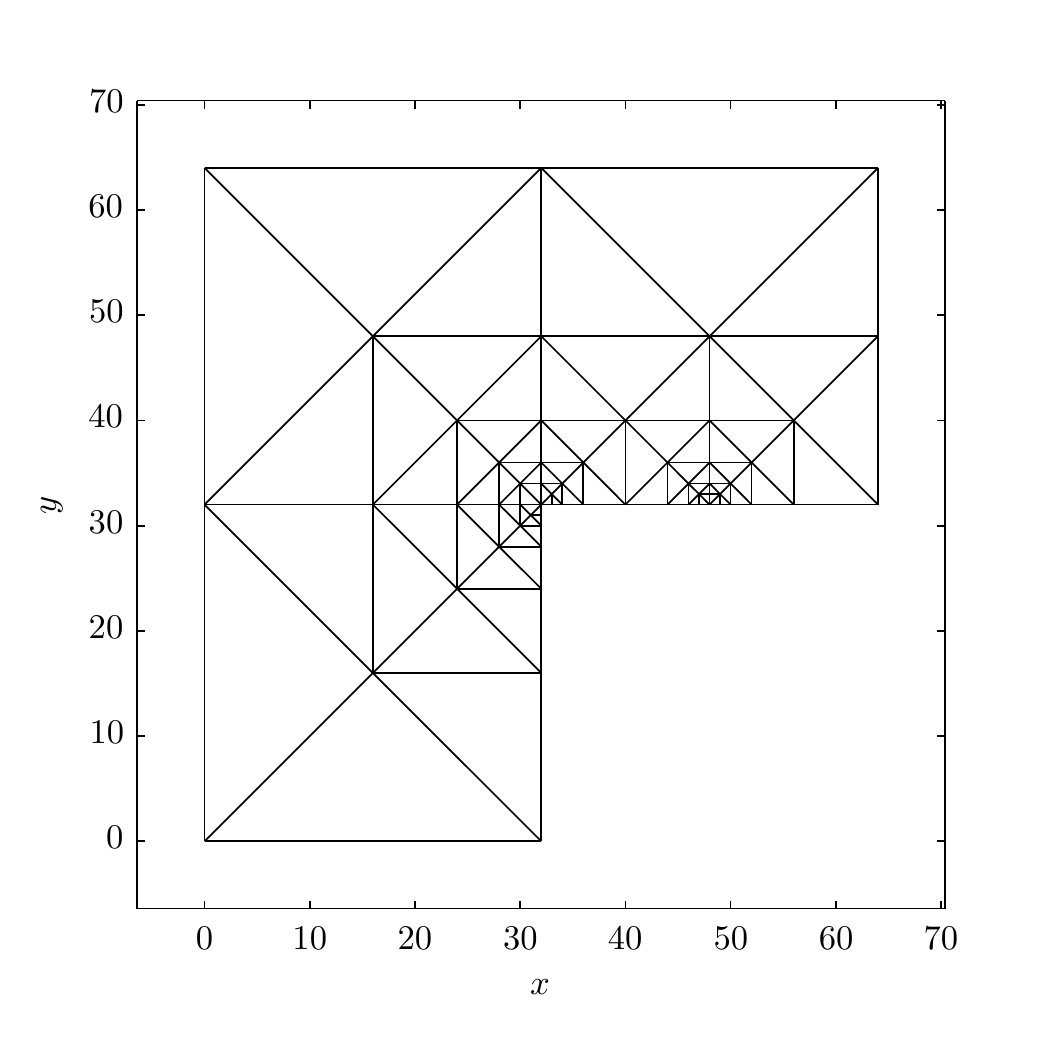}\label{SubSect:SimpleEx:Fig:5e}}\hspace{0.2em}
	\subfloat[$u_\mathrm{D} = 7$]{\includegraphics[scale=0.5]{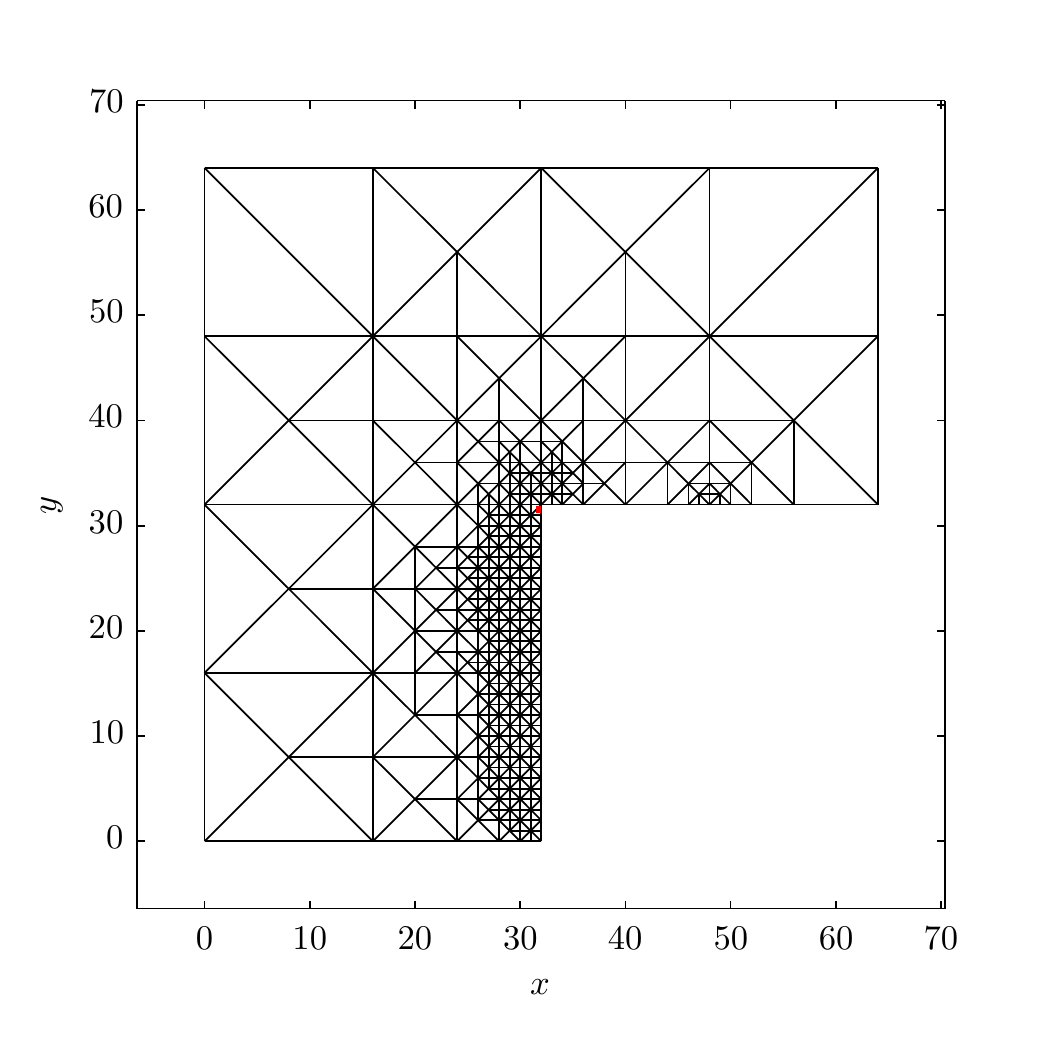}\label{SubSect:SimpleEx:Fig:5f}}\hspace{0.2em}
	\subfloat[$u_\mathrm{D} = 14$]{\includegraphics[scale=0.5]{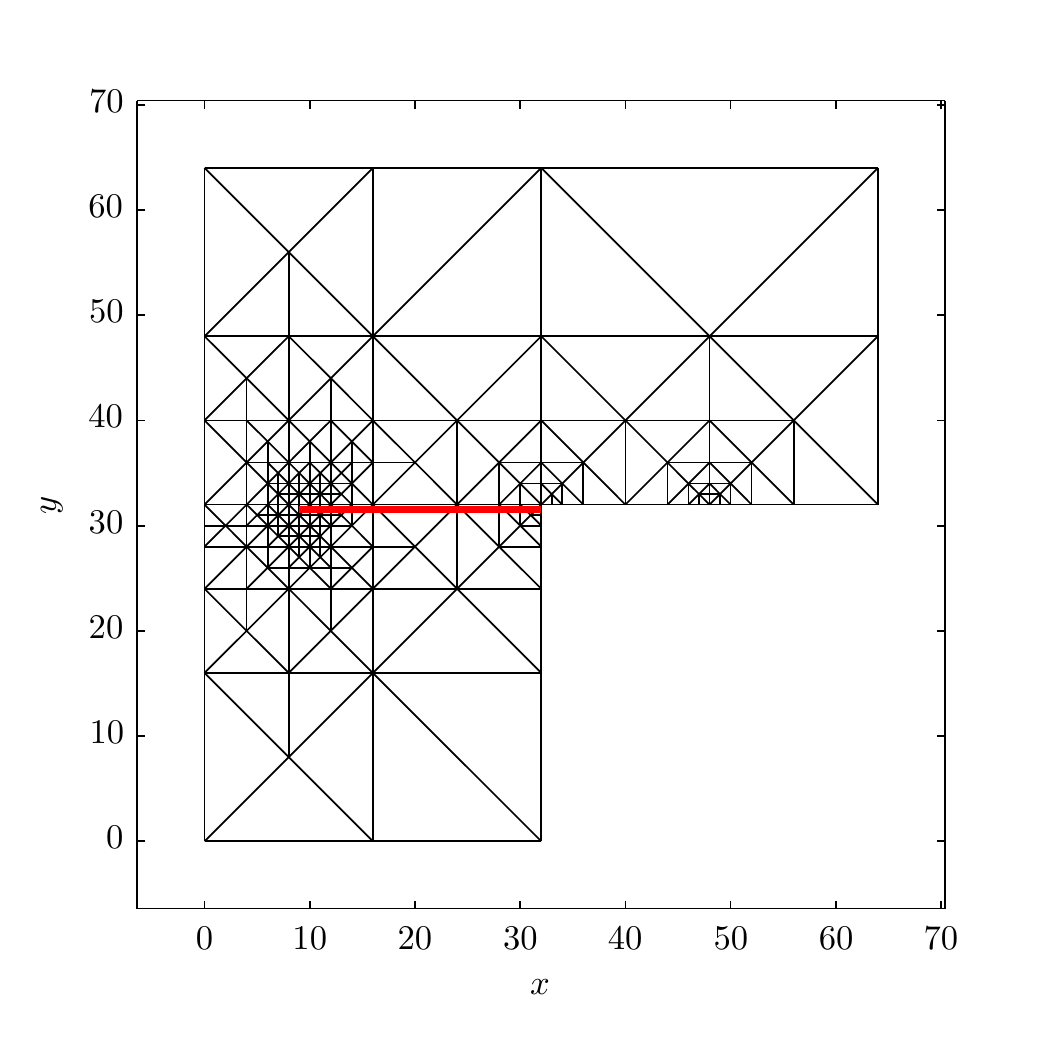}\label{SubSect:SimpleEx:Fig:5g}}\hspace{0.2em}
	\subfloat[$u_\mathrm{D} = 21$]{\includegraphics[scale=0.5]{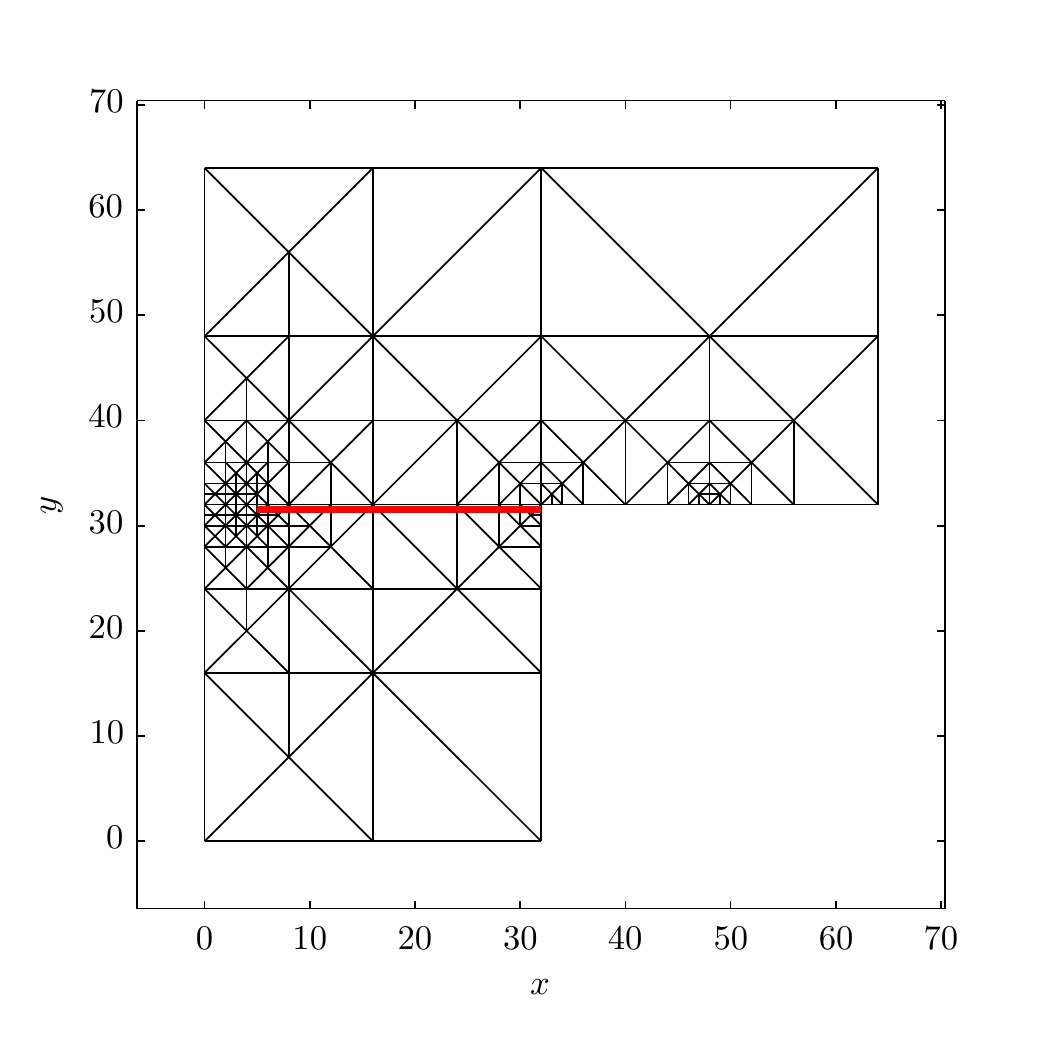}\label{SubSect:SimpleEx:Fig:5h}}
	\caption{Eight triangulations for the L-shaped plate test: (a)~-- (d) the progressive adaptive QC approach, (e)~-- (h) the progressive X-QC approach. The red line indicates the crack. The relative numbers of repatoms and sampling atoms corresponding to these meshes are shown in Fig.~\ref{SubSect:SimpleEx:Fig:3}.}
	\label{SubSect:SimpleEx:Fig:5}
\end{figure}

Finally, Figs.~\ref{SubSect:SimpleEx:Fig:4} and~\ref{SubSect:SimpleEx:Fig:5} present some of the meshes of the X-QC and the adaptive QC frameworks associated with Fig.~\ref{SubSect:SimpleEx:Fig:3}. Although at the initial stages the progressive and moderate X-QC meshes differ substantially, they are almost identical when the crack localizes. The effect of coarsening in both cases is evident. Note that in the progressive case coarsening occurs not only along the crack path, but also along the right edge of the lower part of the plate. This explains the particularly large gain in number of repatoms in this case, i.e. the factor of four visible in Fig.~\ref{SubSect:SimpleEx:Fig:3a}.
%
%
\subsection{Antisymmetric Four-Point Bending Test}
\label{SubSect:Ex2}
\begin{figure}
	\centering
	\includegraphics[scale=1]{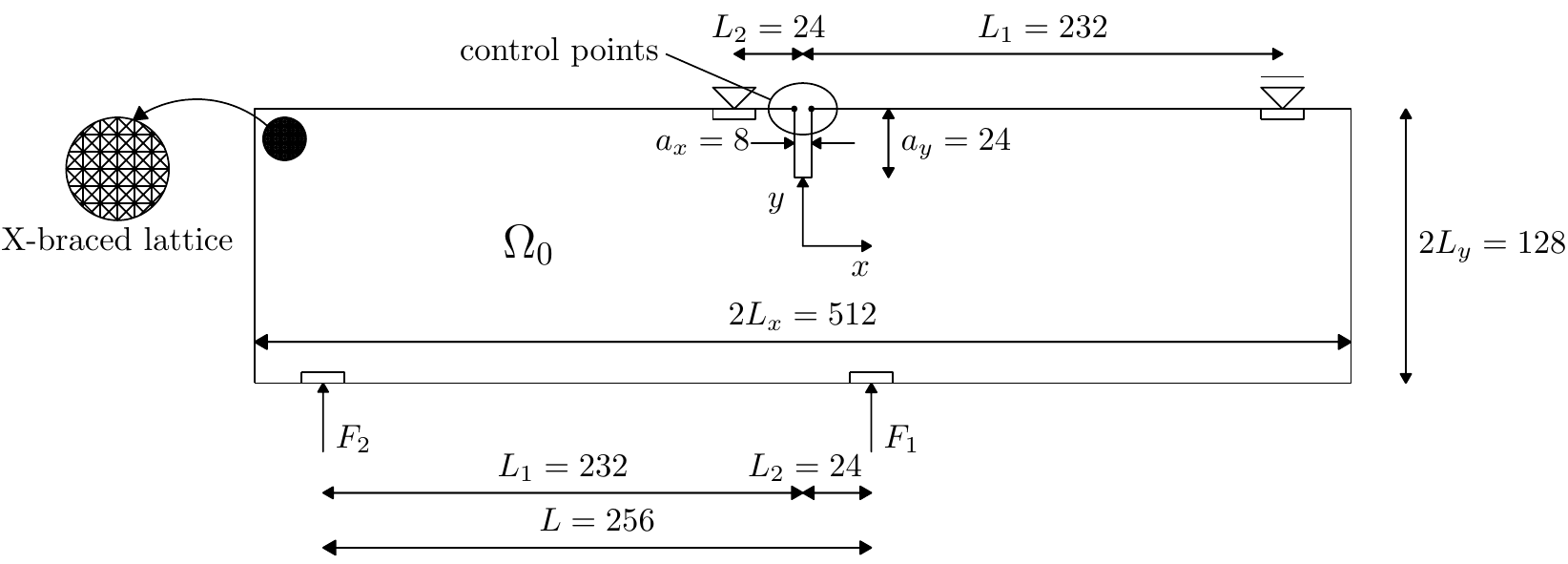}
	\caption{Sketch of the four-point bending test: geometry and boundary conditions.}
	\label{SubSect:ComplexEx:Fig:1}
\end{figure}
\begin{figure}
	\centering
	\subfloat[load control parameters]{\includegraphics[scale=1]{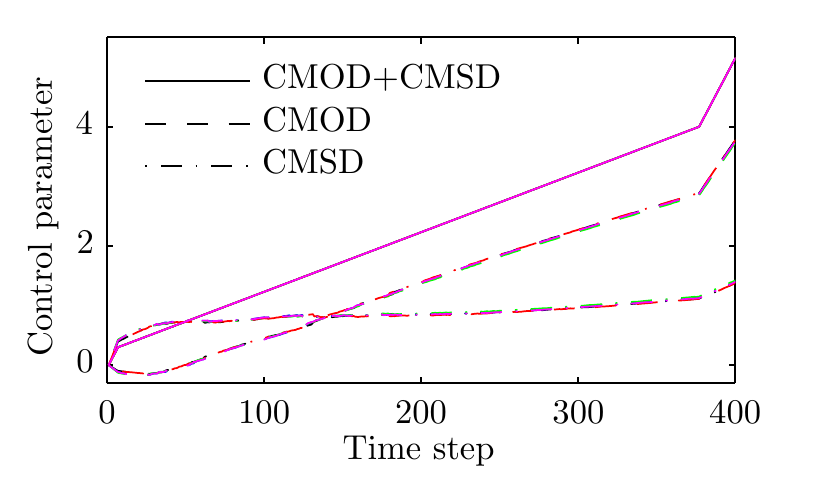}\label{SubSect:ComplexEx:Fig:2a}}\hspace{0.2em}
	\subfloat[$\bs{r}(t_k)$ for~$\mathrm{CMOD}+\mathrm{CMSD} = 5.5$]{\includegraphics[scale=1]{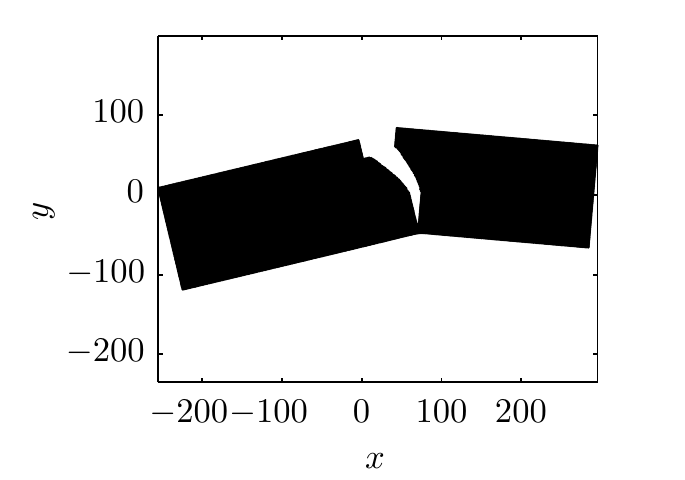}\label{SubSect:ComplexEx:Fig:2b}}
	\caption{Four-point bending test: (a)~evolution of CMOD, CMSD, and their sum (the applied loading program); black~-- the moderate adaptive QC, blue~-- the progressive adaptive QC, green~-- DNS, red~-- moderate X-QC, magenta~-- progressive X-QC (note that they are almost indistinguishable). (b)~The deformed configuration for the DNS; displacements are magnified by a factor of~$10$.}
	\label{SubSect:ComplexEx:Fig:2}
\end{figure}

The second example is a four-point bending test, presented e.g. in~\cite{SchlangenThesis}, cf. also Fig.~\ref{SubSect:ComplexEx:Fig:1}. The rectangular reference domain~$\Omega_0$ is pre-notched at the top to initiate a crack. The otherwise homogeneous X-braced lattice is again locally stiffened where prescribed displacements or forces are applied (the Young's modulus is~$1,000$ times larger than elsewhere and the limit elastic strain~$\varepsilon_0$ is infinite to prevent damage). The domain comprises~$66,009$ atoms connected by~$262,040$ interactions and is loaded by a pair of vertical forces~$F_1$ and~$F_2$, given by
\begin{equation}
F_1 = \frac{L_1}{L}\lambda, \quad F_2 = \frac{L_2}{L}\lambda,
\label{SubSect:ComplexEx:Eq1}
\end{equation}
where~$\lambda$ represents an additional parameter used for indirect displacement solution control, see e.g.~\citealt{Jirasek:2002}, Section~22.2.3. Unlike the L-shaped plate example, the sum of CMOD and CMSD (Crack Mouth Sliding Displacement) is now used to control the simulation. In Fig.~\ref{SubSect:ComplexEx:Fig:2a} the loading program is specified.

Again two X-QC systems are studied: the moderate X-QC with~$\theta_\mathrm{r} = 0.5$, and the progressive X-QC with~$\theta_\mathrm{r} = 0.25$. The results are again compared to those of the adaptive QC (with the same values of~$\theta_\mathrm{r}$) and to those of the DNS.

The deformed configuration predicted by the DNS at~$\mathrm{CMOD} + \mathrm{CMSD} = 5.5$ is depicted in Fig.~\ref{SubSect:ComplexEx:Fig:2b}. The X-QC crack paths, shown in Fig.~\ref{SubSect:ComplexEx:Fig:3a}, are practically identical to those of the adaptive QC (the maximum distance amounts to one lattice spacing). They also match the crack path predicted by the DNS, where the distance does not exceed four lattice spacings in all cases. The same quality of approximation is achieved also for the total applied force, $F_1 + F_2$, plotted in Fig.~\ref{SubSect:ComplexEx:Fig:3b} against~$\mathrm{CMOD}+\mathrm{CMSD}$. Some minor differences can be observed between the responses of the X-QC and the adaptive QC around the peak load (when the crack localizes). At this stage, the X-QC meshes start to coarsen. At later stages, all curves become practically indistinguishable as the specimen becomes stress-free.
\begin{figure}
	\centering
	\begin{tikzpicture}
	\linespread{1}
	
		\node[inner sep=0pt] (Bcrack) at (0,0) {
			\subfloat[crack paths]{\includegraphics[scale=1]{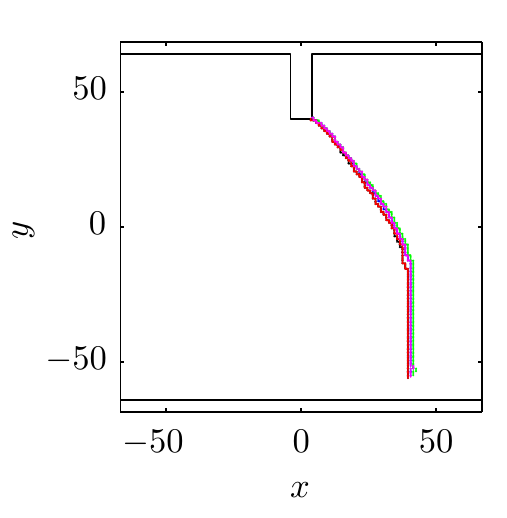}\label{SubSect:ComplexEx:Fig:3a}}
		};
		\node[inner sep=0pt] (Breact) at (9,0.2) {
			\subfloat[force-opening diagram]{\includegraphics[scale=1]{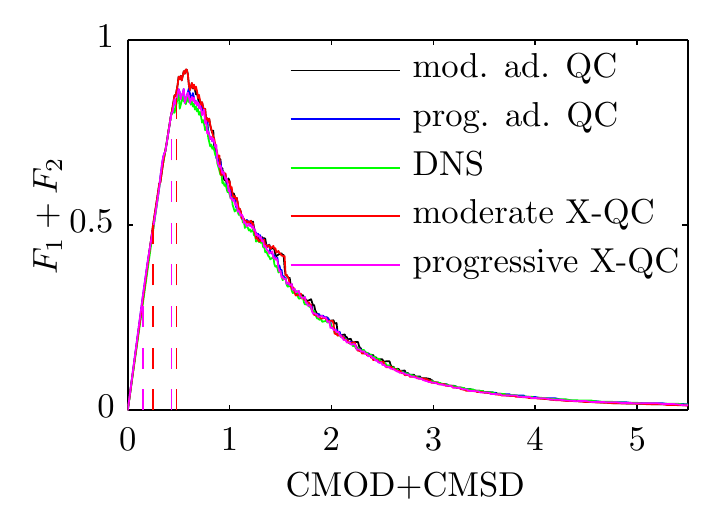}\label{SubSect:ComplexEx:Fig:3b}}
		};
		\node[inner sep=0pt] (BreactZoom) at (4.2,0.7) {
			\includegraphics[scale=1]{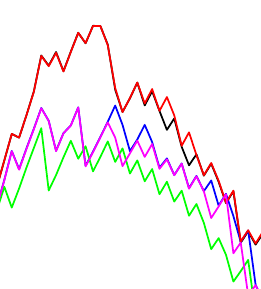}
		};

		\draw[black, thick, dashed, rounded corners] (7,2.3) rectangle (7.7,1.5);

		\draw[black] (2.8,-0.8) rectangle (5.6,2);

		\draw[black, thick, dashed] (7.35,2.3) -- (5.6,2);
		\draw[black, thick, dashed] (7.35,1.5) -- (5.6,-0.8);

		\node (text) at (4.2,2.2) {\footnotesize Zoom in};
	\end{tikzpicture}
	\caption{Four-point bending test: (a)~crack paths, and~(b) force-opening diagram, i.e.~$\lambda = F_1 + F_2$ versus~$\mathrm{CMOD}+\mathrm{CMSD}$. Note that the colour scheme applies to both diagrams. The instants at which mesh refinement and coarsening occur for the first time are indicated by vertical dashed lines of the respective colours.}
	\label{SubSect:ComplexEx:Fig:3}
\end{figure}
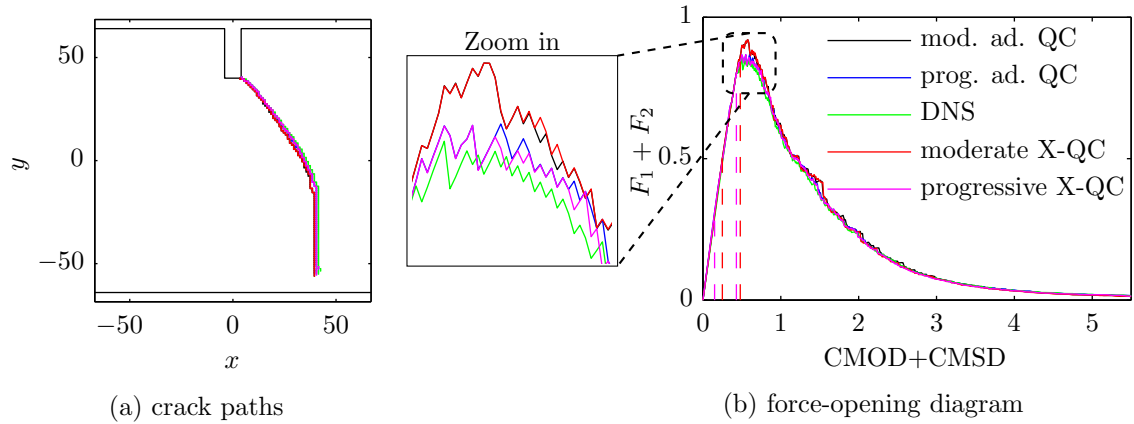
\begin{figure}
	\centering
	\subfloat[energy evolutions]{\includegraphics[scale=1]{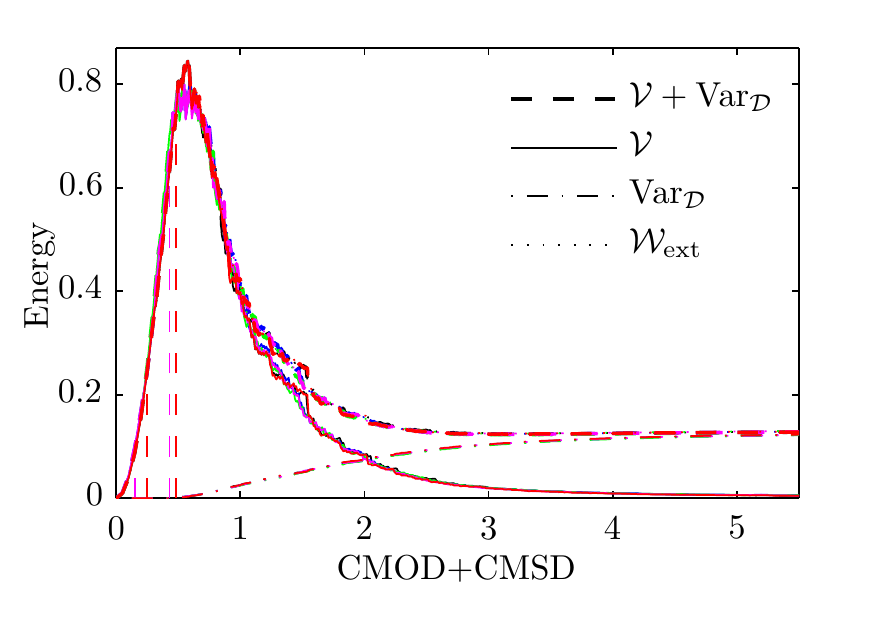}\label{SubSect:ComplexEx:Fig:4a}}\hfill
	\subfloat[energy components for moderate approach]{\includegraphics[scale=1]{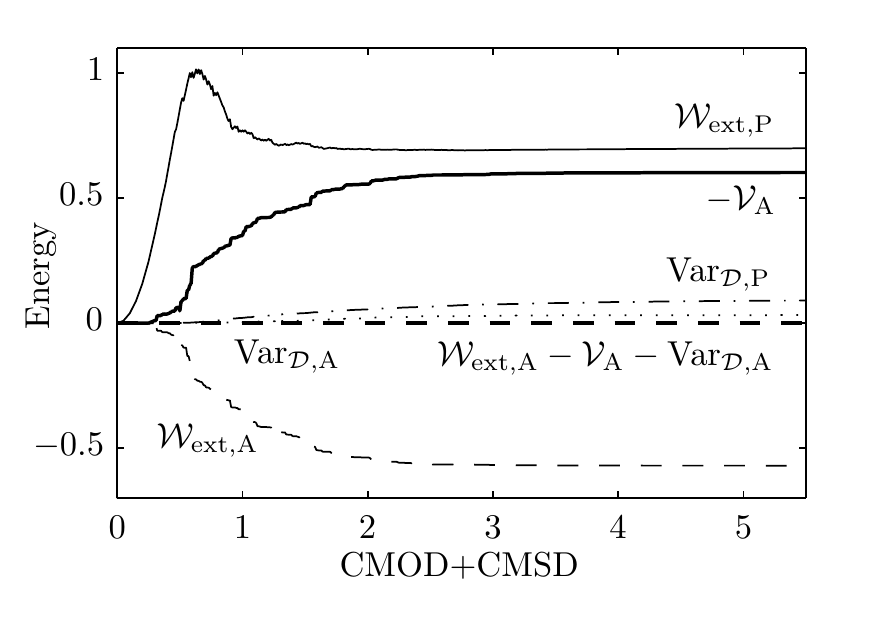}\label{SubSect:ComplexEx:Fig:4b}}
	\caption{Four-point bending test. (a)~Energy evolutions (black~-- the moderate adaptive QC, blue~-- the progressive adaptive QC, green~-- DNS, red~-- moderate X-QC, magenta~-- progressive X-QC). (b)~Energies exchanged during mesh refinement and coarsening for the moderate QC approach, cf. Section~\ref{SubSect:EnImplications}.}
	\label{SubSect:ComplexEx:Fig:4}
\end{figure}

The energy evolutions for all employed approaches are presented in Fig.~\ref{SubSect:ComplexEx:Fig:4a}. The energy balance~\eqref{E} clearly holds. One can also notice that the X-QC energies match the adaptive QC and DNS energies well. In Fig.~\ref{SubSect:ComplexEx:Fig:4b} the energy exchanges of the moderate X-QC approach are shown. Again, like in the previous example, their contributions are substantial.

\begin{figure}
	\centering
	\includegraphics[scale=1]{legend.pdf}\\
	\subfloat[relative number of repatoms]{\includegraphics[scale=1]{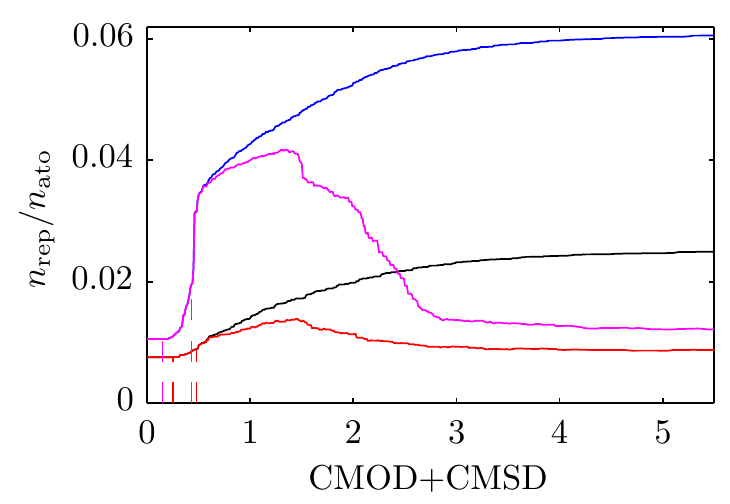}\label{SubSect:ComplexEx:Fig:5a}}\hspace{0.5em}
	\subfloat[relative number of sampling atoms]{\includegraphics[scale=1]{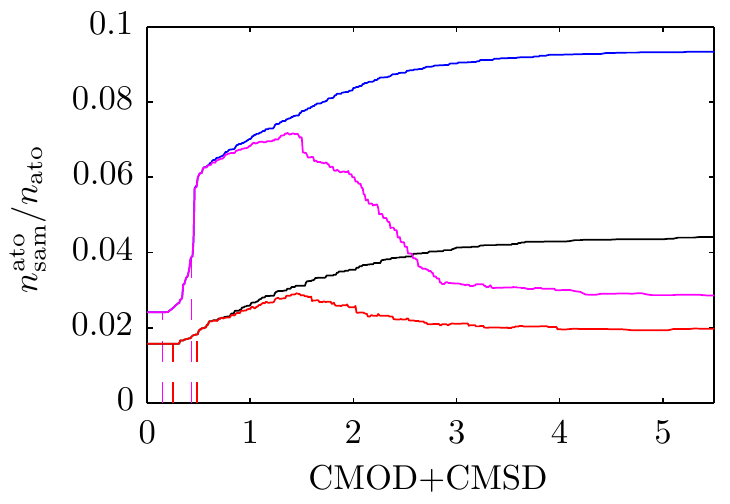}\label{SubSect:ComplexEx:Fig:5b}}
	\caption{(a)~Relative number of repatoms~$n_\mathrm{rep}/n_\mathrm{ato}$ as a function of~$\mathrm{CMOD}+\mathrm{CMSD}$, and (b)~relative number of sampling atoms.}
	\label{SubSect:ComplexEx:Fig:5}
\end{figure}

\begin{figure}
	\centering
	\subfloat[moderate QC, $\mathrm{CMOD}+\mathrm{CMSD} = 0$]{\includegraphics[scale=0.5]{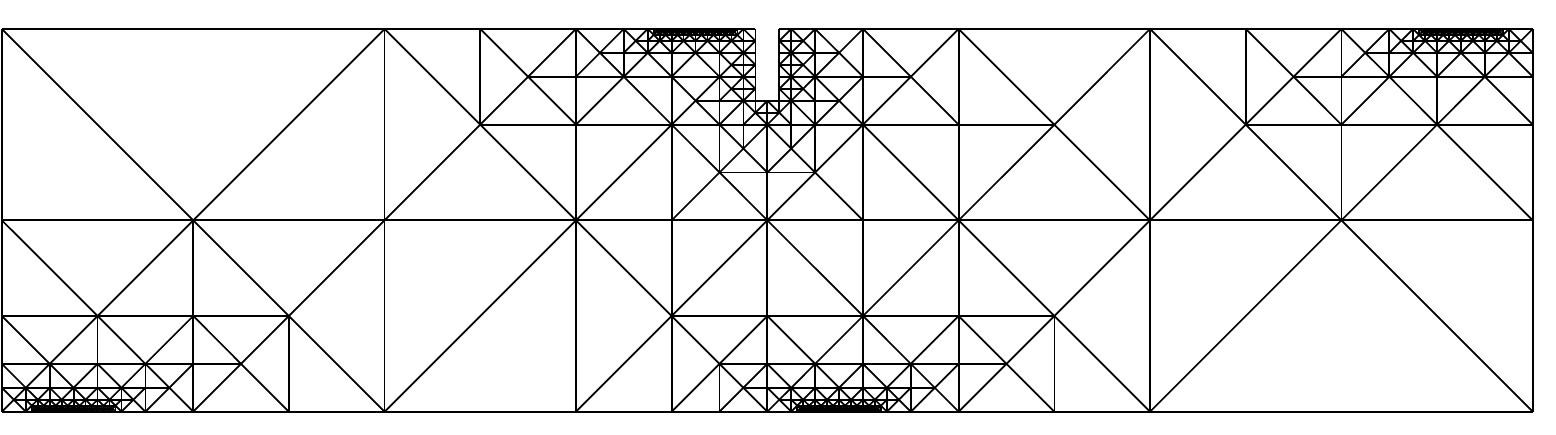}\label{SubSect:ComplexEx:Fig:6a}}\hspace{0.2em}
	\subfloat[moderate X-QC, $\mathrm{CMOD}+\mathrm{CMSD} = 0$]{\includegraphics[scale=0.5]{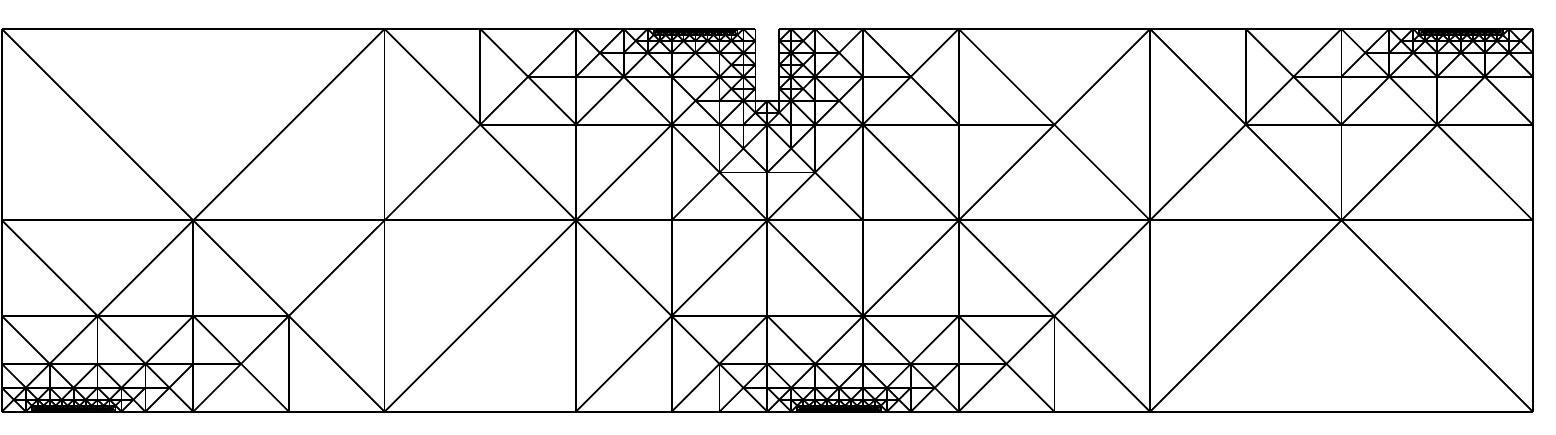}\label{SubSect:ComplexEx:Fig:6e}}\\
	\subfloat[moderate QC, $\mathrm{CMOD}+\mathrm{CMSD} = 1$]{\includegraphics[scale=0.5]{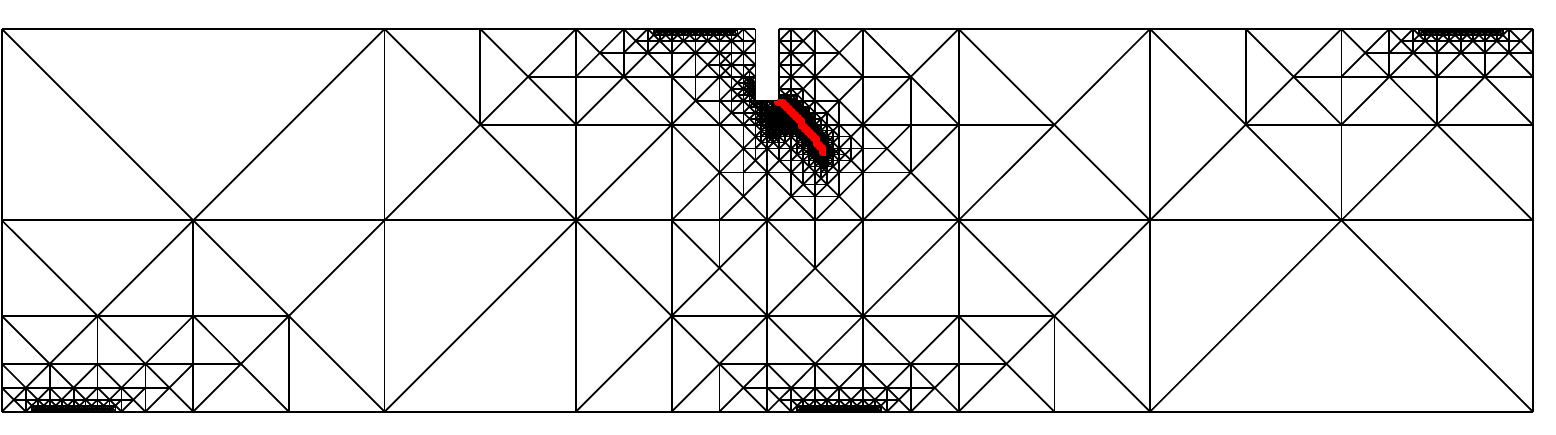}\label{SubSect:ComplexEx:Fig:6b}}\hspace{0.2em}
	\subfloat[moderate X-QC, $\mathrm{CMOD}+\mathrm{CMSD} = 1$]{\includegraphics[scale=0.5]{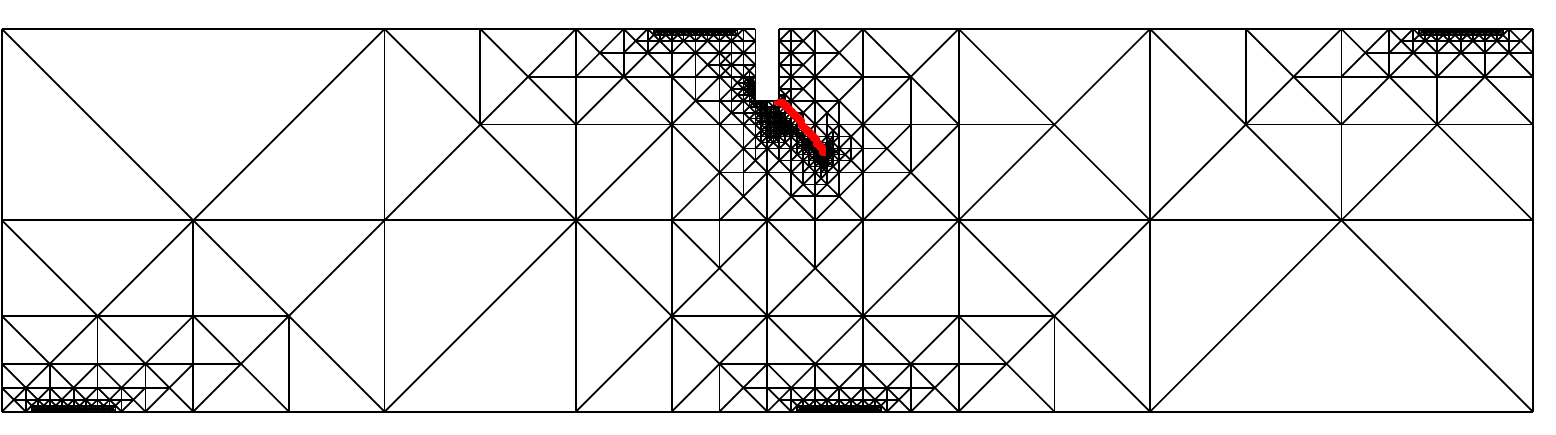}\label{SubSect:ComplexEx:Fig:6f}}\\
	\subfloat[moderate QC, $\mathrm{CMOD}+\mathrm{CMSD} = 2$]{\includegraphics[scale=0.5]{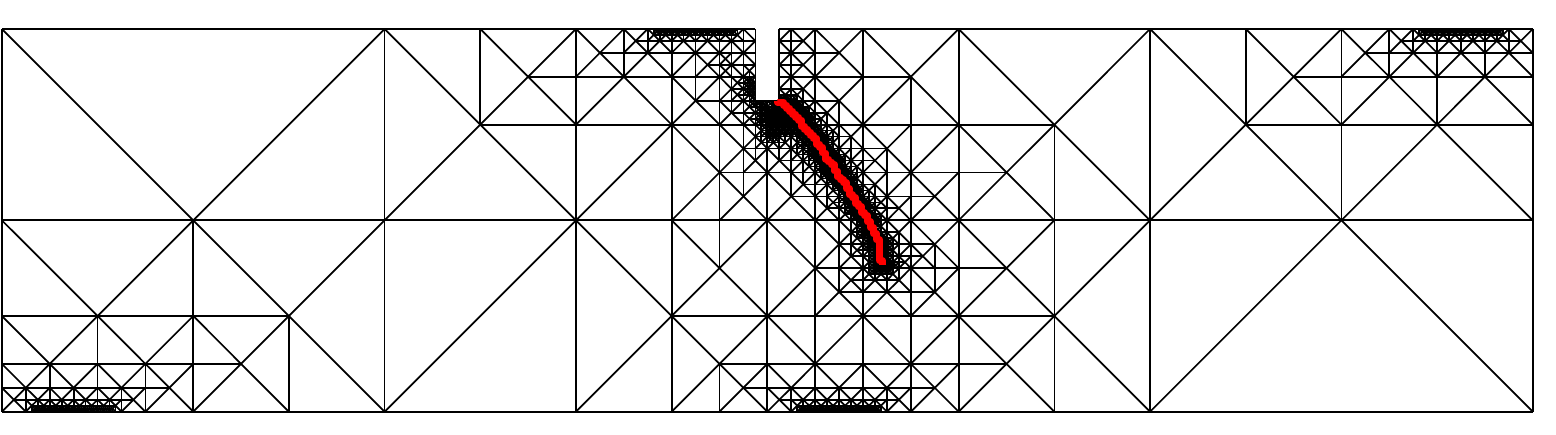}\label{SubSect:ComplexEx:Fig:6c}}\hspace{0.2em}
	\subfloat[moderate X-QC, $\mathrm{CMOD}+\mathrm{CMSD} = 2$]{\includegraphics[scale=0.5]{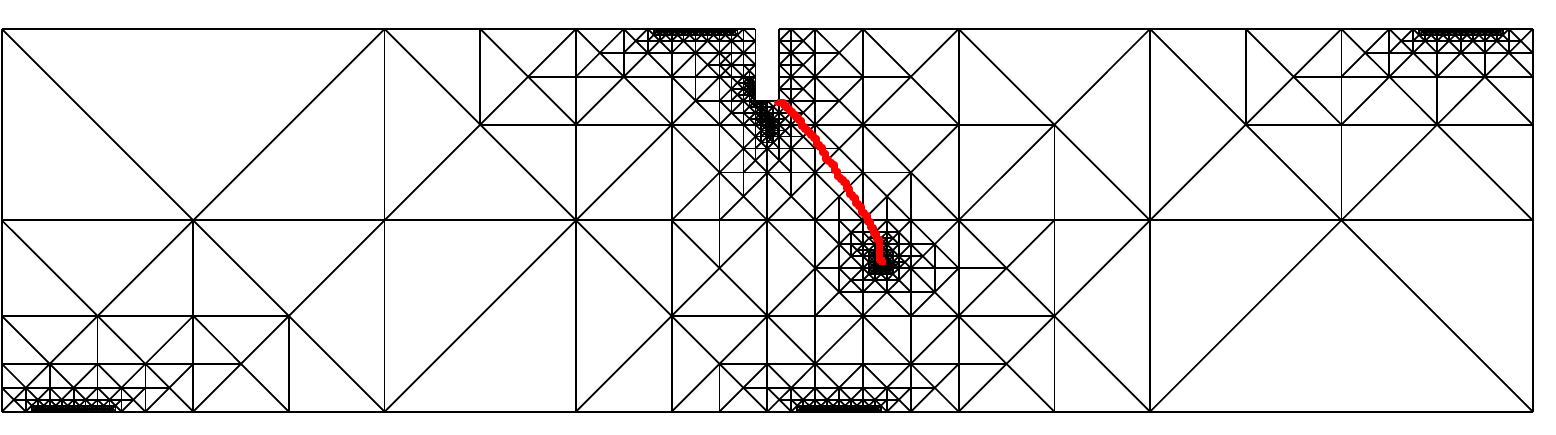}\label{SubSect:ComplexEx:Fig:6g}}\\
	\subfloat[moderate QC, $\mathrm{CMOD}+\mathrm{CMSD} = 5.5$]{\includegraphics[scale=0.5]{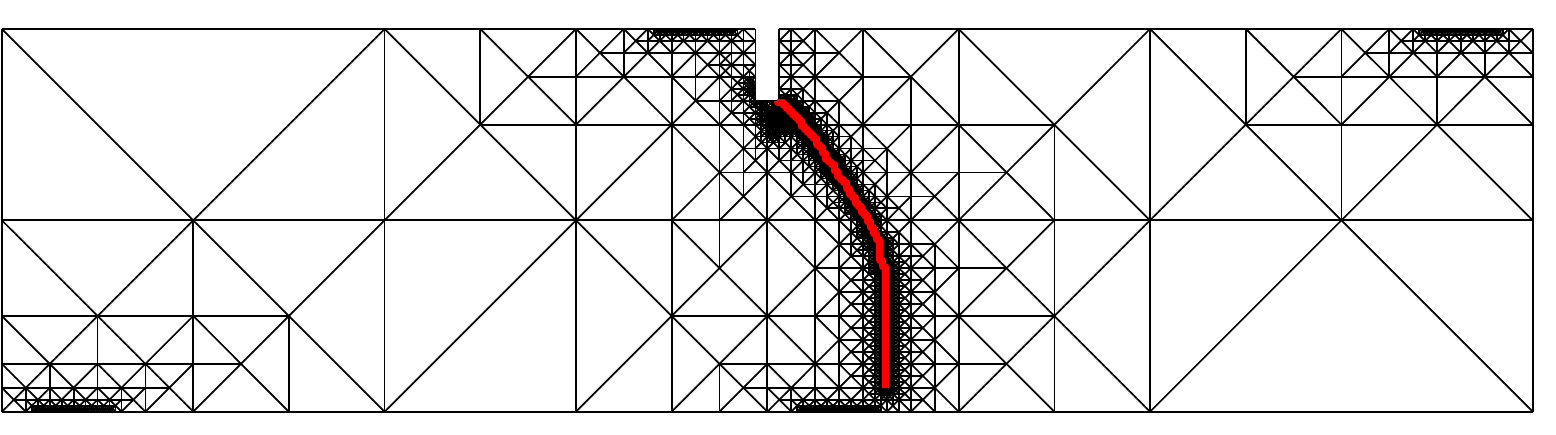}\label{SubSect:ComplexEx:Fig:6d}}\hspace{0.2em}
	\subfloat[moderate X-QC, $\mathrm{CMOD}+\mathrm{CMSD} = 5.5$]{\includegraphics[scale=0.5]{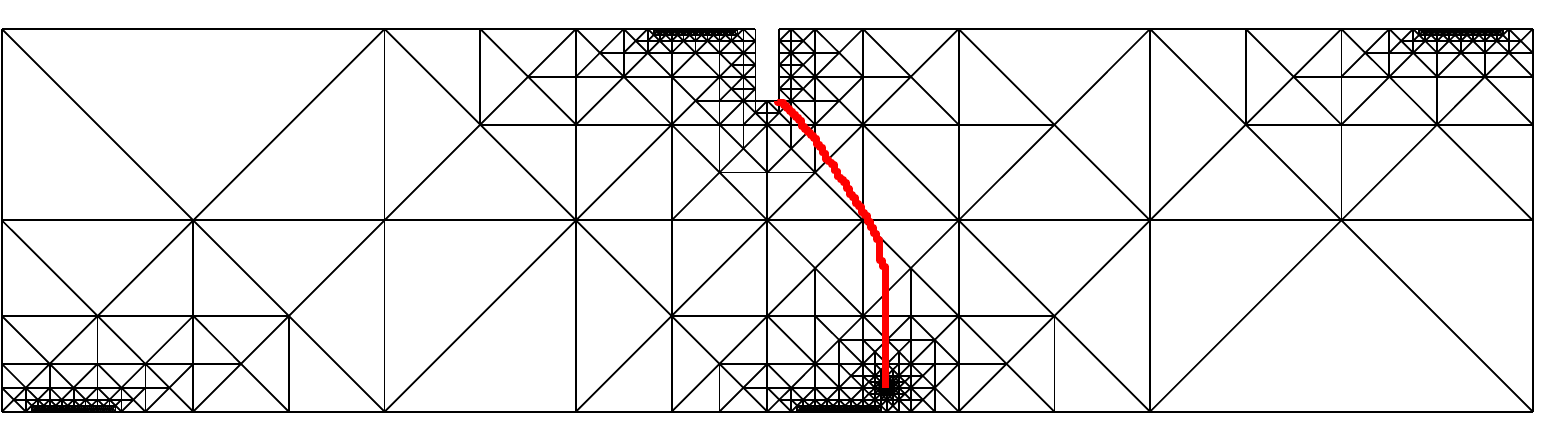}\label{SubSect:ComplexEx:Fig:6h}}
	\caption{Eight triangulations for four-point bending test: (a), (c), (e), (g) correspond to the moderate adaptive QC approach, and~(b), (d), (f), (h) to the moderate X-QC approach. The red line indicates the crack. The relative numbers of repatoms and sampling atoms are shown in Fig.~\ref{SubSect:ComplexEx:Fig:5}.}
	\label{SubSect:ComplexEx:Fig:6}
\end{figure}

\begin{figure}
	\centering
	\subfloat[progressive QC, $\mathrm{CMOD}+\mathrm{CMSD} = 0$]{\includegraphics[scale=0.5]{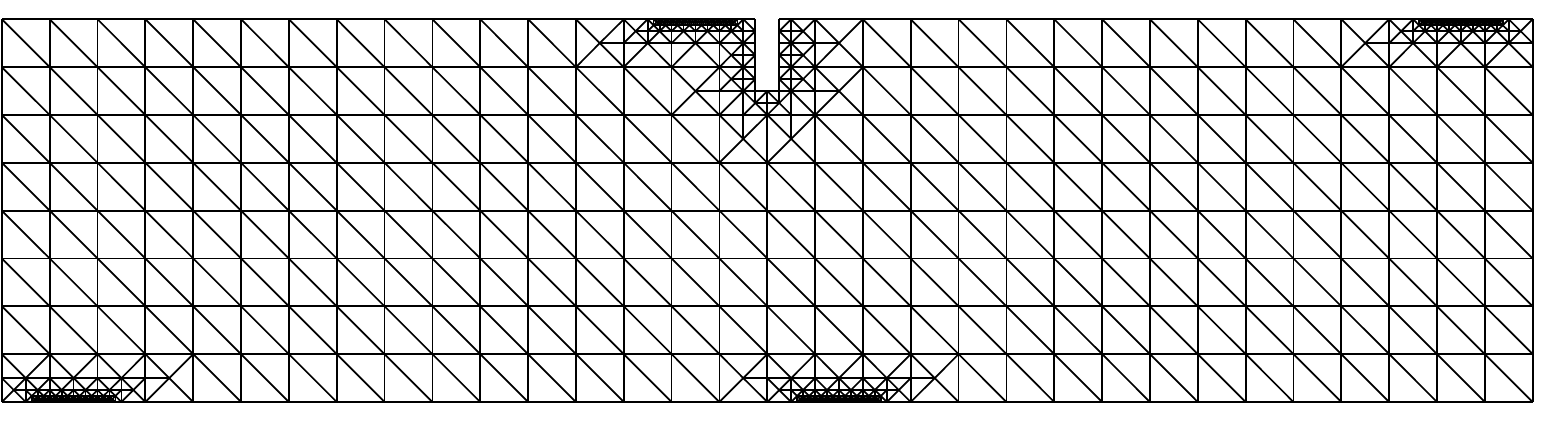}\label{SubSect:ComplexEx:Fig:7a}}\hspace{0.2em}
	\subfloat[progressive X-QC, $\mathrm{CMOD}+\mathrm{CMSD} = 0$]{\includegraphics[scale=0.5]{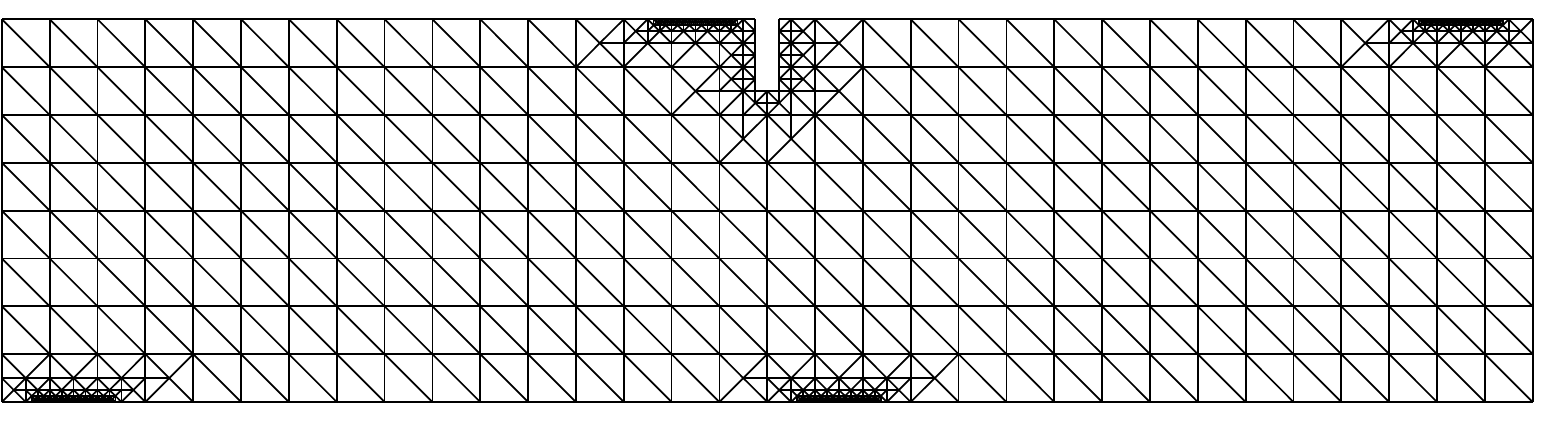}\label{SubSect:ComplexEx:Fig:7e}}\\
	\subfloat[progressive QC, $\mathrm{CMOD}+\mathrm{CMSD} = 1$]{\includegraphics[scale=0.5]{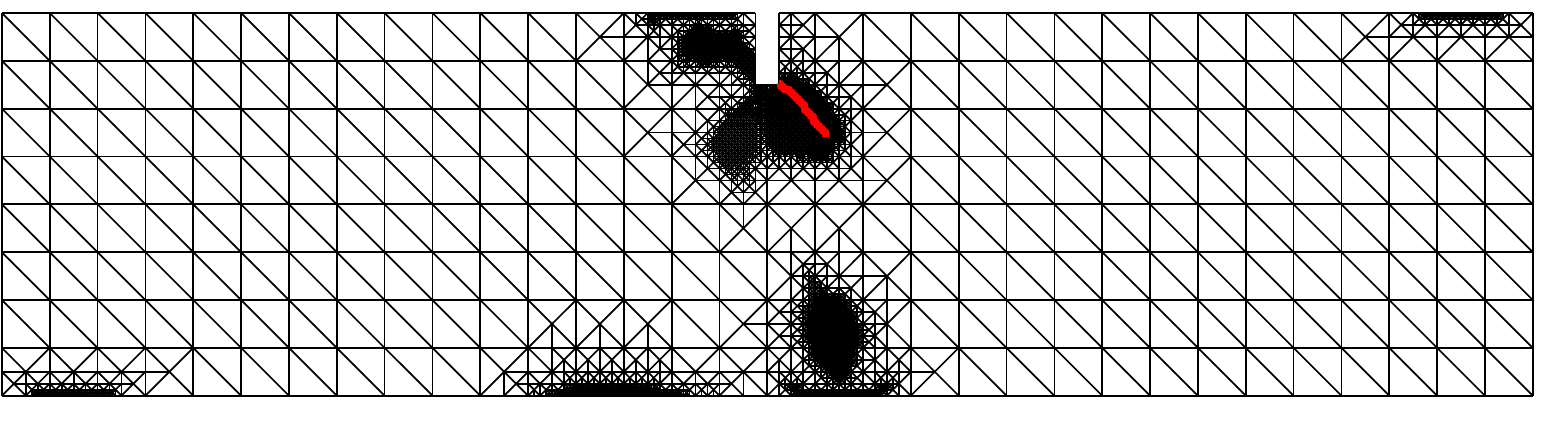}\label{SubSect:ComplexEx:Fig:7b}}\hspace{0.2em}
	\subfloat[progressive X-QC, $\mathrm{CMOD}+\mathrm{CMSD} = 1$]{\includegraphics[scale=0.5]{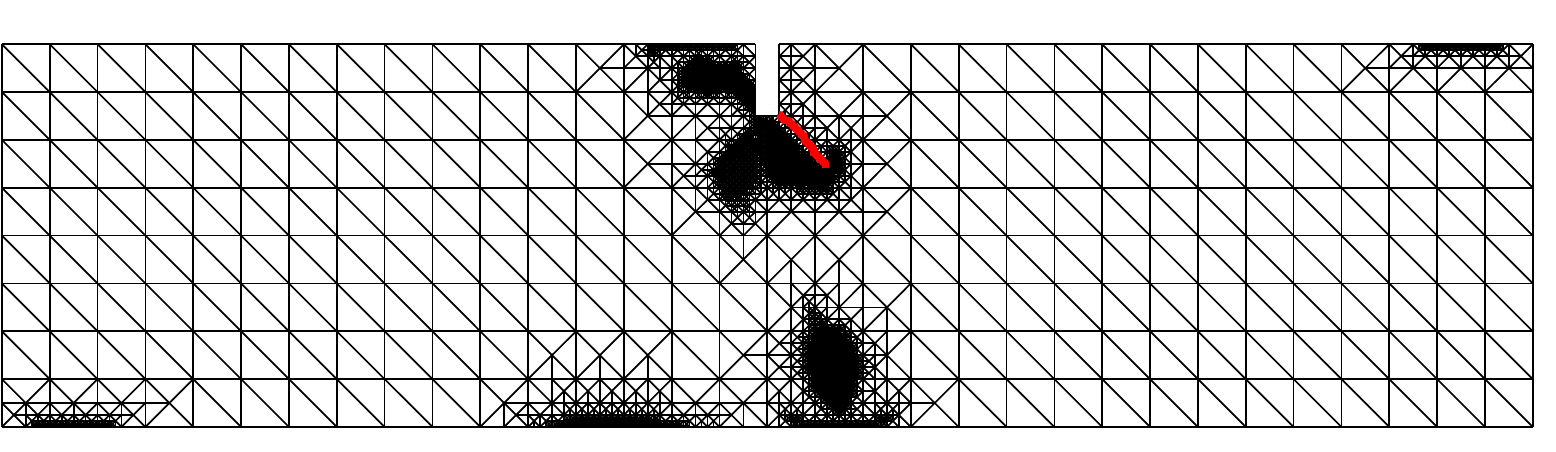}\label{SubSect:ComplexEx:Fig:7f}}\\
	\subfloat[progressive QC, $\mathrm{CMOD}+\mathrm{CMSD} = 2$]{\includegraphics[scale=0.5]{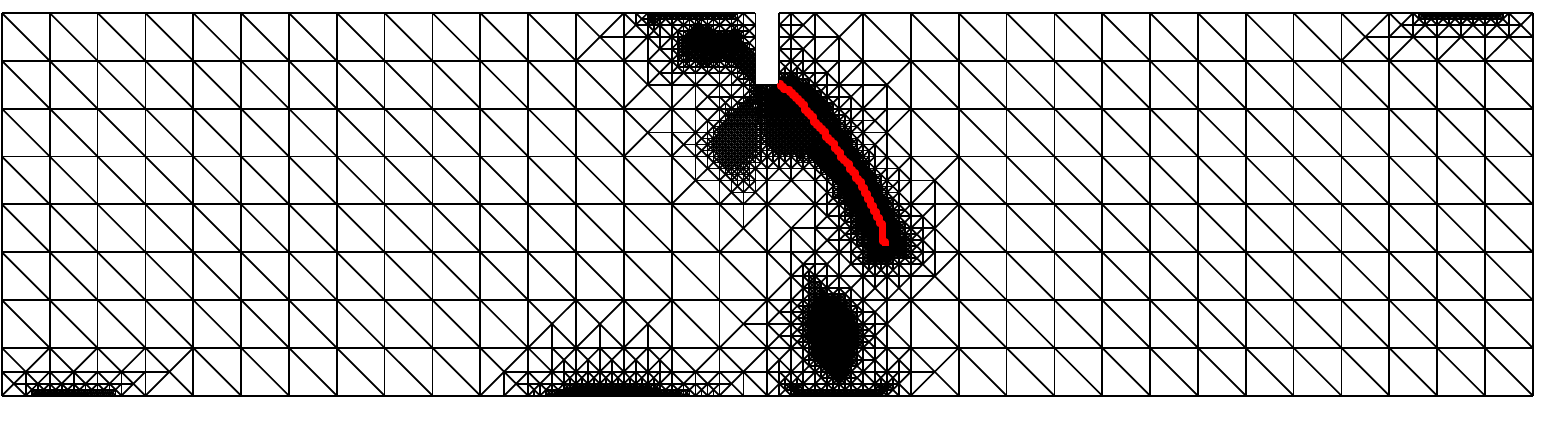}\label{SubSect:ComplexEx:Fig:7c}}\hspace{0.2em}
	\subfloat[progressive X-QC, $\mathrm{CMOD}+\mathrm{CMSD} = 2$]{\includegraphics[scale=0.5]{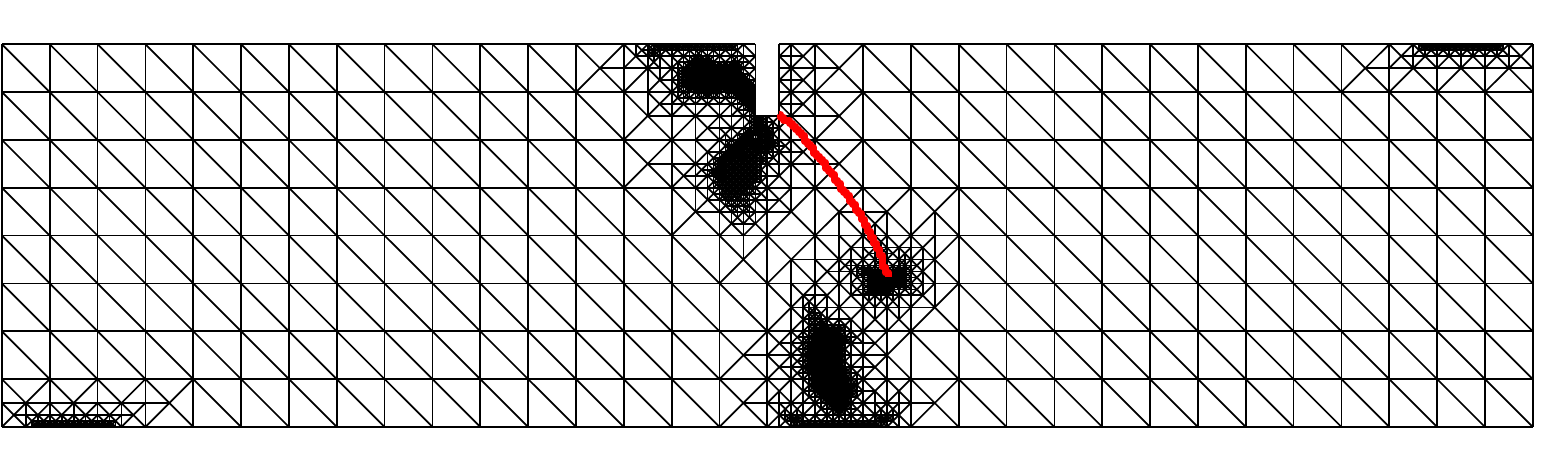}\label{SubSect:ComplexEx:Fig:7g}}\\
	\subfloat[progressive QC, $\mathrm{CMOD}+\mathrm{CMSD} = 5.5$]{\includegraphics[scale=0.5]{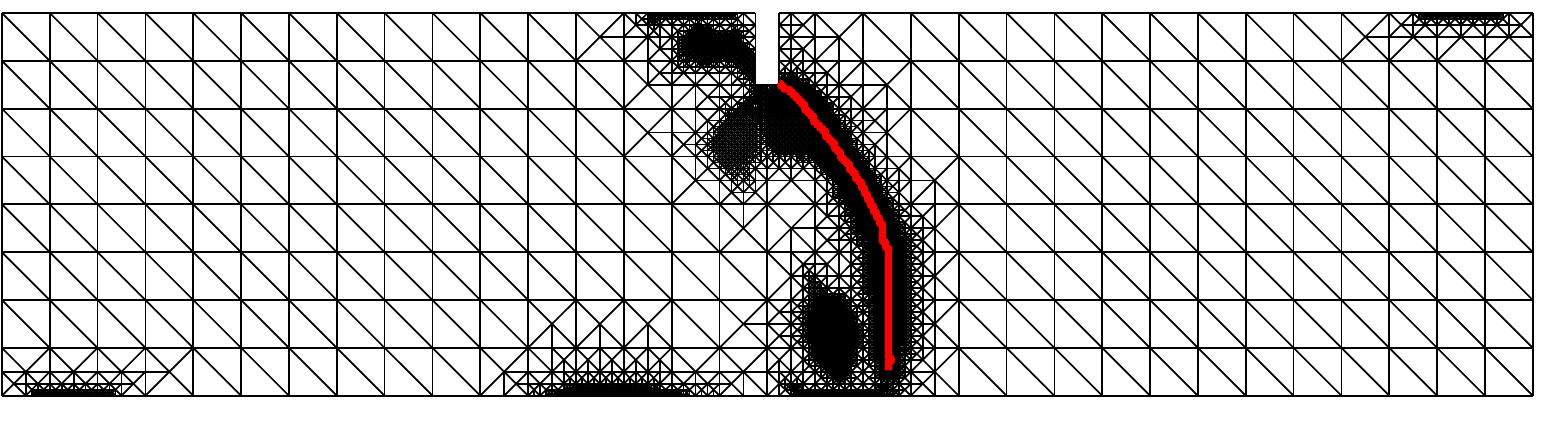}\label{SubSect:ComplexEx:Fig:7d}}\hspace{0.2em}
	\subfloat[progressive X-QC, $\mathrm{CMOD}+\mathrm{CMSD} = 5.5$]{\includegraphics[scale=0.5]{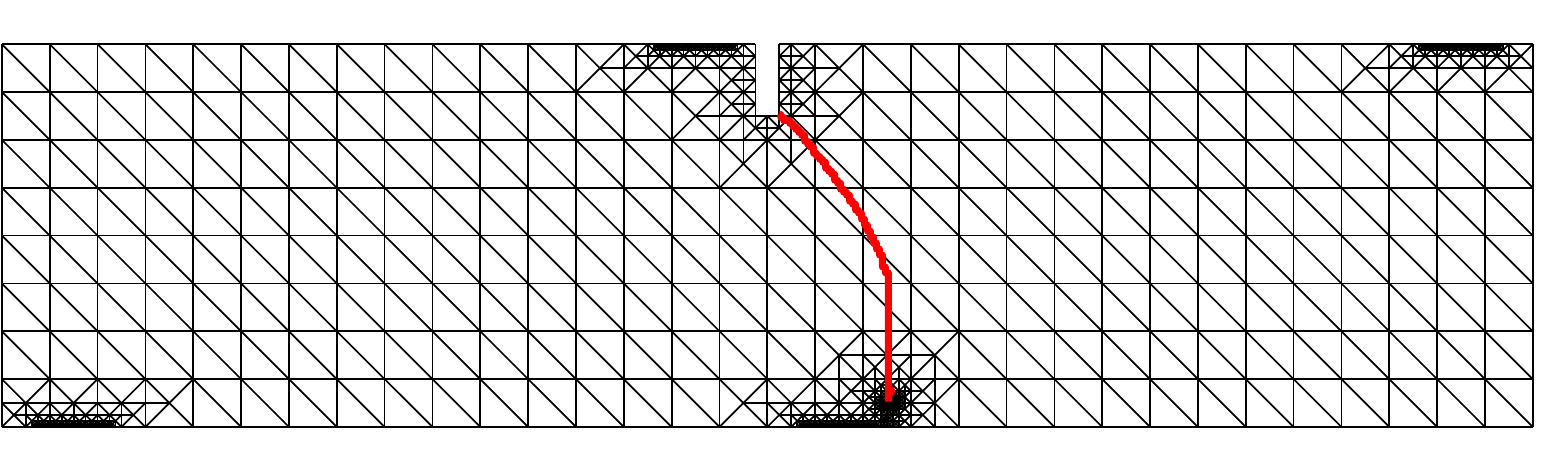}\label{SubSect:ComplexEx:Fig:7h}}
	\caption{Eight triangulations for four-point bending test: (a), (c), (e), (g) correspond to the adaptive progressive QC approach, and~(b), (d), (f), (h) to the progressive X-QC approach. The red line indicates the crack. The relative numbers of repatoms and sampling atoms are shown in Fig.~\ref{SubSect:ComplexEx:Fig:5}.}
	\label{SubSect:ComplexEx:Fig:7}
\end{figure}

In Fig.~\ref{SubSect:ComplexEx:Fig:5a}, the relative number of repatoms of the QC schemes are shown as a function of~$\mathrm{CMOD}+\mathrm{CMSD}$. The number of enriched repatoms is again negligible compared to the number of repatoms ($n^\star/n_\mathrm{rep} < 0.03$). Before the crack localizes, the number of repatoms rapidly increases. But it drops significantly after crack localization, and converges to approximately~$1\,\%$ for both X-QC approaches. Similar behaviour can be observed for the relative number of sampling atoms shown in Fig.~\ref{SubSect:ComplexEx:Fig:5b}. This result, in combination with the accuracy captured by Figs.~\ref{SubSect:ComplexEx:Fig:3} and~\ref{SubSect:ComplexEx:Fig:4a}, clearly demonstrates the strength of the X-QC framework. Finally, in Figs.~\ref{SubSect:ComplexEx:Fig:6} and~\ref{SubSect:ComplexEx:Fig:7} several snapshots capturing the evolution of the triangulations~$\mathcal{T}_k$ are presented. 
%
%
%
\section{Summary and Conclusion}
\label{Sect:Conclusion}
In this contribution, an adaptive mesh algorithm for the variational QC for lattice networks with localized damage is developed, which refines the mesh where necessary, but also coarsens it where possible. To achieve the latter, XFEM-like enrichment functions are required. It is shown that the efficiency of the QC framework applied to lattices with localised damage increases significantly. The main results can be summarized as follows:
\begin{enumerate}
	\item The two QC steps, interpolation and summation, were generalized to incorporate special enrichment functions, reflecting cracks within the framework of the fully-nonlocal variational QC.

	\item To determine regions to be coarsened, a heuristic marking strategy was proposed with a variable parameter controlling the mesh coarseness.

	\item The mesh refinement and coarsening were discussed from an energetic point of view.

	\item For the investigated examples the extended QC methodology, including coarsening, reduces the number of repatoms and sampling atoms with at least~$50\,\%$ of that of a simpler adaptive QC method with refinement only. The accuracy remains the same, however.
\end{enumerate}

One can note that the employed coarsening procedure neglected the diffusive dissipation. This was justified for the application to localized damage, but if more ductile behaviour is modelled (due to elastic-plastic models with damage), this must be incorporated. Therefore, a further generalization is required. Moreover, instead of the heuristic marking strategies used here, goal-oriented error estimators may be used to further improve the accuracy and efficiency of the QC methodology. The variational foundation of the method provides an ideal platform for this, cf. e.g.~\cite{Radovitzky:1999}. The same X-QC framework can also be employed for the description of heterogeneities in otherwise homogeneous lattice networks. These aspects are, however, outside the scope of this contribution and will be reported separately.
%
%
%
%
%
%
%
\section*{Acknowledgements}
This work was supported by the Czech Science Foundation (GA\v{C}R), through project No.~14-00420S. In addition, JZ acknowledges a partial support by the Czech Science Foundation, through project No.~16-34894L.
%
%
%


\end{document}